\documentclass[letterpaper]{article} 
\usepackage{aaai2026}  
\usepackage{times}  
\usepackage{helvet}  
\usepackage{courier}  
\usepackage[hyphens]{url}  
\usepackage{graphicx} 
\usepackage{natbib}  
\usepackage{caption} 
\usepackage{algorithm}

\usepackage{caption}
\usepackage{subcaption}
\usepackage{amsmath}
\usepackage{amssymb}
\usepackage{multirow}
\usepackage{array}
\usepackage{algpseudocode}
\usepackage{booktabs}
\usepackage{tabularx}
\usepackage{makecell}

\usepackage{threeparttable}

\usepackage{tikz} 
\usetikzlibrary{positioning} 

\usepackage{newfloat}
\usepackage{listings}
\DeclareCaptionStyle{ruled}{labelfont=normalfont,labelsep=colon,strut=off} 
\floatstyle{ruled}
\newfloat{listing}{tb}{lst}{}
\floatname{listing}{Listing}
\title{Decoding Cortical Microcircuits: A Generative Model for Latent Space Exploration and Controlled Synthesis}
\author {
    Xingyu Liu\textsuperscript{\rm 1},
    Yubin Li\textsuperscript{\rm 2},
    Guozhang Chen\textsuperscript{\rm 1}
}
\affiliations {
    \textsuperscript{\rm 1}National Key Laboratory for Multimedia Information Processing,\\ School of Computer Science, Peking University, Beijing, China\\
    \textsuperscript{\rm 2}Yuanpei College, Peking University, Beijing, China\\
    \textsuperscript{\rm 2}School of Computer Science and Engineering, University of Electronic Science and \\Technology of China, Chengdu, Sichuan, China\\
    guozhang.chen@pku.edu.cn
}

\usepackage{bibentry}

\begin{document}

\maketitle

\begin{abstract}
A central idea in understanding brains and building brain neural networks is that structure determines function. However, the brain's connectome is a massively high-dimensional graph, making the direct investigation of its structure-function relationships computationally intractable. Therefore, identifying a compact, low-dimensional representation that captures the connectome's essential structural organization is crucial for elucidating these relationships. The existence of such a representation is biologically plausible: the "genomic bottleneck" theory provides a strong basis for such a compressed developmental blueprint. We introduce a generative model to learn this underlying representation from detailed connectivity maps of mouse cortical microcircuits. Our model successfully captures the essential structural information of these circuits within a compressed latent space. We then associate specific network structures, as encoded in this latent space, with computational functions using reservoir computing tasks. Building on this, our methodology allows for the controllable generation of novel, synthetic microcircuits with desired structural features by navigating the learned latent space. This research paradigm establishes a computational testbed to systematically investigate the brain's inherent structure-function relationships. The ability to generate diverse, bio-plausible circuits could inform the development of more advanced artificial neural networks.
\end{abstract}


\section{Introduction}
The relationship between structure and function is a core idea in both neuroscience and the development of artificial intelligence~\cite{Gradientsofstructure-functiontetheringacrossneocortex,NAS1,NAS2}. The brain, with its extraordinary capabilities, inspires us to build better AI systems~\cite{neuronAI1,neuronAI2,hassabis2017,lecun2015}. The maps of brain connections are called \textit{connectomes}~\cite{connectome_def1,connectome_def2_original,connectome_def3}, whose structure is incredibly complex~\cite{complex_connectome1,complex_connectome2}. Studying connectomes is important because they hold clues about how the brain processes information and learns~\cite{HCP,USbrain1}. However, given that the brain's connectome is a massively high-dimensional graph, it is impractical to directly investigate which structures within the connectome itself yield specific functional performance. Therefore, the objective of this paper is to find a low-dimensional representation that captures the key features of the connectome. This new paradigm leverages this low-dimensional representation as a novel tool to investigate structure-function relationships.

The existence of such a low-dimensional representation is also supported by the 'genomic bottleneck' theory. An animal's genes guide the development of its nervous system~\cite{Genetic_influences_on_hub_connectivity_of_the_human_connectome,Genetic_architecture_of_the_structural_connectome}. However, the amount of information genes can hold is much smaller than what would be needed to explicitly list every single connection in a fully formed brain~\cite{info_capacity1,info_capacity2}. This observation, known as the 'genomic bottleneck'~\cite{pnas_genetic_btnk}, suggests (without negating the combined influence of genetics and environment) that there must be a simpler, more compact set of rules or a low-dimensional representation that guides how brain networks grow and organize themselves, for which the finite information capacity of the genome provides primary evidence.

In this work, we focus on \textit{cortical microcircuits}, which are small, repeating patterns of cortical connections that can be thought of as fundamental building blocks and basic computational units of the brain~\cite{microcircuits_douglas,microcircuits_MILLER201675}. We primarily implement this new paradigm for investigating structure-function relationships using data from the MICrONS program~\cite{microns}, which has produced a large and detailed connectome dataset from the visual cortex of a mouse. This dataset provides an unprecedented opportunity to study the organization of these microcircuits. As a supplementary study, we also conduct similar investigations on the FlyWire connectome dataset from the fruit fly (See Appendix~\ref{sec:fly}).

\textbf{Our primary contributions are as follows:}
\begin{enumerate}
    \item We propose a new paradigm for investigating the structure-function relationships of the brain connectome, which we implement on connectome data from both mice and fruit flies. While the core paradigm remains consistent, the specific datasets and model architectures differ between species. Due to space constraints, this paper primarily details the model for the mouse connectome, with results from the fruit fly experiments presented in the Appendix~\ref{sec:fly}.
    \item We introduce a Variational Autoencoder (VAE)-based generative model specifically designed to learn a compressed latent representation of microcircuit topology from this mouse visual cortex data.
    \item Crucially, we demonstrate that specific aspects of this learned latent space show strong, understandable relationships with key structural properties of the microcircuits, such as how densely connected they are or how they form clusters. This means we can find meaningful ways to describe the core variations in circuit structure.
    \item Building on this, we propose a method for the controlled generation of microcircuits. By carefully adjusting these meaningful aspects in the latent space, our method can create new, artificial network structures that have specific desired characteristics. 
    \item We investigated the influence of connectivity patterns on network function across various tasks. Utilizing controllably generated networks, we constructed reservoir networks and demonstrated that networks generated by the VAE to emulate brain-like connectivity patterns exhibited enhanced task performance compared to randomly connected networks of the same density. Furthermore, we found that alterations in specific structural features had an approximately monotonic effect on the performance of certain tasks.
\end{enumerate}

To our knowledge, the application of VAE for controllable generation of high-resolution structural microcircuits and for the understanding of structural-function relationship has not been done before. Furthermore, understanding the brain's structure could eventually help in designing more efficient and capable artificial neural networks.

\section{Problem Definition}
Our overarching goal is to develop a computational framework that learns a low-dimensional generative representation of brain microcircuit topology. This representation is intended to serve as a new paradigm for investigating structure-function relationships, providing a basis for controllably generating novel microcircuits with desirable properties. We define two primary objectives:

\subsection{Learning Compact Generative Representations of Microcircuit Topology}
\label{sec:problem_representation_learning}

Given a dataset of $N$ biological microcircuit graphs, denoted as $\mathbf{G}_{\text{data}} = \{\mathcal{G}^{(1)}, \mathcal{G}^{(2)}, \dots, \mathcal{G}^{(N)}\}$, the first objective is to learn a low-dimensional latent representation $\mathbf{z}^{(i)} \in \mathbb{R}^{d_z}$ for each microcircuit $\mathcal{G}^{(i)}$. Each microcircuit is represented as a graph $\mathcal{G}^{(i)} = (\mathbf{X}^{(i)}, \mathbf{A}^{(i)})$, where $\mathbf{X}^{(i)} \in \mathbb{R}^{n_i \times d_v}$ is a matrix of node features for its $n_i$ neurons (with $d_v$ feature dimensionality), and $\mathbf{A}^{(i)} \in \{0,1\}^{n_i \times n_i}$ is its adjacency matrix indicating synaptic connections. In this work, since the number of nodes varies across the graph data, we pad each adjacency matrix to a fixed size of 100×100. For consistent processing, a canonical node ordering $\pi$ is assumed for inputs to certain model components, thus we may refer to an ordered graph as $\mathcal{G}_\pi^{(i)}$.

This learned latent representation $\mathbf{z}^{(i)}$ must be \textit{generative}. That is, we aim to learn a probabilistic model $p_{\theta}(\mathcal{G}_\pi|\mathbf{z})$ capable of generating realistic microcircuit graphs from these latent codes. The quality of this representation will be assessed by its ability to:
\begin{enumerate}
    \item Faithfully reconstruct observed graph structures and their topological properties from their latent codes.
    \item Generate novel, diverse graphs that capture the statistical characteristics of the biological training data.
\end{enumerate}

\subsection{Controllable Generation of Microcircuits with Target Properties}
\label{sec:problem_controlled_generation}

Building upon the learned latent space $\mathcal{Z}$ (the space of all $\mathbf{z}$) and the generative model $p_{\theta}(\mathcal{G}_\pi|\mathbf{z})$ from Section~\ref{sec:problem_representation_learning}, our second objective is to enable the \textit{controlled generation} of novel microcircuits. 

Specifically, given a target structural, dynamical, or functional property $P$ (e.g., a specific mean degree, clustering coefficient, or level of assortativity), and a desired value $t_{\text{target}}$ for this property, the goal is to synthesize new connectome graphs $\mathcal{G}_{\pi, \text{new}}$. These generated graphs should:
\begin{enumerate}
    \item Optimally satisfy the specified constraint, meaning the property $P$ for $\mathcal{G}_{\pi, \text{new}}$ should be close to $t_{\text{target}}$. We denote this constraint as $\mathcal{T}$.
    \item Preserve general topological and dynamical characteristics inherent to biological neural microcircuits, ensuring they remain plausible.
\end{enumerate}
Formally, this requires effectively sampling from, or being guided by, a conditional probability distribution $p(\mathbf{z}|\mathcal{T})$ in the latent space. Latent vectors $\mathbf{z}_{\text{new}}$ drawn from this distribution are then decoded using $p_{\theta}(\mathcal{G}_\pi|\mathbf{z}_{\text{new}})$ to produce the desired microcircuits.

\section{Related Work}

Analyzing the structural organization of brain connectomes is fundamental to understanding the intricate relationship between brain structure and function~\cite{ref1, ref2}. A number of methods have been used to address this link, including statistical models~\cite{statistical_models}, communication models~\cite{communication_model1}, and biophysical models~\cite{biophysical_models1,biophysical_models2}. By modeling brain networks as graphs, researchers employ graph-theoretic approaches to reveal key topological properties that underlie efficient communication and cognitive processes~\cite{ref3, ref4}. Understanding how these structural features relate to functional dynamics is a central goal in connectomics.

To investigate the principles governing connectome formation and organization, generative models have been developed. Early approaches typically relied on a small set of predefined wiring rules or biological principles, such as cost-efficiency or growth mechanisms, to replicate observed network features in silico~\cite{Generativemodelsofthehumanconnectome, Modellingthedevelopmentofcorticalsystemsnetworks, ASimpleGenerativeModel}. While successful in capturing certain global properties, these rule-based methods often lack flexibility and depend on manually specified or constrained generative factors, limiting their ability to explore the full complexity of biological variability.

More recently, deep generative models, such as Variational Autoencoders (VAEs), have emerged as powerful tools for analyzing and synthesizing complex data like brain networks~\cite{ref5, ref6}. These models are particularly well-suited for learning a low-dimensional latent representation of the network data, effectively performing dimensionality reduction. Furthermore, by manipulating or sampling from this learned latent space, these models enable the controlled synthesis of novel, biologically plausible network configurations, offering new avenues for exploring connectome variability and its functional implications~\cite{ref7, ref8}. Prior works have also applied graph VAEs to macro-scale functional connectomes for discriminative tasks or developmental modeling~\cite{graph_vae_BEHROUZI2022108202}.

\section{Method}
\label{sec:method}

\subsection{MICrONS Dataset and Preprocess}
 We adopt the IARPA MICrONS dataset~\cite{microns}, which encompasses a 1.4 × 0.87 × 0.84 mm volume of cortex from a P87 mouse, containing neurons within the area and their interconnections, which is the sole source for such high-resolution connectivity. Given the cortex's division into six distinct layers, each associated with a specific level of information processing~\cite{layers1,layers2,layer3}, our study concentrates on cortical microcircuits organized as vertically oriented functional columns that traverse all laminar layers~\cite{circuit1,circuit2}. To model these microcircuits, we extracted cylindrical subvolumes oriented perpendicularly to the cortical layers. Each cylindrical unit comprises 80–100 neurons (both excitatory and inhibitory neurons), preserving intra-column connectivity while excluding connections extending beyond the columnar boundary. Our VAE was trained on 3,285 circuits from MICRONS dataset. The test set was from a spatially disjoint region to prevent data leakage.

 Focusing only on network topology, we treat each circuit as a binary directed graph. We establish a canonical ordering rule, $\pi$, sorting neurons by their y-coordinates (reflecting cortical depth). This depth-based ordering is a deliberate choice, as it is both biologically significant (the y-coordinate inherently encodes the cortical layer) and technically crucial. Without this consistent ordering, determining the correct correspondence for element-wise comparison between original and generated adjacency matrices becomes intractable. Neuron type and synapse weight information are temporarily disregarded, as this work focuses solely on the topological structure.

 We applied a consistent methodology to the fruit fly connectome. While the overall paradigm of using a generative model to learn a low-dimensional topological representation remains the same, the specific VAE architecture was adapted to accommodate the different structural properties of the fly brain. 

\subsection{Connectome Graph Variational Autoencoder}
As the goal is to find latent representations similar to ``genomic bottleneck'', we need a generative model with ``information bottleneck''. Thus, adopting a variational autoencoder (VAE)~\cite{VAE} on the connectome data can be formulated as follows: 
Assume a set of training connectome graphs $\mathbf{G}=\{\mathcal{G}_\pi^{(i)}\}$ is generated from the distribution of a set of unobserved latent representation $\textbf{z}=(z_1,...,z_m)$, where $z^{(i)}\sim p_{\theta^*}(z)$ and training data is sampled from true conditional $p_{\theta^*}(\mathcal{G}_\pi|z^{(i)})$. The data generation process is denoted by $p_\theta(\mathcal{G}_\pi|\mathbf{z})$. Following the standard VAE setting~\cite{VAE}, we approximate the intractable posterior by $q_\phi(\mathbf{z}|\mathcal{G}_\pi)\approx p_\theta(\mathbf{z}|\mathcal{G}_\pi)$ and maximize the evidence lower bound on the marginal likelihood of graph $\mathcal{G}^{(i)}$:

\begin{align*}
& \quad \log p_{\theta}(\mathcal{G}_\pi^{(i)}) \geq\mathcal{L}(\phi,\theta;\mathcal{G}_\pi^{(i)}) \\
&= \mathbb{E}_{q_{\phi}(\mathbf{z}|\mathcal{G}_\pi^{(i)})}\left[\log p_{\theta}(\mathcal{G}_\pi^{(i)}|\mathbf{z})\right] - \beta\mathrm{KL}\left[q_{\phi}(\mathbf{z}|\mathcal{G}_\pi^{(i)})||p_{\theta}(\mathbf{z})\right],
\end{align*}

where $\beta$ is a hyperparameter to balance the reconstruction loss and KL-divergence loss during the training process~\cite{betaVAE}.

The proposed VAE model for connectome graphs comprises four main components. First, a node feature encoder employs a multi-layer, multi-head Graph Attention Network (GAT)~\cite{GAT,attentionisallyouneed} to compute an embedding vector for each node. Second, a graph global encoder, which is a transformer encoder augmented with a special token~\cite{BERT,pigvae,attentionisallyouneed}, extracts an embedding for the entire sequence of node embeddings. This extracted sequence embedding serves as the graph-level embedding, upon which a 32-dimensional mean and variance are computed. The reparameterization of a standard VAE is applied to this 32D latent embedding representing the whole graph. Third, a node feature decoder, which is a transformer decoder, takes the graph-level embedding as input to reconstruct node features necessary for subsequent edge prediction. Finally, the edge predictor utilizes these decoded node features to predict the edges of the graph.

The overall architecture of the VAE model is illustrated in Figure~\ref{fig:overall_structure}. The details of the 4 components are in Appendix~\ref{sec:model_details}. The detailed encoder structure is depicted in Appendix Figure~\ref{fig:encoder}, and the detailed decoder structure is depicted in Appendix Figure~\ref{fig:decoder}.

\begin{figure*}[htbp]
\centering
\includegraphics[width=0.7\textwidth]{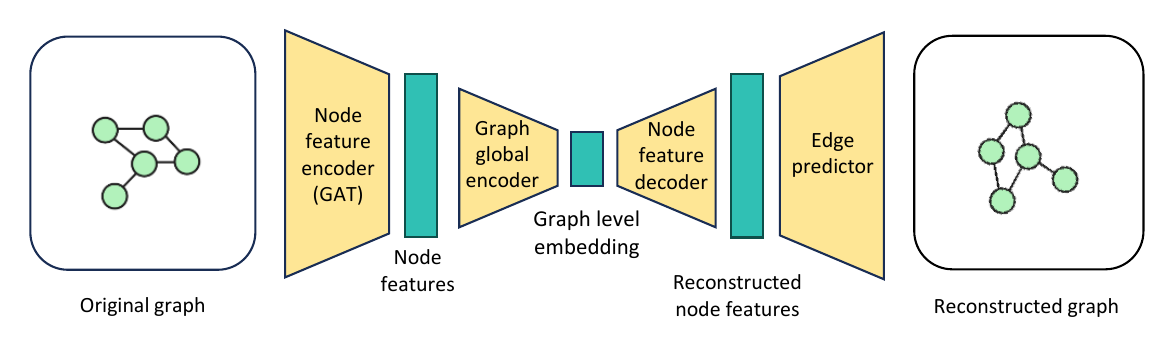} 
\caption{Overall model structure.}
\label{fig:overall_structure}
\end{figure*}

\subsection{Controllable Connectome Generation by Sampling from the Latent Space}
\label{sec:method_controlled_generation}

Investigating the intricate interplay between structure, dynamics, and function in brain neural microcircuits necessitates the capability to generate novel networks in a controlled fashion. Specifically, given a target network property $\mathcal{T}$ (which can be structural, dynamical, or functional), our objective is to synthesize novel connectomes $\mathcal{G}_\pi$ that optimally satisfy the specified constraint $\mathcal{T}$ while preserving general topological and dynamical characteristics inherent to biological neural microcircuits. Formally, the goal is to generate new samples from the conditional probability distribution $p(\mathcal{G}_\pi|\mathcal{T})$.

Given the computational cost associated with large-scale graph generation via brute-force sampling, we propose leveraging latent space properties to inform our sampling strategy. By employing geometric characteristics of the latent manifold as a sampling heuristic, this approach significantly reduces computational overhead compared to the brute-force generate-then-filter paradigm. Specifically, we aim to find the probability distribution of the latent vector $\mathbf{z}$ conditioned on the target property $\mathcal{T}$, denoted as $p(\mathbf{z}|\mathcal{T})$, which can be represented as an energy model:
\begin{equation}
p(\mathbf{z}|\mathcal{T})=\frac{1}{Z}p(\mathbf{z})^{{1}/{\tau}}\exp(\lambda S(\mathcal{T},\mathbf{z})),
\end{equation}
where $S(\mathcal{T},\mathbf{z})$ is a condition indicator function, defined as $S(\mathcal{T},\mathbf{z})=1 \text{ if } \mathbf{z}\in\Omega_\mathcal{T} \text{ else } 0$.
$\Omega_\mathcal{T}$ is a subset of latent space where the generated graphs are predicted to be close to the target. $Z=\int p(\mathbf{z})^{{1}/{\tau}}\exp(\lambda S(\mathcal{T},\mathbf{z}))d\mathbf{z}$ is the normalization factor, and $\lambda$ is a tuning parameter which determine the constraining strength of $\mathcal{T}$ and $\tau$ is a temperature parameter which balance exploitation and exploration based on prior latent distribution of $p(\mathbf{z})$.

When $\lambda\to\infty$, the energy model degenerates into a strictly constrained version (by $\mathcal{T}$). In this case we have:
\begin{equation}
p(\mathbf{z}|\mathcal{T})=\frac{p(\mathbf{z})^{{1}/{\tau}}}{p(\Omega_\mathcal{T})}\cdot\mathbb{I}(\mathbf{z}\in\Omega_\mathcal{T}),
\label{equation:condition_dist}
\end{equation}
where the new normalization factor becomes $p(\Omega_\mathcal{T})=\int_{\Omega_\mathcal{T}}p(\mathbf{z})^{{1}/{\tau}}d\mathbf{z}$.

\subsection{Reservoir Network with Structured Connectivity}
We define reservoir network~\cite{esn} with a $D_r$ dimensional reservoir $\mathbf{R}$, input projection parameterized by $W_{in}\in \mathbb{R}^{D_r\times D}$ and output projection parameterized by $W_{out}\in \mathbb{R}^{D\times D_r}$. Reservoir $\mathbf{R}$ stores a reservoir state $\mathbf{r}(t)$ with dimension $D_r$, and the recurrent connection is represented by an adjacency matrix $\mathbf{A}\in\mathbb{R}^{D_r\times D_r}$, whose elements are 0, 1 or -1, corresponding to non-connection, excitatory connection and inhibitory connection. We scale $\mathbf{A}$ to satisfy the pre-specified spectral radius $sr$. Given input vector $\mathbf{u}(t)$, the reservoir state is updated by $\mathbf{r}(t+1)=(1-\alpha)\mathbf{r}(t)+\alpha\cdot\tanh(\mathbf{A} \cdot \mathbf{r}(t) + W_{in} \cdot \mathbf{u}(t))$, where $\alpha$ is the leaking rate. $\mathbf{r}(t + 1)$ is then flowed to the output coupler, where a mapping is done to obtain $\mathbf{v}(t+1)=W_{out}\cdot\mathbf{r}(t+1)$. During the training process, we kept the recurrent connection weights of the reservoir fixed, training only the input and readout layers. This approach allows us to study the computational functions that arise from the connection structure itself.

\section{Experiments}
\label{sec:experiments}
\subsection{Evaluation Metrics}
We selected several graph theory metrics (descriptors) commonly employed in the analysis of connectomes~\cite{brain_graph_metrics}, including mean degree, efficiency, transitivity, clustering coefficient, modularity, and assortativity. The detailed definitions of these metrics are provided in the Appendix~\ref{sec:graph_metrics}.

\subsection{Microcircuit Reconstruction and Generation Results}
Examples of original and reconstructed graphs obtained through the VAE are provided in the Appendix~\ref{sec:recon_results}. To validate the model's generative capabilities, we evaluated several graph metrics on the generated samples. We employed the maximum mean discrepancy (MMD)~\cite{MMD} to compare the distributions of these graph statistics between an equal number of generated and test graphs. Following the standard protocol established by GraphRNN~\cite{you2018b_mmd}, we specifically measured the distributions of degree, clustering coefficient and spectrum. For MMD computation, we utilized both the Gaussian Earth Mover's Distance (EMD) kernel and the total variation (TV) distance~\cite{liao2019}. Given the absence of existing generative models specifically designed for connectome graphs, we selected three alternative models commonly used for molecule generation or general graph generation~\cite{GDSS, DisCo, EDGE} as baselines. The following Table~\ref{tab:comparison} reports the MMD values for different graph metrics evaluated across these models and our proposed approach. To further check the diversity and novelty of the generated results, we verified that 2,000 generated samples are all unique among themselves and not present in the training data. 

\begin{figure*}[htbp]
    \centering
    \begin{subfigure}[b]{0.09\textwidth}
        \includegraphics[width=\textwidth]{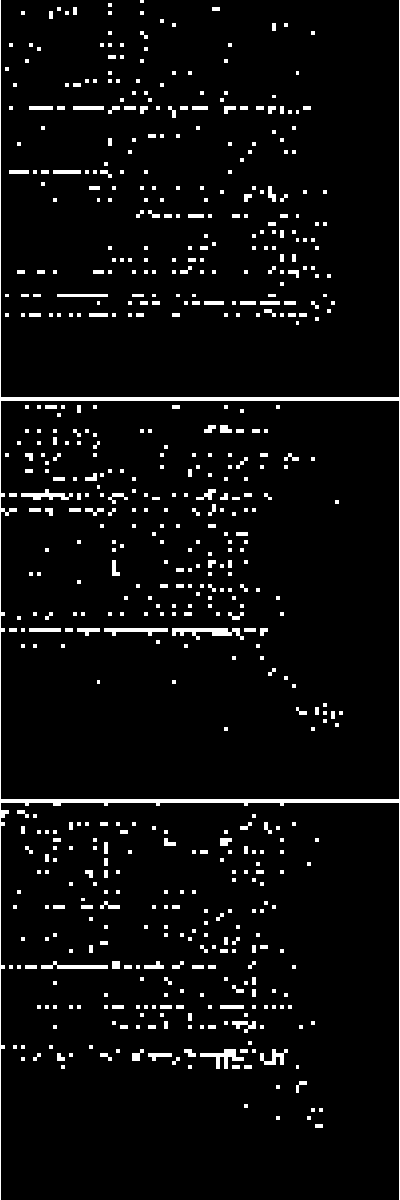}
        \caption{Ori.}
        \label{fig:ori_adj}
    \end{subfigure}
    \begin{subfigure}[b]{0.09\textwidth}
        \includegraphics[width=\textwidth]{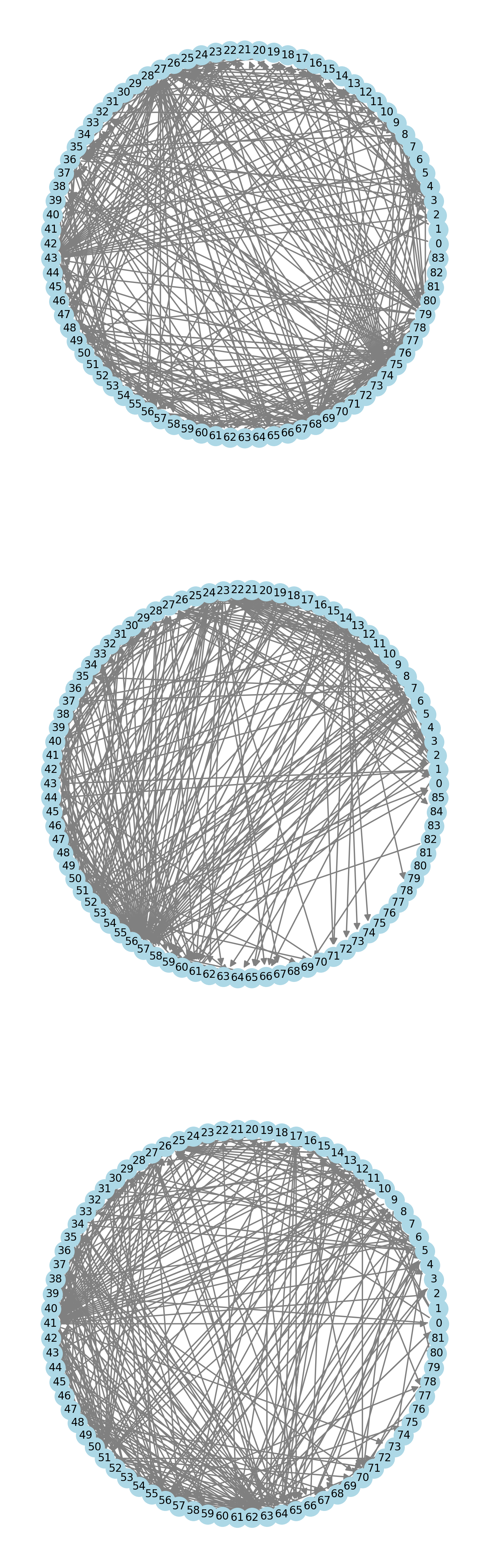}
        \caption{Ori.}
        \label{fig:ori_net}
    \end{subfigure}
    \begin{subfigure}[b]{0.09\textwidth}
        \includegraphics[width=\textwidth]{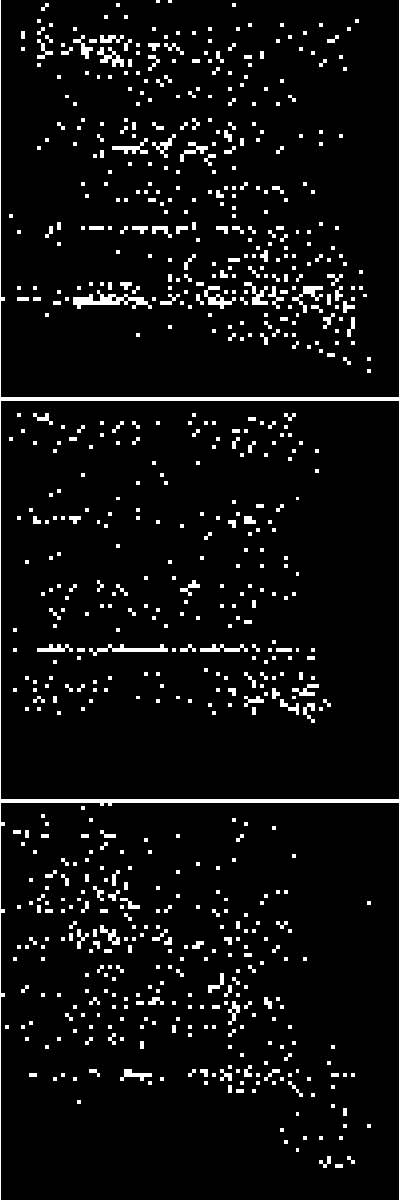}
        \caption{Ours}
        \label{fig:ours_adj}
    \end{subfigure}
    \begin{subfigure}[b]{0.09\textwidth}
        \includegraphics[width=\textwidth]{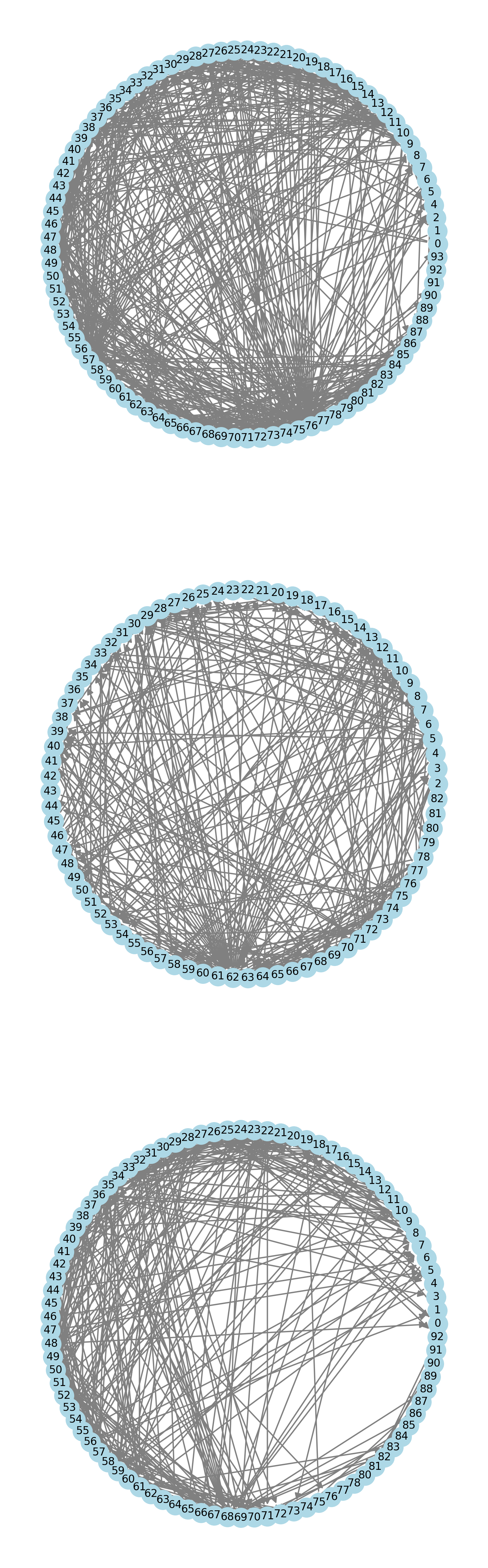}
        \caption{Ours}
        \label{fig:ours_net}
    \end{subfigure}
    \begin{subfigure}[b]{0.09\textwidth}
        \includegraphics[width=\textwidth]{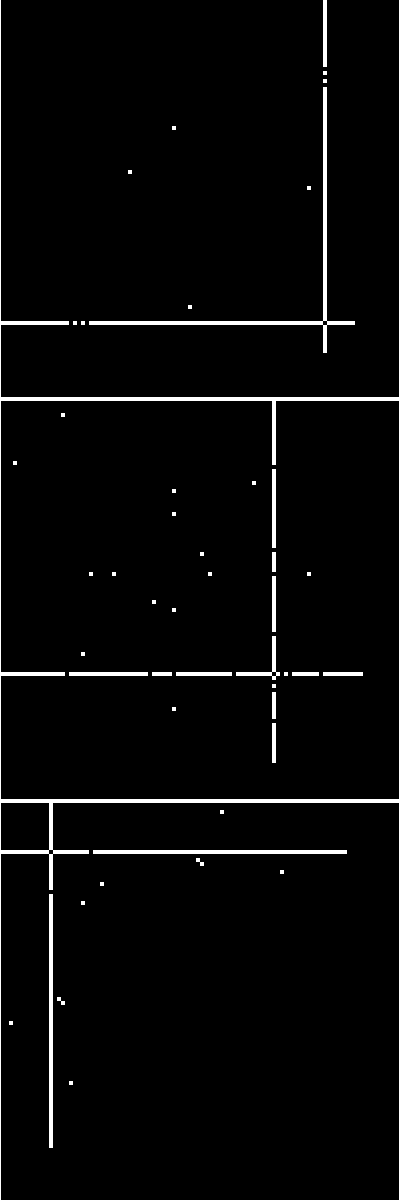}
        \caption{GDSS}
        \label{fig:gdss_adj}
    \end{subfigure}
    \begin{subfigure}[b]{0.09\textwidth}
        \includegraphics[width=\textwidth]{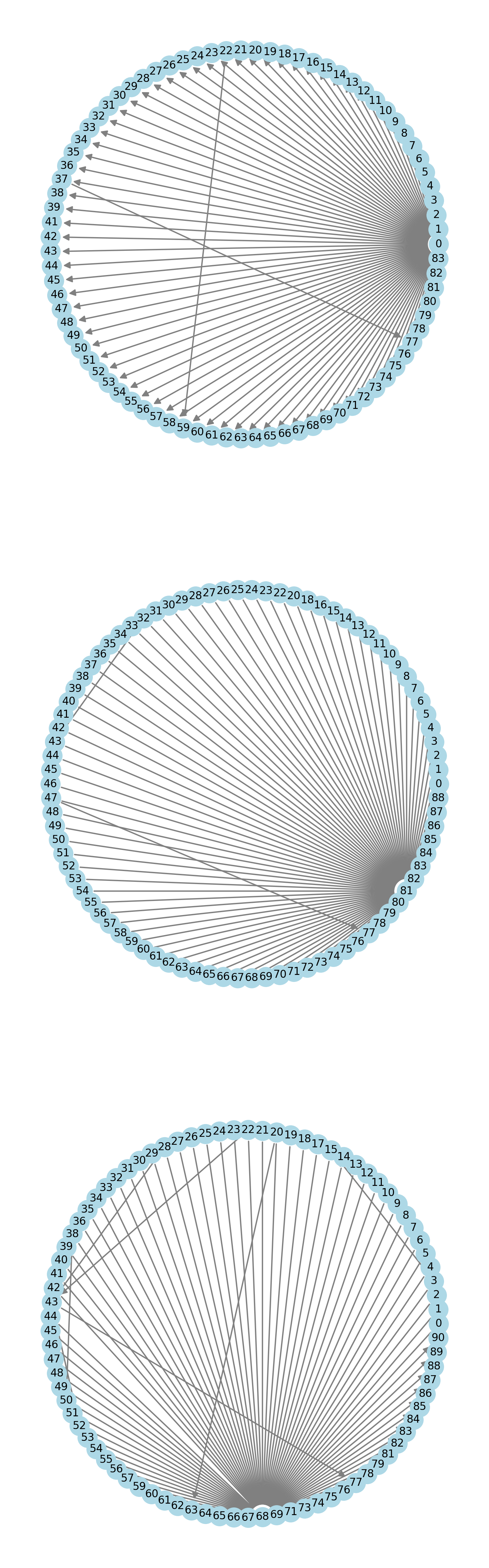}
        \caption{GDSS}
        \label{fig:gdss_net}
    \end{subfigure}
    \begin{subfigure}[b]{0.09\textwidth}
        \includegraphics[width=\textwidth]{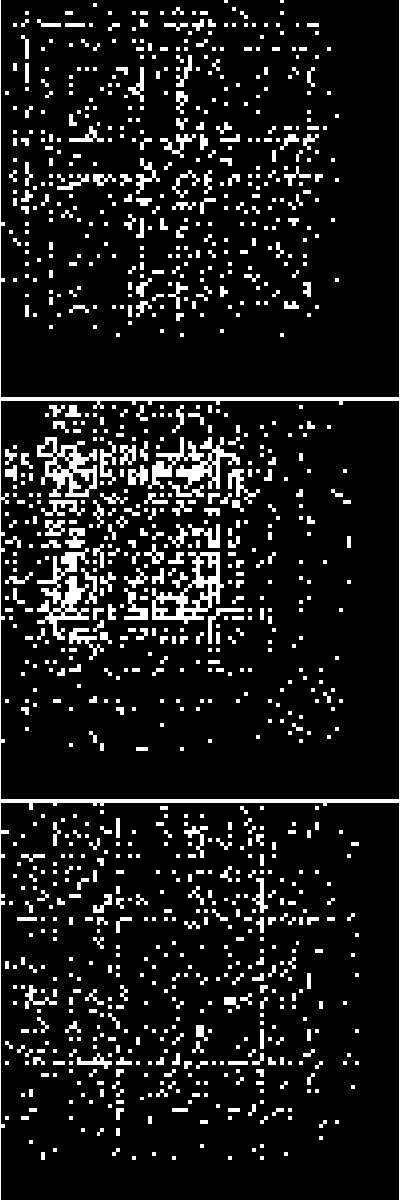}
        \caption{EDGE}
        \label{fig:edge_adj}
    \end{subfigure}
    \begin{subfigure}[b]{0.09\textwidth}
        \includegraphics[width=\textwidth]{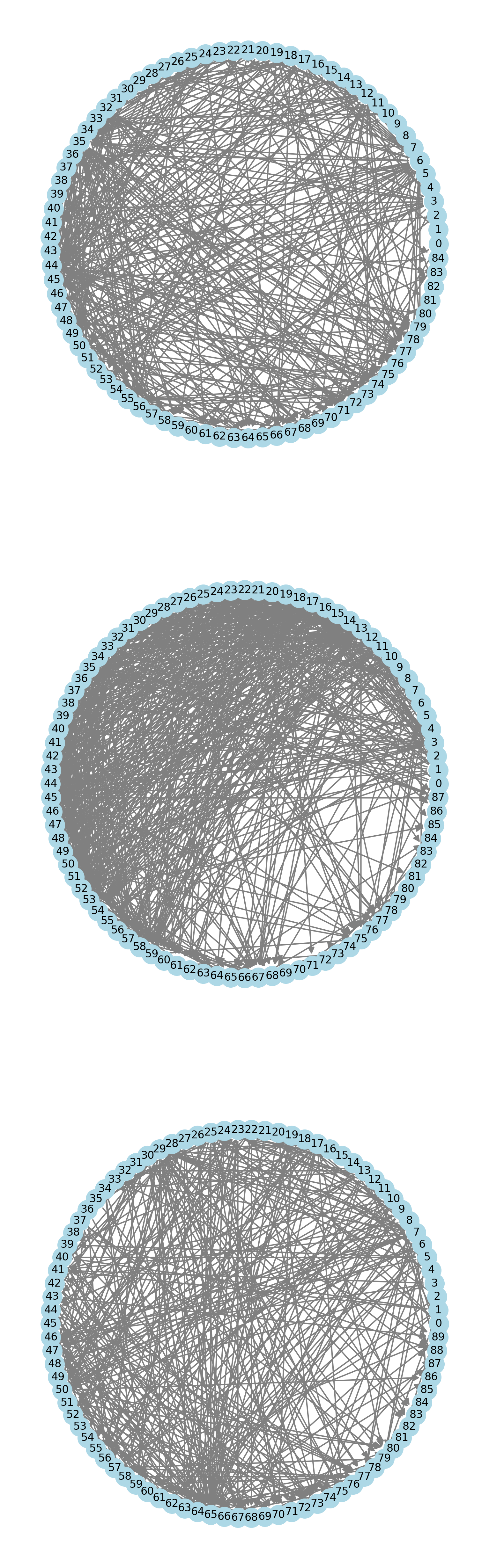}
        \caption{EDGE}
        \label{fig:edge_net}
    \end{subfigure}
    \begin{subfigure}[b]{0.09\textwidth}
        \includegraphics[width=\textwidth]{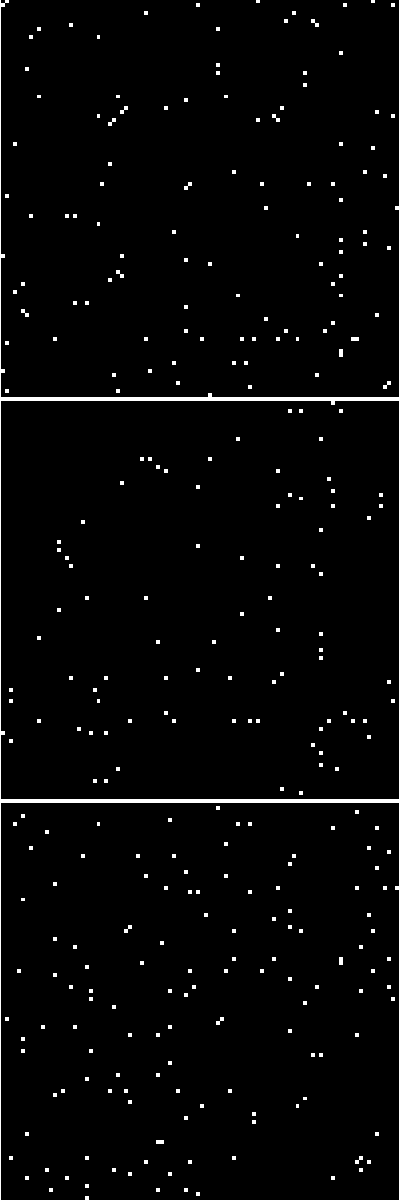}
        \caption{DisCo}
        \label{fig:disco_adj}
    \end{subfigure}
    \begin{subfigure}[b]{0.09\textwidth}
        \includegraphics[width=\textwidth]{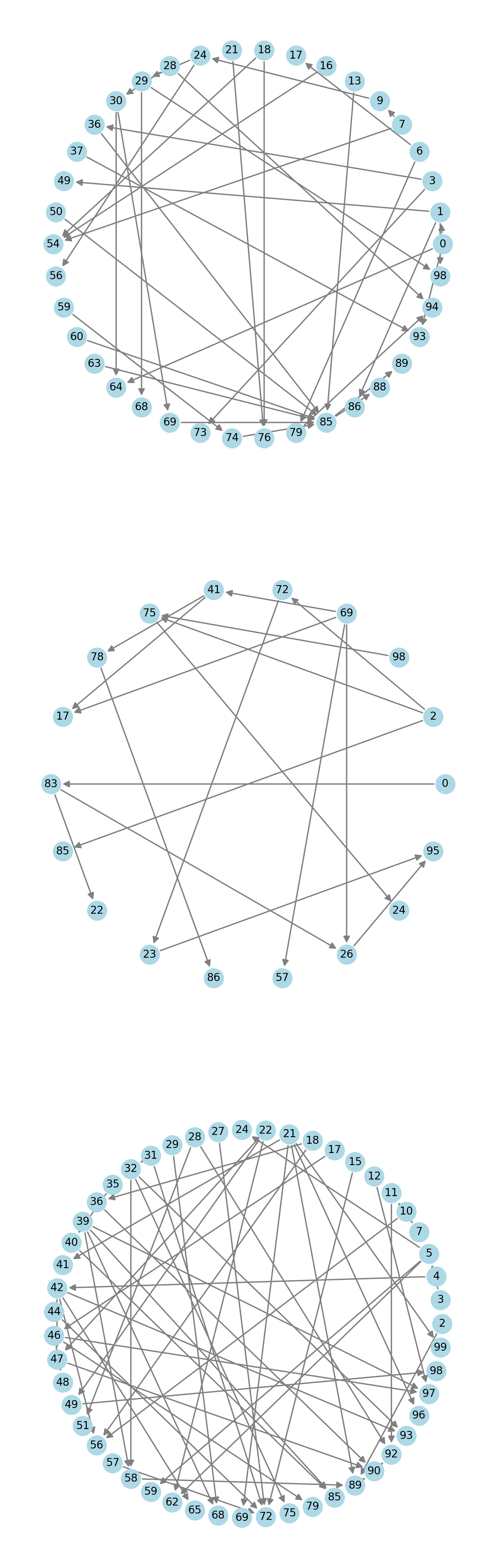}
        \caption{DisCo}
        \label{fig:disco_net}
    \end{subfigure}
    \caption{Generated Results Demonstrate the Authenticity of Our Model}
    \label{fig:generated_result_comparison}
\end{figure*}

Figure~\ref{fig:generated_result_comparison} compares graphs generated by our proposed model and other models with the original data (column Ori.).

\begin{table*}[htbp]
\centering
\begin{tabular}{c|cc|cc|cc}
\hline
\multirow{2}{*}{Model} & \multicolumn{2}{c|}{Deg.} & \multicolumn{2}{c|}{Clus. Coef.} & \multicolumn{2}{c}{Spec.} \\
\cline{2-7}
& EMD $\downarrow$ & TV $\downarrow$ & EMD $\downarrow$ & TV $\downarrow$ & EMD $\downarrow$ & TV $\downarrow$ \\
\hline
GDSS~\cite{GDSS} & 1.138 & 0.493 & 1.381 & 0.959 & 0.325 & 0.515  \\
DisCo~\cite{DisCo} & 1.047 & 0.304 & 1.315 & 0.550 & 0.094 & 0.334  \\
EDGE~\cite{EDGE} & 0.660 & 0.041 & 0.343 & 1.016 & 0.972 & 0.111  \\
Ours & $\boldsymbol{0.164}$ & $\boldsymbol{0.028}$ & $\boldsymbol{0.154}$  & $\boldsymbol{0.021}$ & $\boldsymbol{0.047}$ & $\boldsymbol{0.007}$ \\
\hline
\end{tabular}
\caption{Demonstrating Model Effectiveness: Comparison with Baseline Models}
\label{tab:comparison}
\end{table*}

\subsection{Latent Representation Analysis}
\subsubsection{SHAP Analysis}
\begin{figure}[htbp]
\centering
\includegraphics[width=0.8\columnwidth]{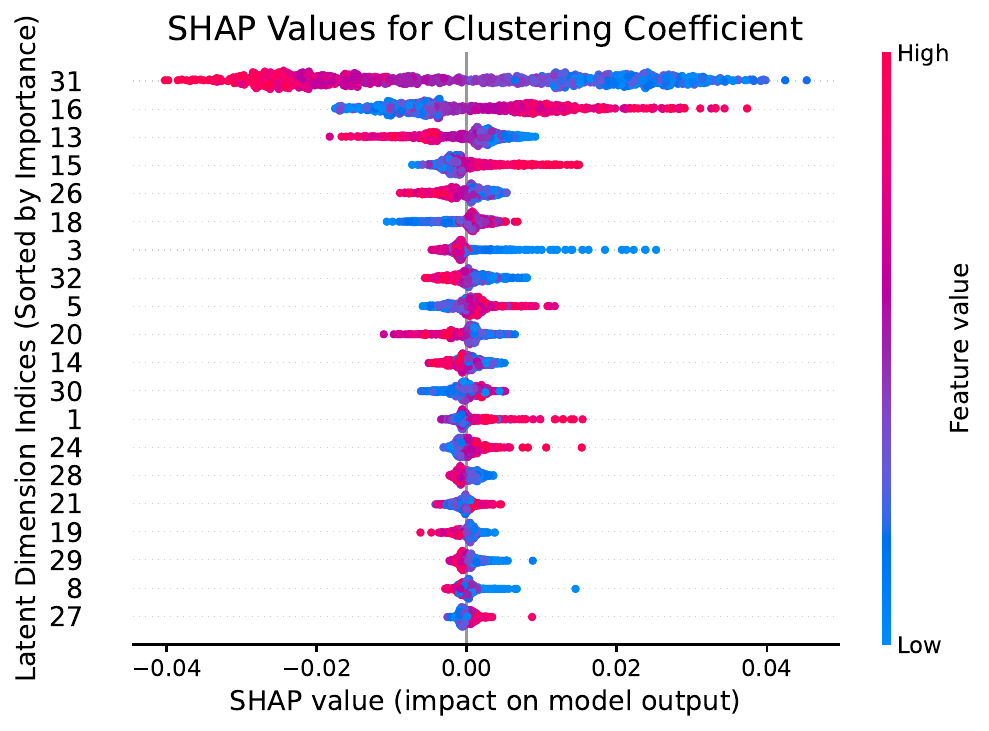}
\caption{SHAP analysis for clustering coefficient. The results of SHAP analysis show the entangled relationship between the latent dimensions and the graph metrics.}
\label{fig:shap_result_clustering}
\end{figure}

Using SHAP analysis~\cite{shap} to interpret feature contributions, we analyzed the relationship between the 32 latent dimensions and generated graph properties. Figure~\ref{fig:shap_result_clustering} illustrates the SHAP results for the clustering coefficient, showing the entangled relationship between the latent dimensions and the graph metrics. Similar analyses were performed for each graph property (see Appendix~\ref{sec:shap_details}). These results motivated our subsequent development of a controllable generation method that operates effectively on such an entangled representation.

\subsubsection{Discovering Key Latent Directions Affecting Generation Metrics}

\begin{figure*}[t]
    \centering 

    \begin{subfigure}{0.15\textwidth}
        \centering
        \includegraphics[width=\linewidth]{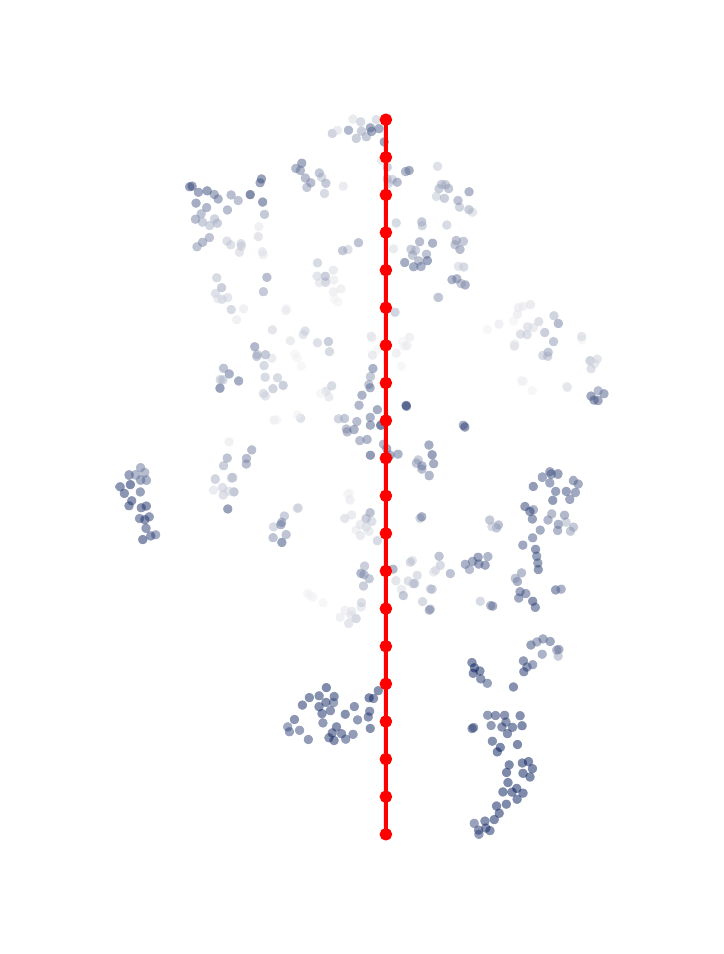}
        \caption{Mean degree}
        \label{fig:tsne_a}
    \end{subfigure}\hfill
    \begin{subfigure}{0.15\textwidth}
        \centering
        \includegraphics[width=\linewidth]{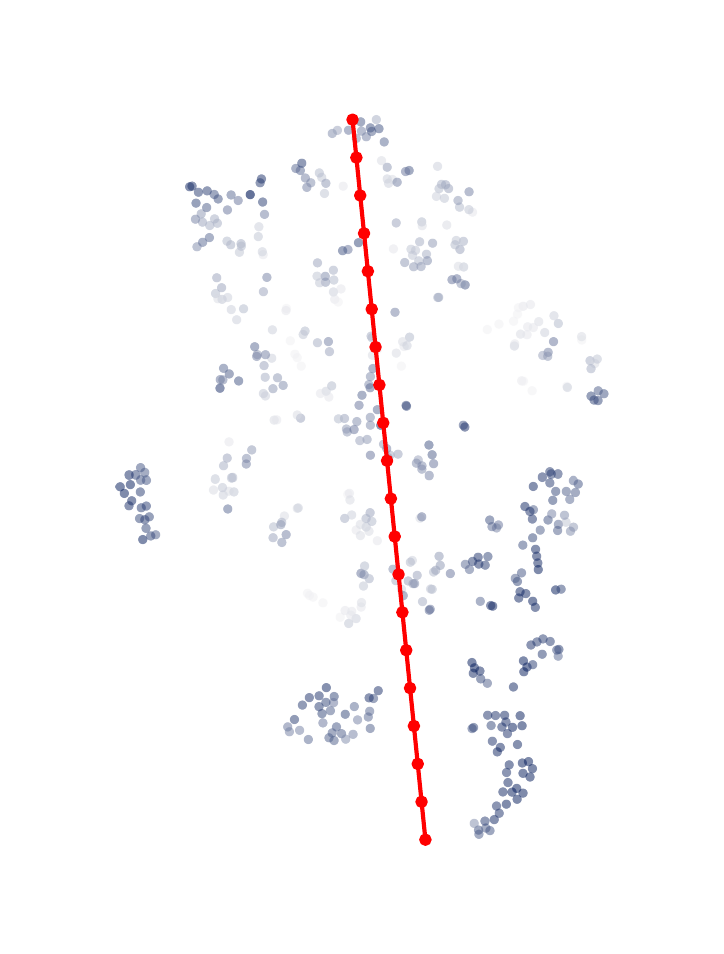}
        \caption{Efficiency}
        \label{fig:tsne_b}
    \end{subfigure}\hfill%
    \begin{subfigure}{0.15\textwidth}
        \centering
        \includegraphics[width=\linewidth]{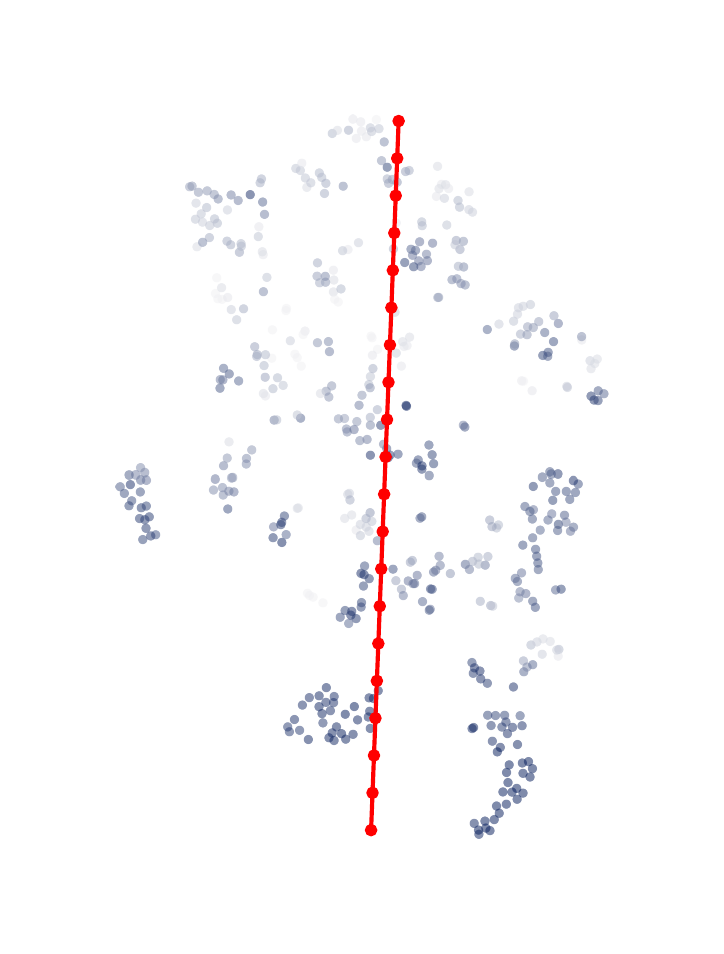}
        \caption{Transitivity}
        \label{fig:tsne_c}
    \end{subfigure}\hfill%
    \begin{subfigure}{0.15\textwidth}
        \centering
        \includegraphics[width=\linewidth]{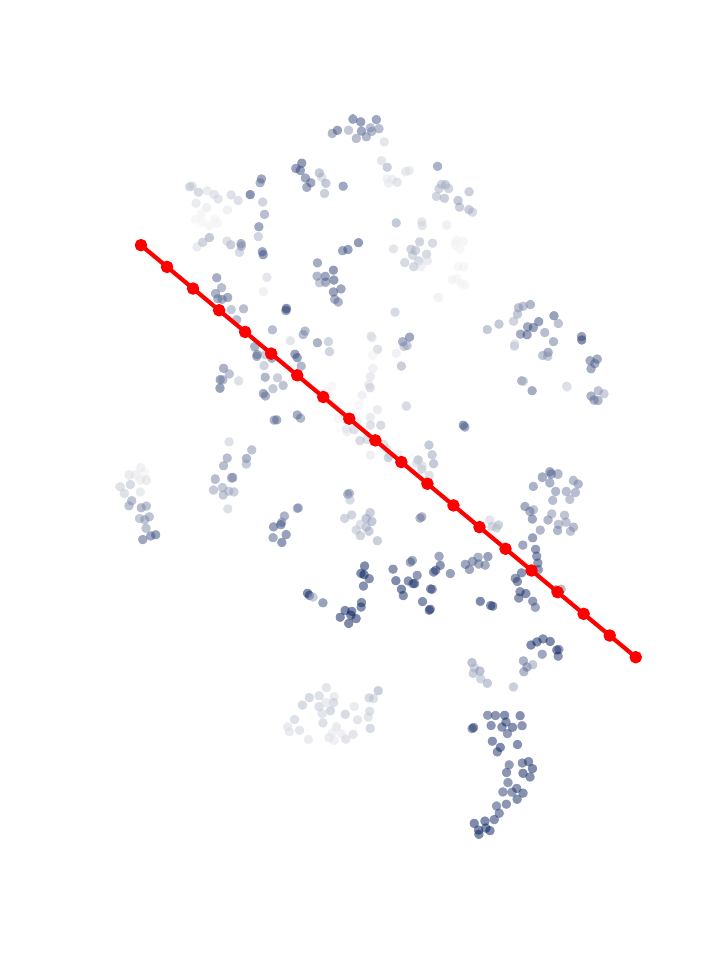}
        \caption{Assortativity}
        \label{fig:tsne_d}
    \end{subfigure}\hfill%
    \begin{subfigure}{0.15\textwidth}
        \centering
        \includegraphics[width=\linewidth]{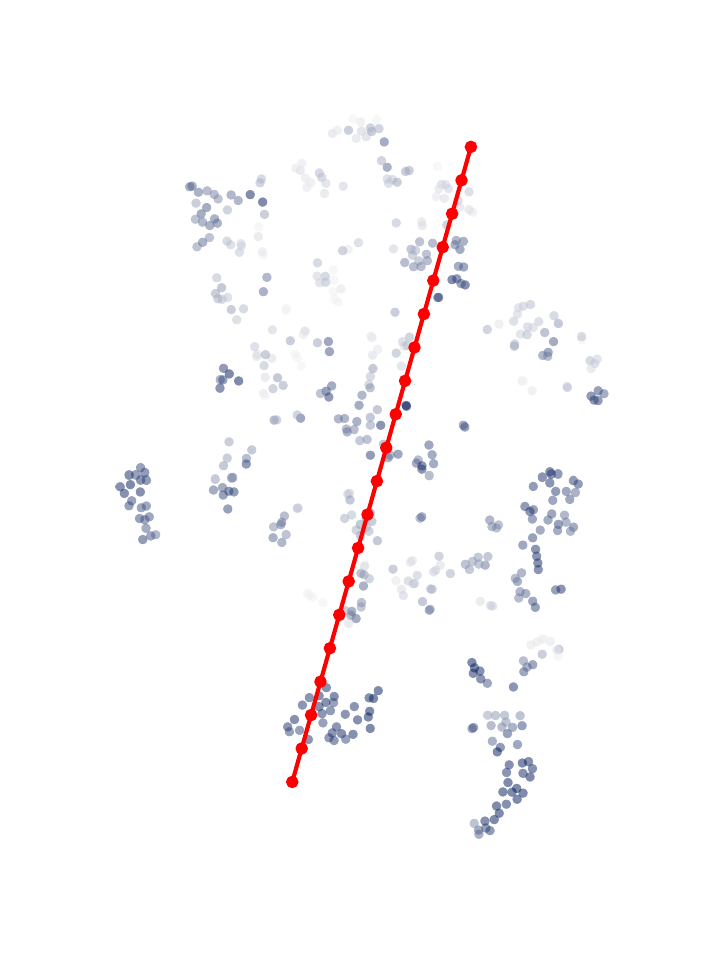}
        \caption{Clus. Coef.}
        \label{fig:tsne_e}
    \end{subfigure}\hfill%
    \begin{subfigure}{0.15\textwidth}
        \centering
        \includegraphics[width=\linewidth]{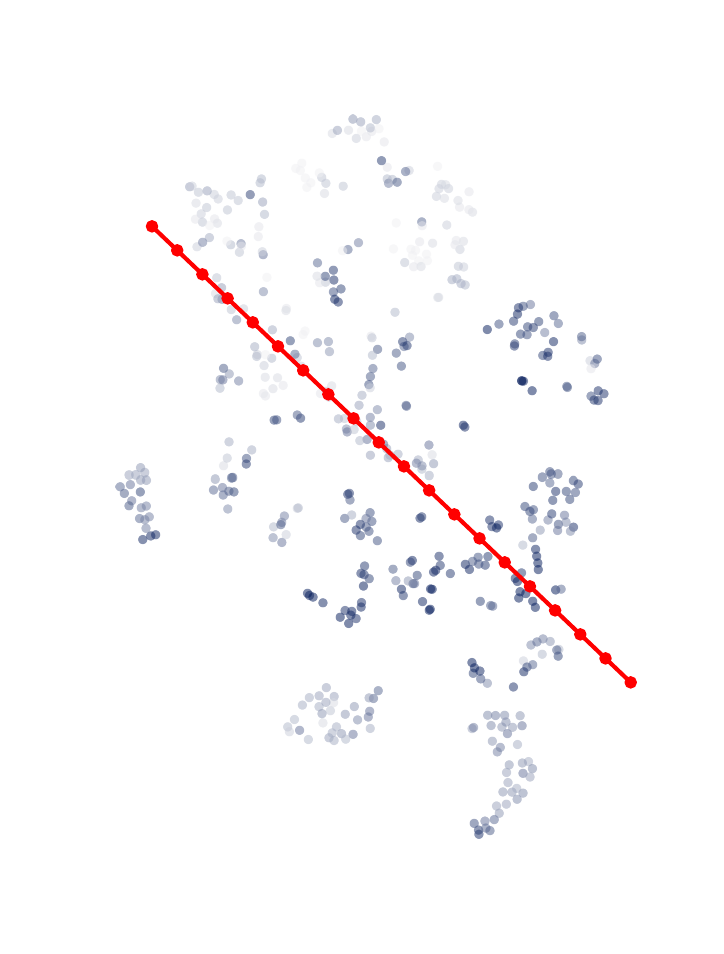}
        \caption{Modularity}
        \label{fig:tsne_f}
    \end{subfigure}
    
    \caption{t-SNE visualizations of the latent space reveal that variations in specific topological metrics correspond to distinct gradient directions within the embeddings. The darkness of each point's color corresponds to the magnitude of its associated bin index for the specific metric being considered. To further aid visualization, the red line overlaid on the t-SNE plot represents the direction of metric variation, obtained by fitting an auxiliary 2D linear regression model to the 2D t-SNE embeddings. This 2D regression is solely for visual convenience and does not replace the 32-dimensional analysis described in the main text.}
    \label{fig:tsne}
\end{figure*}

We first visualized the microcircuit topology with t-SNE (Figure~\ref{fig:tsne}). To identify a latent direction for each graph metric, we trained a linear model to predict its value bin (discretized into 20 quantiles from 500 test graphs) from the 32D latent vectors. The resulting constant gradient defines each direction (Appendix~\ref{sec:linear_regression}), and the model's high $R^2$ scores confirm a strong fit (Figure~\ref{fig:correlations}, left).



We then verified that these directions correspond to distinct metrics. The pairwise correlations between the metrics themselves (Figure~\ref{fig:correlations}, middle) closely matched the cosine similarities of their corresponding latent directions (Figure~\ref{fig:correlations}, right), confirming their specificity (Appendix~\ref{sec:correlations_details}).

Finally, a traversal experiment validated these directions. Moving along a specific gradient from a mean latent vector and then decoding controllably manipulated the corresponding metric in the generated circuits, confirming that the directions directly govern structural features (Appendix~\ref{sec:gradient_direction_moving}).

\subsection{A Pipeline for Controllable Microcircuit Generation With MCMC Sampling}
For the investigation of structure-function relationships in brain networks, the ability to generate network samples that closely match specific target descriptor values would be invaluable. Here, we propose a general pipeline to achieve this goal, leveraging our latent VAE connectome generator in conjunction with Markov Chain Monte Carlo (MCMC) sampling~\cite{MCMC}.

Given a target metric value $t$, the linear regression model $y=f(\mathbf{z})=\mathbf{w}^T\mathbf{z}+b$ previously fitted on the 32-dimensional latent space can be leveraged as a heuristic for search. Our objective is to sample $N$ latent vectors $\mathbf{z}=(z_1, z_2, ..., z_{32})$ that satisfy the condition $\mathcal{T}$:$|f(\mathbf{z})-t|<\epsilon$, where $\epsilon$ defines a tolerance around the target value. This inequality defines a feasible region in the latent space, denoted as $\Omega_\mathcal{T} \subset \mathbb{R}^{32}$. However, since the domain of the linear regression model is unbounded, there are infinitely many solutions to this inequality distributed throughout the space. Many of these solutions might lie far from the true latent space distribution, leading to invalid or unrealistic generated graphs. Therefore, we should impose constraints on the distribution of the sampled latent vectors, aiming for the sampled points to follow the the dataset's inherent latent distribution to the maximum extent.
We approximate the distribution of the latent space as a 32-dimensional joint distribution by fitting a multivariate Gaussian distribution. Our goal is to sample $N$ latent vectors from the conditioned distribution $p(\mathbf{z}|\mathcal{T}), \text{ where } \textbf{z}\in \Omega_\mathcal{T}$. According to Equation~\ref{equation:condition_dist}, this conditioned probability distribution can be represented as $p(\mathbf{z}|\mathcal{T})=\frac{p(\mathbf{z})^{{1}/{\tau}}}{p(\Omega_\mathcal{T})}\cdot\mathbb{I}(\mathbf{z}\in\Omega_\mathcal{T})$, where $\mathbb{I}(\mathbf{z}\in\Omega_\mathcal{T})$ is an indicator function that is 1 if $\mathbf{z}$ belongs to the feasible region $\Omega_\mathcal{T}$ and 0 otherwise, and $p(\Omega_\mathcal{T})=\int_{\Omega_\mathcal{T}}p(\mathbf{z})^{{1}/{\tau}}d\mathbf{z}$ is the probability mass of the feasible region.

As direct integration to compute $p(\Omega_\mathcal{T})$ is not feasible, we utilize Markov Chain Monte Carlo (MCMC)~\cite{MCMC} to sample from the target conditional probability distribution despite the unknown denominator. We define a log-target-density function (LTD) to evaluate the likelihood of a proposed latent vector $\mathbf{z}$:
\begin{equation}
    \text{LTD}(\textbf{z})=\displaystyle\frac{1}{\tau}\text{log}p(\textbf{z}),
\end{equation}
Thus, the target conditional distribution for $\mathbf{z}$ should be proportional to $\text{exp}(\text{LTD}(\mathbf{z}))\cdot \mathbb{I}(\mathbf{z}\in\Omega_\mathcal{T})$. We utilize the Metropolis-Hastings acceptance rule to generate a sequence of $N$ sampled latent vectors. The detailed steps and parameters of the sampling algorithm are provided in the Appendix~\ref{sec:mcmc_details}.

\begin{figure*}[!htbp]
    \centering
    \begin{subfigure}[b]{0.2\textwidth}
        \centering
        \includegraphics[width=\textwidth]{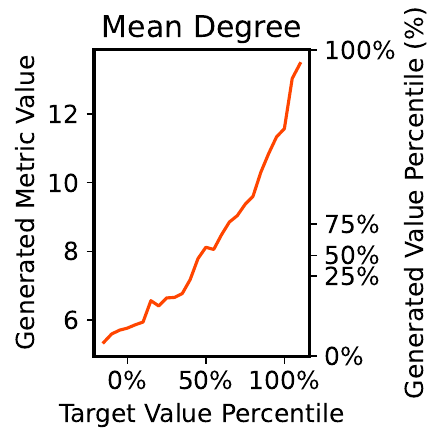}
        \caption{}
        \label{fig:mean_degree}
    \end{subfigure}
    \hfill%
    \begin{subfigure}[b]{0.2\textwidth}
        \centering
        \includegraphics[width=\textwidth]{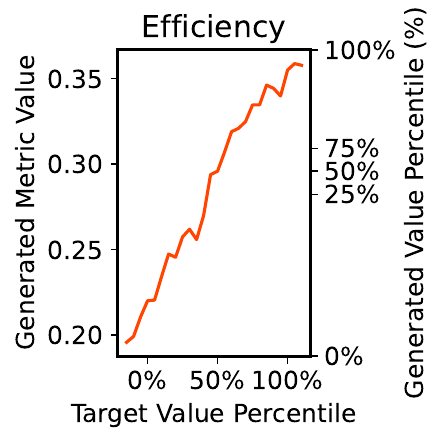}
        \caption{}
        \label{fig:efficiency}
    \end{subfigure}
    \hfill%
    \begin{subfigure}[b]{0.2\textwidth}
        \centering
        \includegraphics[width=\textwidth]{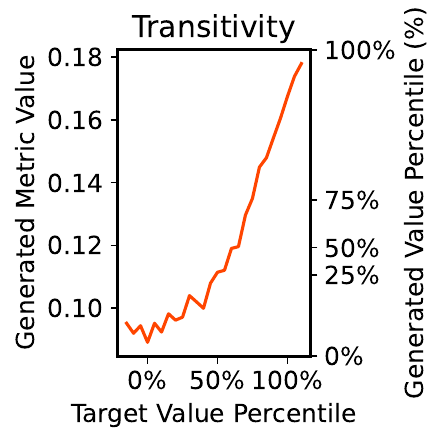}
        \caption{}
        \label{fig:transitivity}
    \end{subfigure}
    \hfill%
    \begin{subfigure}[b]{0.2\textwidth}
        \centering
        \includegraphics[width=\textwidth]{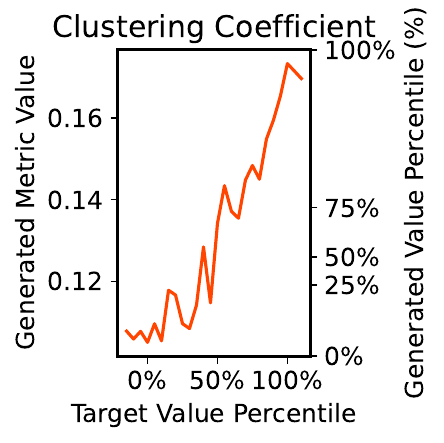}
        \caption{}
        \label{fig:clustering}
    \end{subfigure}

    \caption{Metrics of generated graphs when setting different targets (4 of 6 target graph metrics).}
    \label{fig:control_gene_curves}
\end{figure*}

To demonstrate the ability to generate graphs satisfying specific properties, we conducted experiments by specifying target percentile values ranging from 0\% to 100\%. We set $N=500$ in the experiment. Figure~\ref{fig:control_gene_curves} shows how the mean values of four key graph metrics evolve with the target percentile; the complete set of six curves is deferred to the appendix (Figure~\ref{fig:control_gene_curves_full}) for brevity. Generated graph metric values that show a roughly monotonically increasing trend with the target bin index demonstrates the effectiveness of this method. 
Furthermore, examples of the generated graphs for different target metrics and target values are presented in the Appendix\ref{sec:controlled_generation_results}.

\begin{figure*}[!h]
    \centering
    \begin{subfigure}[b]{0.13\textwidth}
        \includegraphics[width=\textwidth]{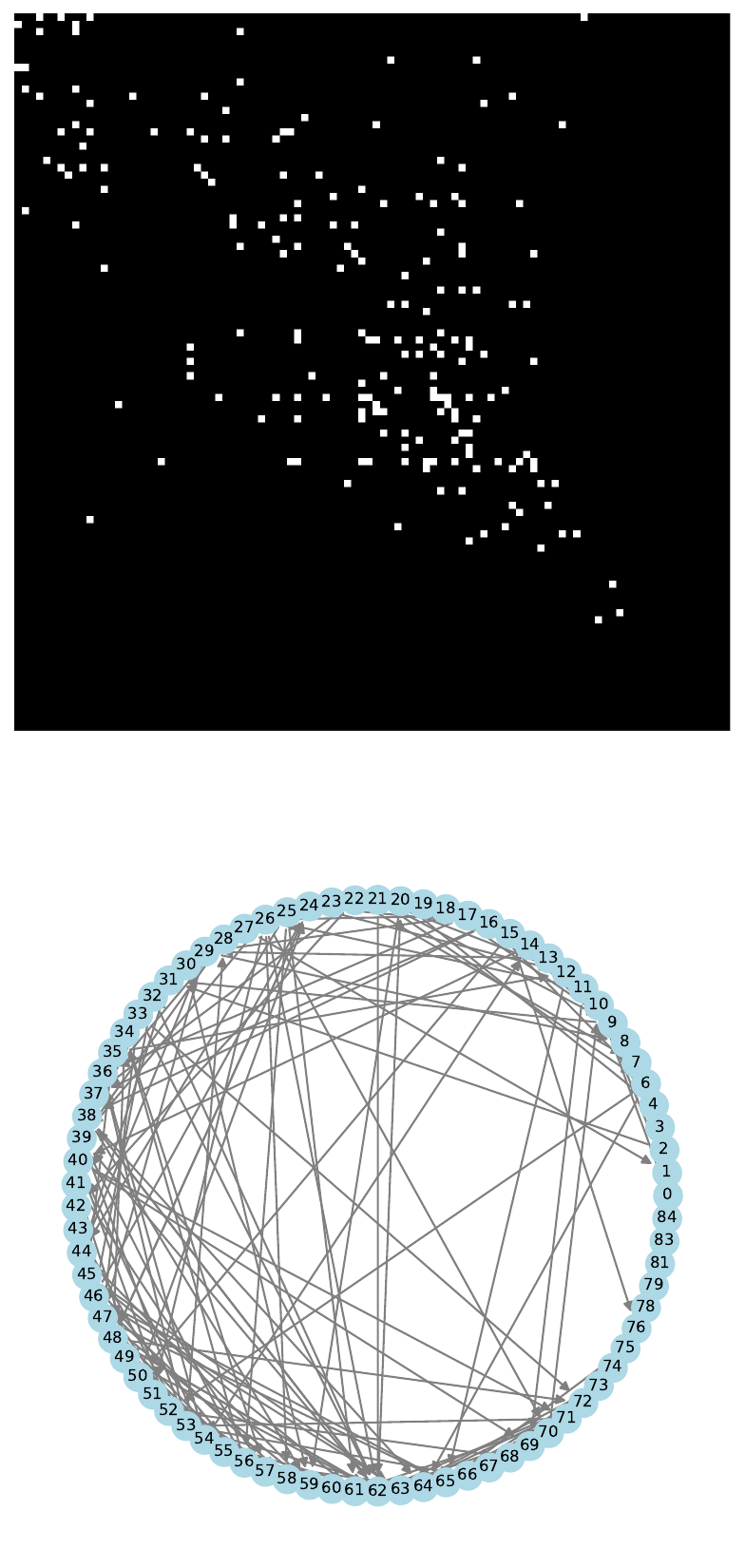}
        \caption{target=0\%}
        \label{fig:0}
    \end{subfigure}
    \begin{subfigure}[b]{0.13\textwidth}
        \includegraphics[width=\textwidth]{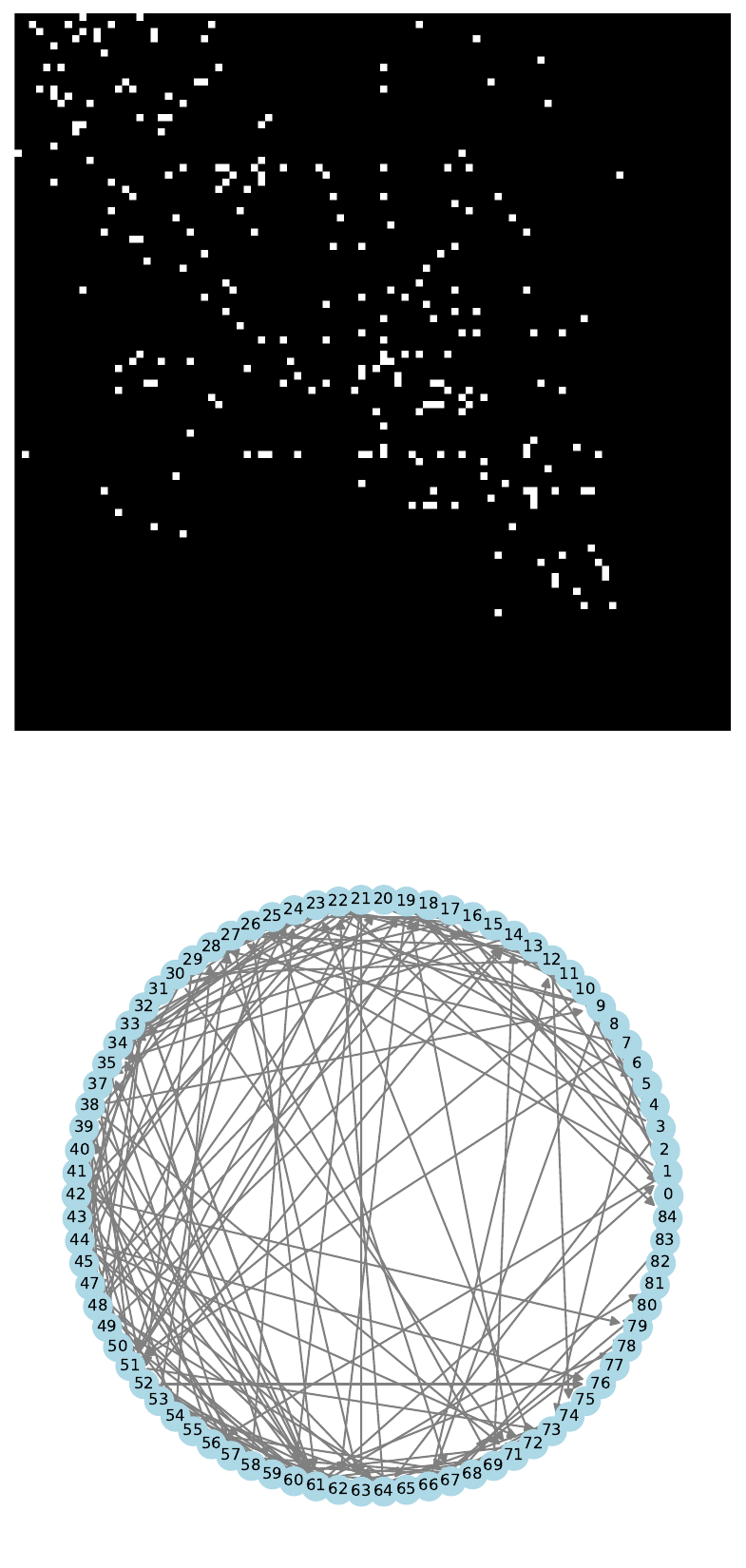}
        \caption{20\%}
        \label{fig:20}
    \end{subfigure}
    \begin{subfigure}[b]{0.13\textwidth}
        \includegraphics[width=\textwidth]{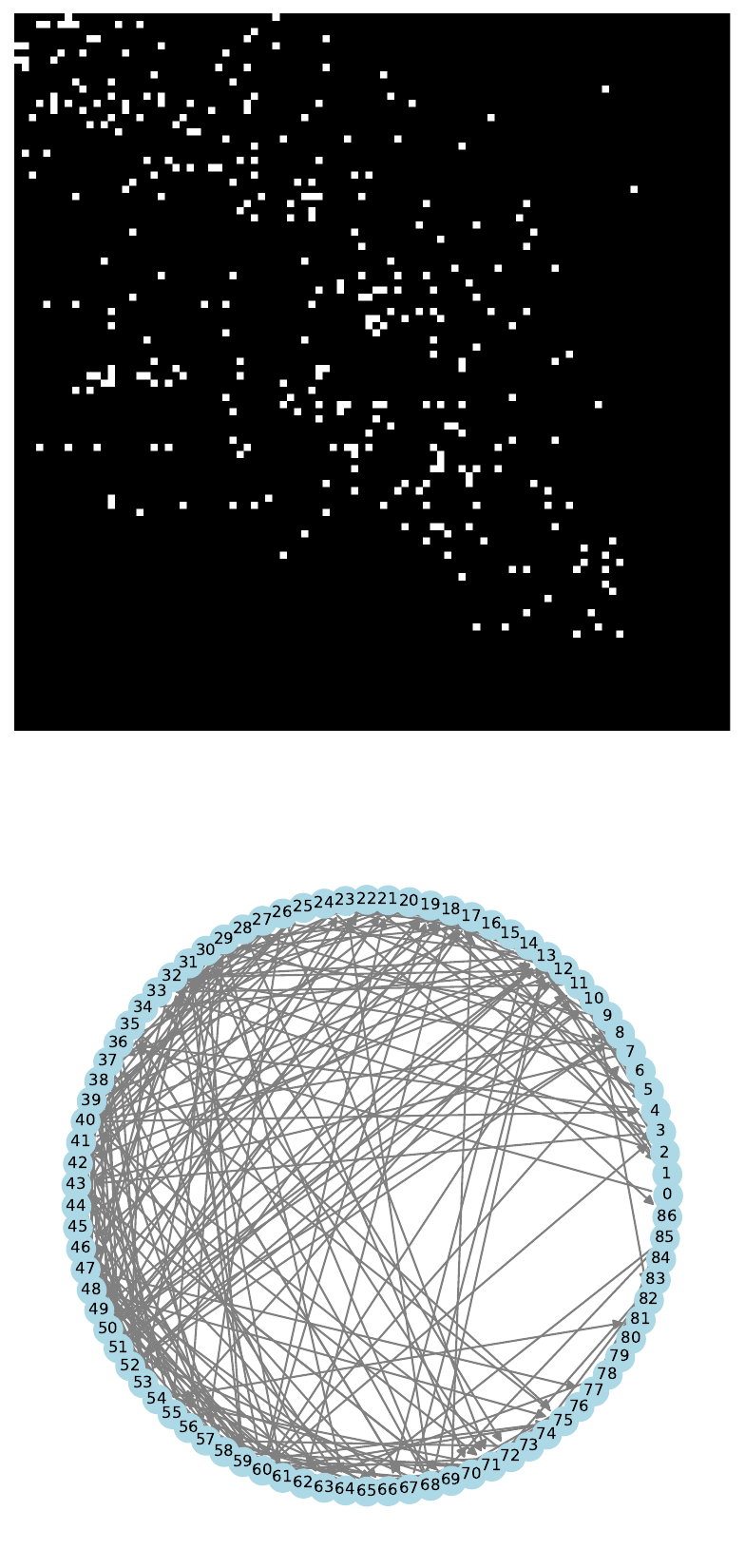}
        \caption{40\%}
        \label{fig:40}
    \end{subfigure}
    \begin{subfigure}[b]{0.13\textwidth}
        \includegraphics[width=\textwidth]{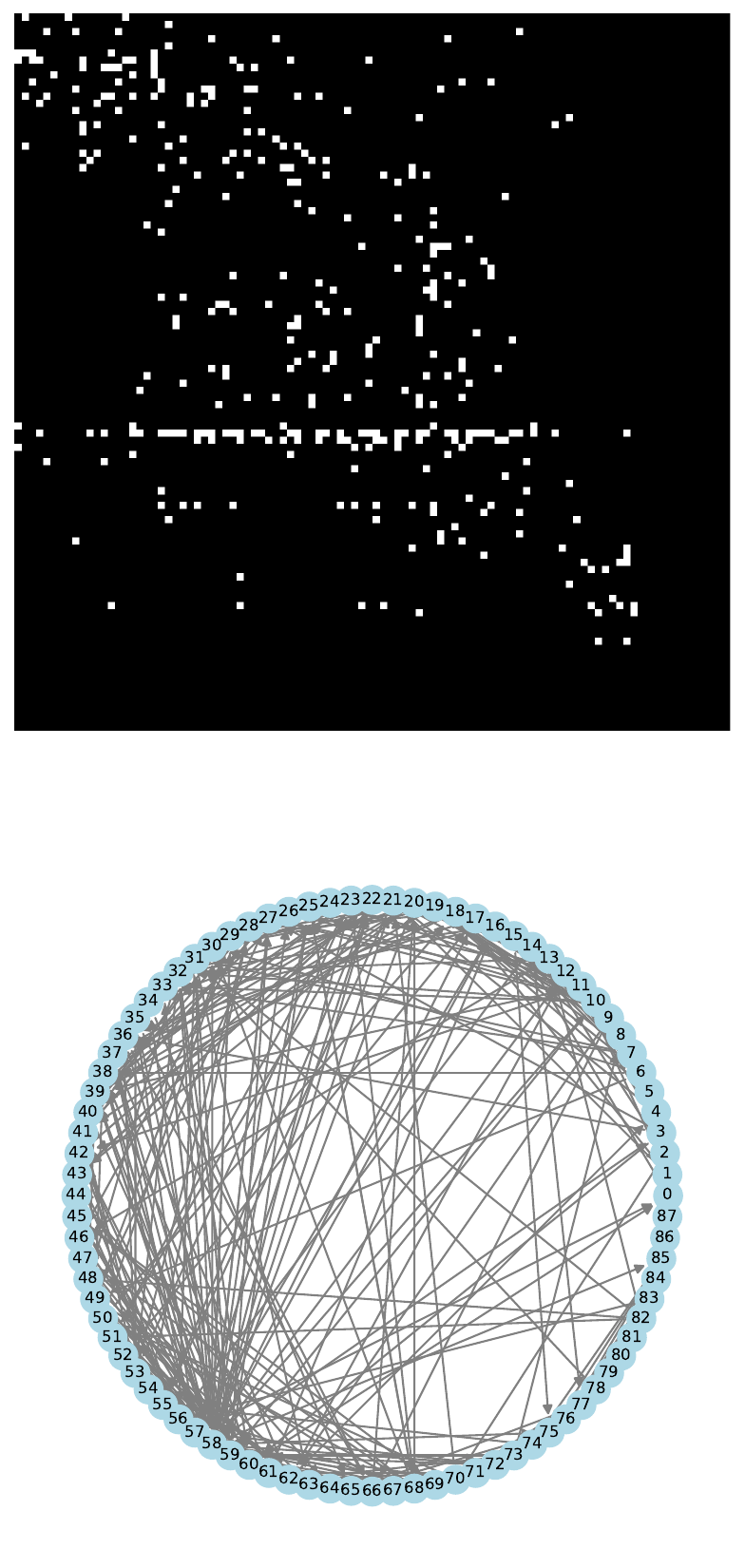}
        \caption{60\%}
        \label{fig:60}
    \end{subfigure}
    \begin{subfigure}[b]{0.13\textwidth}
        \includegraphics[width=\textwidth]{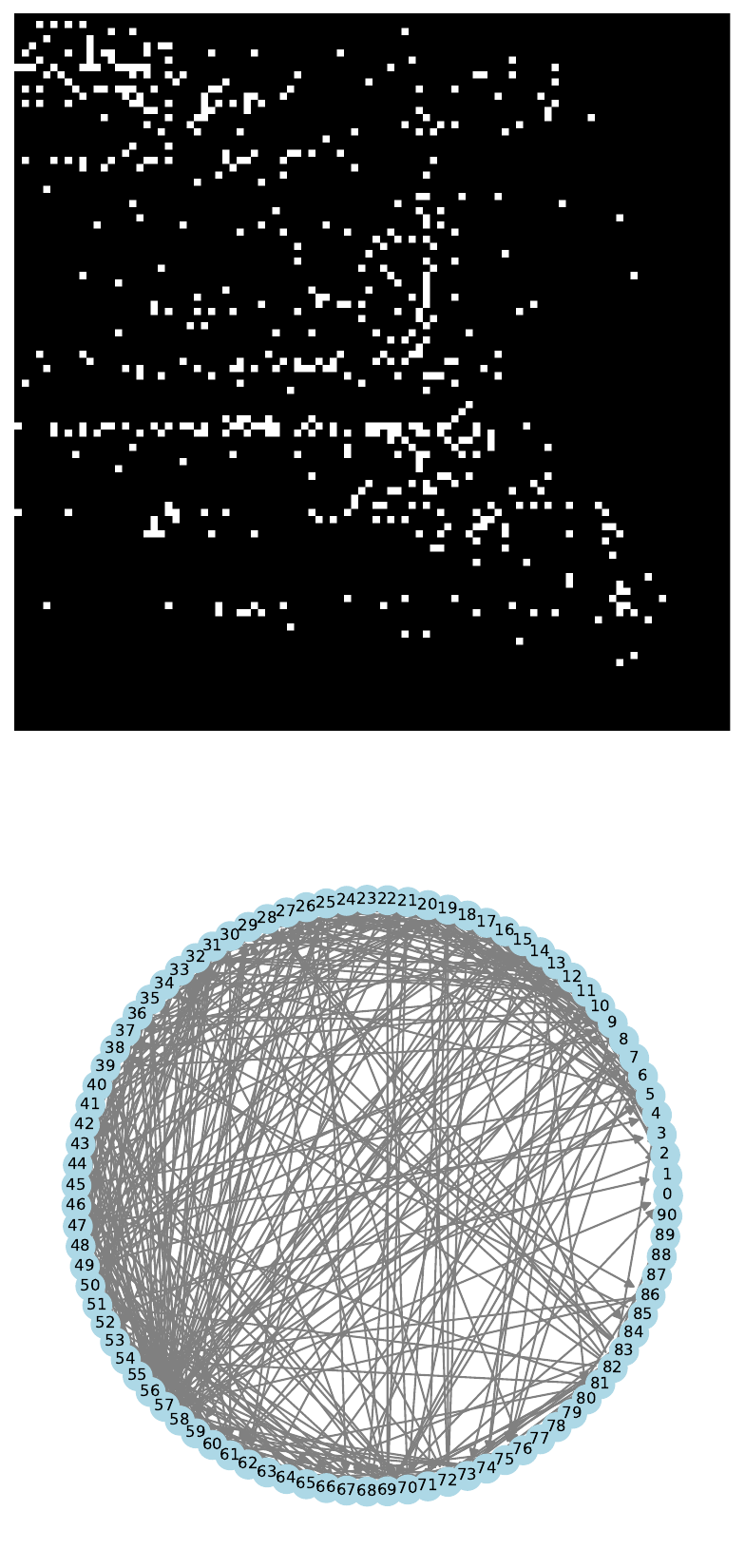}
        \caption{80\%}
        \label{fig:80}
    \end{subfigure}
    \begin{subfigure}[b]{0.13\textwidth}
        \includegraphics[width=\textwidth]{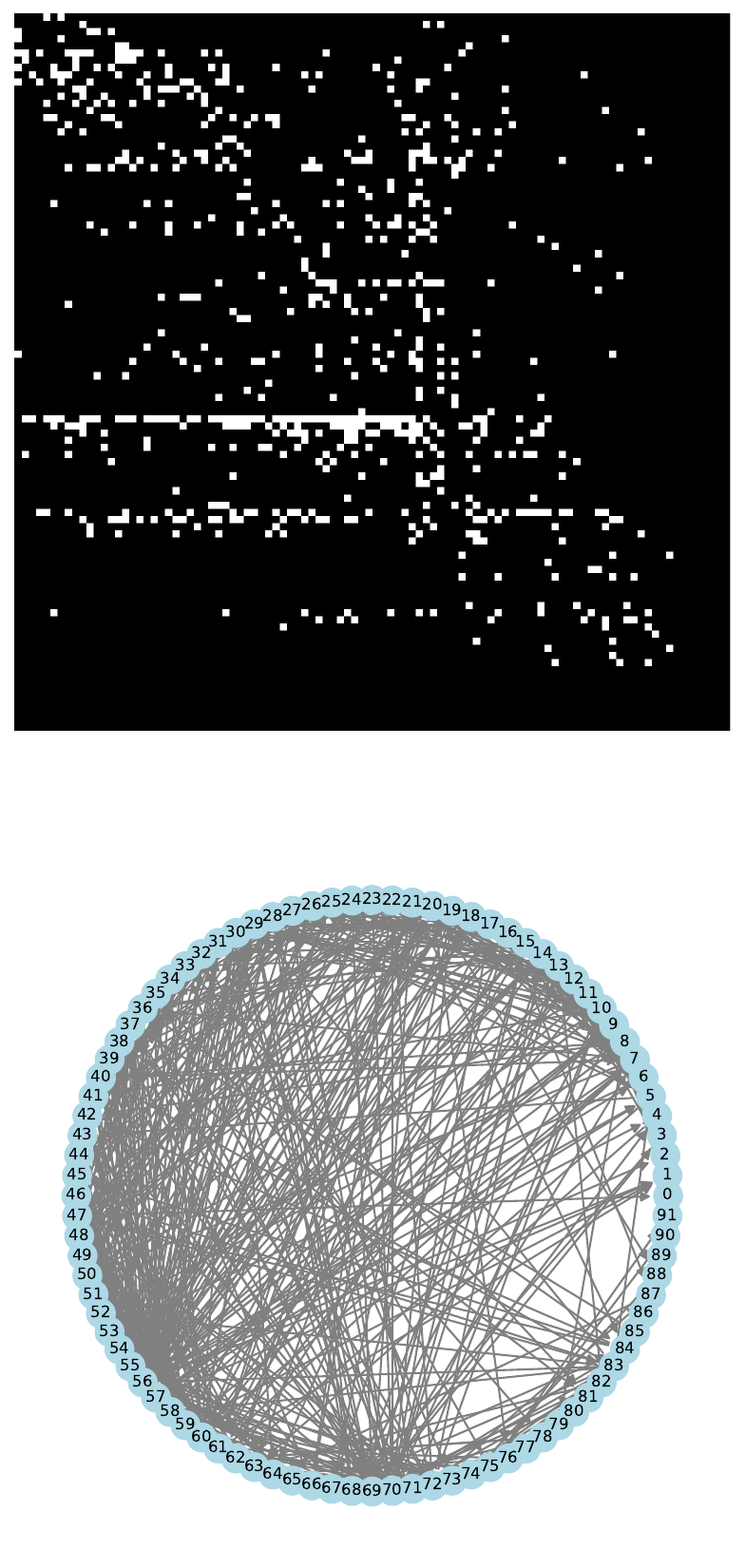}
        \caption{100\%}
        \label{fig:100}
    \end{subfigure}
    \caption{Generated graph examples targeting different mean degree percentile ranges.}
    \label{fig:control_gene_results_mean_degree_small}
\end{figure*}

Figure~\ref{fig:control_gene_results_mean_degree_small} presents examples of controlled graph generation where the mean degree is the targeted property, showing samples generated by setting different target percentile ranges for this metric. Detailed results for controlling other graph properties are provided in the Appendix~\ref{sec:controlled_generation_results}.

\subsection{Explore the Relationship between Structure and Function}
Given that cortical microcircuits inherently operate as recurrent neural networks (RNNs), we employ reservoir computing (RC) to assess their functional capabilities. We use performance on classic tasks, specifically memory (copying) and classification, as a proxy for the network's functional features. To investigate this, we constructed reservoir networks using connectome architectures from the VAE and tested them on copying and classification tasks (Appendix~\ref{sec:detail_esn}). We primarily investigated two questions: 1. Do these connectome-based reservoirs outperform randomly connected ones? 2. Are there graph features that, when systematically varied, predictably modulate task-specific performance? All of the hyperparameters of the experiments are stated in Appendix~\ref{sec:detail_esn}.

\begin{figure*}[!htbp] 
    \centering 

    \begin{subfigure}{0.2\textwidth}
        \centering
        \includegraphics[width=\linewidth]{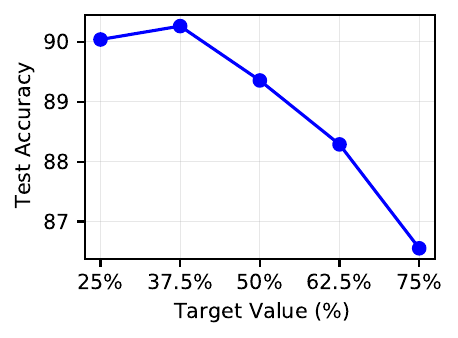} 
        \caption{Efficiency}
        \label{fig:sub1}
    \end{subfigure}
    \hfill 
    \begin{subfigure}{0.2\textwidth}
        \centering
        \includegraphics[width=\linewidth]{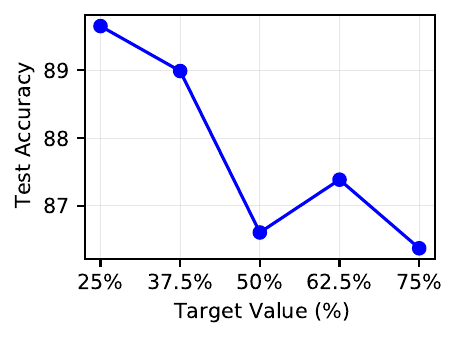} 
        \caption{Mean Degree}
        \label{fig:sub2}
    \end{subfigure}
    \hfill 
    \begin{subfigure}{0.2\textwidth}
        \centering
        \includegraphics[width=\linewidth]{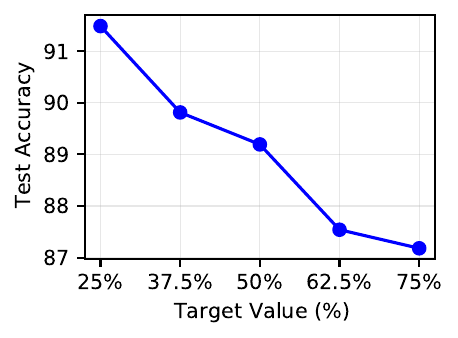} 
        \caption{Transitivity}
        \label{fig:sub3}
    \end{subfigure}
    \hfill 
    \begin{subfigure}{0.2\textwidth}
        \centering
        \includegraphics[width=\linewidth]{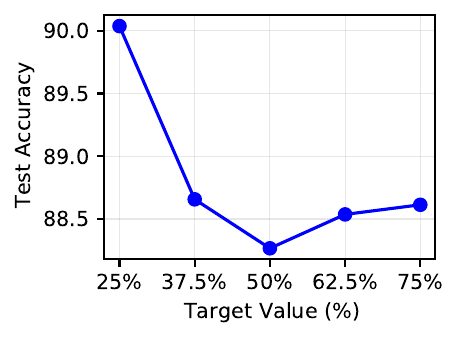} 
        \caption{Clustering Coefficient}
        \label{fig:sub4}
    \end{subfigure}
    
    \caption{4 of the 6 metrics have a significant impact on task performance.}
    \label{fig:4_of_6_significant_metrics}
\end{figure*}

\paragraph{Copy Task}
Following~\cite{wavernn}, we use a copy task to assess network memory retention (details in Appendix~\ref{sec:detail_copy}). We specified target values for a set of features at five distinct percentiles (25\%, 37.5\%, 50\%, 62.5\%, 75\%) of their distributions. For each condition, we constructed Reservoir Networks from three VAE-generated connectome graphs, yielding an average test loss $L_{gen}$. As a baseline, we repeated this process using density-matched random graphs to obtain a corresponding loss $L_{random}$. The performance gain of the VAE-generated structures over their random counterparts is quantified by the percentage decrease in test loss, $\delta$, calculated as follows:
\begin{equation}
    \delta=\displaystyle\frac{L_{random}-L_{gen}}{L_{random}}\times100\%
\end{equation}
The results are documented in Table~\ref{tab:copy_results}:
\begin{table}[!h] 
    \centering
    \begin{tabularx}{\columnwidth}{l*{6}{>{\centering\arraybackslash}X}}
        \toprule
        \makecell{\textbf{Target} \\ \textbf{Percentile}} & \textbf{25\%} & \textbf{37.5\%} & \textbf{50\%} & \textbf{62.5\%} & \textbf{75\%} & \textbf{Ave.} \\
        \midrule
        Mean Deg. & 39.94 & 54.32 & 55.23 & 50.34 & 55.36 & 51.04 \\
        Clus. Coef. & 52.67& 52.69& 52.45& 53.45& 49.03 & 52.06 \\
        Efficiency & 47.64& 59.29& 59.31& 54.61& 52.00& 54.57 \\
        Transitivity & 52.23& 45.53& 56.05& 58.93& 51.15 & 52.78 \\
        Assortativity & 49.68& 57.55& 52.01& 45.28& 53.71 & 51.64 \\
        Modularity & 63.98& 60.03& 58.42& 49.82& 57.20 & 57.89 \\
        \bottomrule
    \end{tabularx}
    \caption{Performance Advantage of VAE-Generated Connectomes over Random Baselines in Reservoir Tasks (Measured by \% Test Loss Decrease)}
    \label{tab:copy_results}
\end{table}

Appendix~\ref{sec:detail_copy} visually compares the copy task outputs from networks with VAE-generated reservoirs against those with random counterparts.

\paragraph{Classification Task}
On the sequential MNIST classification task, we found a clear monotonic or near-monotonic relationship between a reservoir network's test accuracy and several of its reservoir's graph features—notably efficiency, mean degree, transitivity, and clustering coefficient (Figure~\ref{fig:4_of_6_significant_metrics}). This result was established by training networks built from reservoir graphs systematically generated to have feature values at five distinct percentiles (25-75\%). Specifically, for each percentile condition, we randomly selected three generated graphs, constructed an network from each, and averaged their final test accuracies (details in Appendix~\ref{sec:detail_class}).

\section{Conclusions and Discussions}
\label{sec:conclusions}
We introduced a VAE-based generative model that learns a compact, interpretable latent space for mouse cortical microcircuit topology. We demonstrated that directions in this space encode specific network properties, enabling the controlled synthesis of novel circuits with desired features via latent navigation and an MCMC pipeline. This approach provides a powerful tool to investigate neural design principles, explore structure-function relationships, and inform advanced AI development by revealing the low-dimensional generative rules behind complex structures.
\paragraph{Limitations and Future Work}
Despite promising results, this work has several limitations. Firstly, our choice of a Variational Autoencoder was deliberate, driven by the primary objective of learning an explicit, interpretable, and low-dimensional latent space. This ``information bottleneck'' is crucial for uncovering the compact generative blueprint hypothesized to underlie brain development and for enabling controlled synthesis. While other modern generative models, such as diffusion models, demonstrate powerful sample generation capabilities, their latent spaces are not always as directly optimized for, or as easily amenable to, the extraction of such a compressed and interpretable code as a VAE's. Future work could, however, investigate adaptations of these models or hybrid approaches to achieve similar goals.
Secondly, our model currently focuses on binary topological structure, neglecting crucial biological details such as neuron types, synaptic weights, and activity dynamics. Incorporating these features is a key direction for future work to enhance biological realism. Thirdly, the study is based on microcircuits from mouse visual cortex; the learned representations and generative rules may need adaptation and validation for other brain regions, larger-scale connectomes, or different species. The current fixed-size input representation (100x100 padding) also poses challenges for direct application to circuits of highly variable sizes without architectural modifications. Furthermore, while we identified interpretable linear directions in the latent space, exploring more complex, non-linear relationships and the full extent of encoded biological constraints warrants further investigation. Finally, while preliminary work has linked generated structures to function, more rigorously validating their functional viability is a key next step to bridge structural generation with functional understanding.



\clearpage

\appendix
\setcounter{secnumdepth}{2}

\section{Training Details}
\label{sec:training_details}
Training was performed with a learning rate (lr) of 0.001. We employed a cyclical beta annealing schedule~\cite{betascheduling} with a cycle length of 600 epochs. In this schedule, $\beta$ was linearly increased from 0 to $1e-6$ during the first half of each cycle and held constant at $1e-6$ for the second half. The model was trained for a total of 10000 epochs on an RTX4090 GPU.

\section{Model Structure Details}
\label{sec:model_details}
\subsection{Model Components}
\paragraph{Node Feature Encoder}
The node feature encoder, utilizing a three-layer multi-head GAT network, transforms 100-dimensional one-hot node representations into 32-dimensional node embeddings. See Appendix~\ref{sec:GAT} for GAT architecture details.

\paragraph{Graph Global Encoder}
Treating the y-ordered nodes as a sequence (analogous to words in a sentence), the graph global encoder transforms node embeddings into a fixed-size latent representation. Following~\cite{BERT, pigvae}, a dummy node $v_0$ is prepended to the sequence to serve as a global embedding. The encoder applies rotational positional encoding (RoPE)~\cite{rope} and several transformer encoder layers to this augmented sequence. The final embedding of $v_0$ is taken as the global graph representation. An MLP is then applied to compute the 32-dimensional mean and variance for sampling the 32-dimensional latent vector $\mathbf{z}$, similar to standard VAEs.

\paragraph{Node Feature Decoder}
The Node Feature Decoder reconstructs individual node embeddings from the global graph embedding using several transformer decoder layers. The input is the global graph embedding, augmented with rotational positional encoding (RoPE)~\cite{rope}. The global graph embedding also serves as memory for the decoder's cross-attention, leveraging this global context during node feature reconstruction.
\paragraph{Edge Predictor}
The edge predictor is a cross-node interaction layer. It takes the node feature decoder output $h\in \mathbb{R}^{n\times d}$, where $n=100$ is the maximum number of nodes and $d$ is the embedding dimension. Edges are predicted using the dot product of embeddings transformed by two distinct linear layers with activation.
\begin{equation}
  \mathbf{A}_{\text{pred}} = \sigma(\text{LeakyReLU}(\mathbf{h}\mathbf{W}_1) (\text{LeakyReLU}(\mathbf{W}_2\mathbf{h})^\top)),
\end{equation}
The output $\mathbf{A}_{\text{pred}}$ provides a probabilistic adjacency matrix where each entry, a floating-point number between 0 and 1, denotes the likelihood of an edge. We then perform a Bernoulli sampling process on each entry using this probability to generate a binary adjacency matrix.

\subsection{GAT Mechanism}
\label{sec:GAT}
In GAT network, the representation of each node is iteratively refined by aggregating information from its neighboring nodes through a message-passing mechanism. Specifically, for each node $v_i$, an attention score $e_{ij}$ is computed with respect to its neighboring nodes $v_j$ by $e_{ij}=a\left(\mathbf{W}\vec{h}_i,\mathbf{W}\vec{h}_j\right)$, where $\vec{h}_i$ represents the current feature vector of node $v_i$, $\vec{h}_i$ represents the current feature vector of a neighboring node $v_j$, and $a$ denotes the attention mechanism, parameterized by the weight matrix $\mathbf{W}$. These attention scores are then normalized across the neighbors of $v_i$ using the softmax function to obtain the attention coefficients $\alpha_{ij}$. Finally, the updated representation $\vec{h}_i^{\prime}$ of node $v_i$ is obtained by aggregating the feature vectors of its neighbors, weighted by the calculated attention coefficients, followed by a non-linear activation function $\sigma$ by $\vec{h}_i^{\prime}=\sigma(\sum_{j\in N_i}\alpha_{ij}W\vec{h}_j)$



\begin{figure*}[tbp] 
  \centering
  \includegraphics[width=1.0\textwidth]{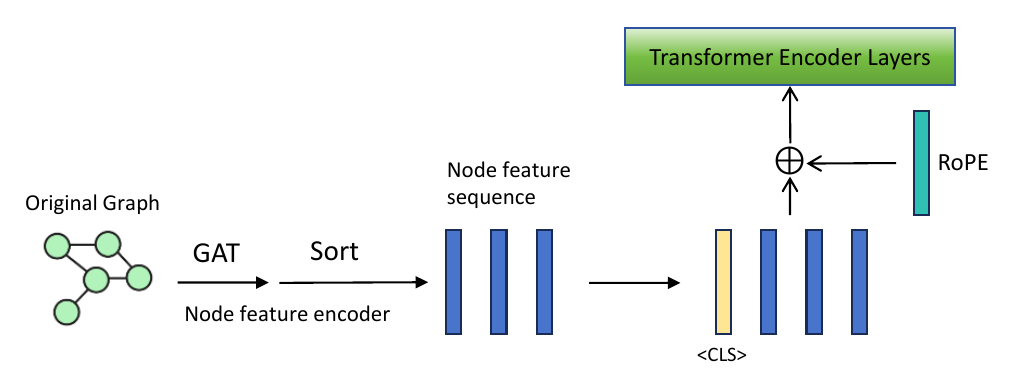}
  \caption{Encoder Structure}
  \label{fig:encoder}
\end{figure*}

\begin{figure*}[tbp] 
  \centering
  \includegraphics[width=1.0\textwidth]{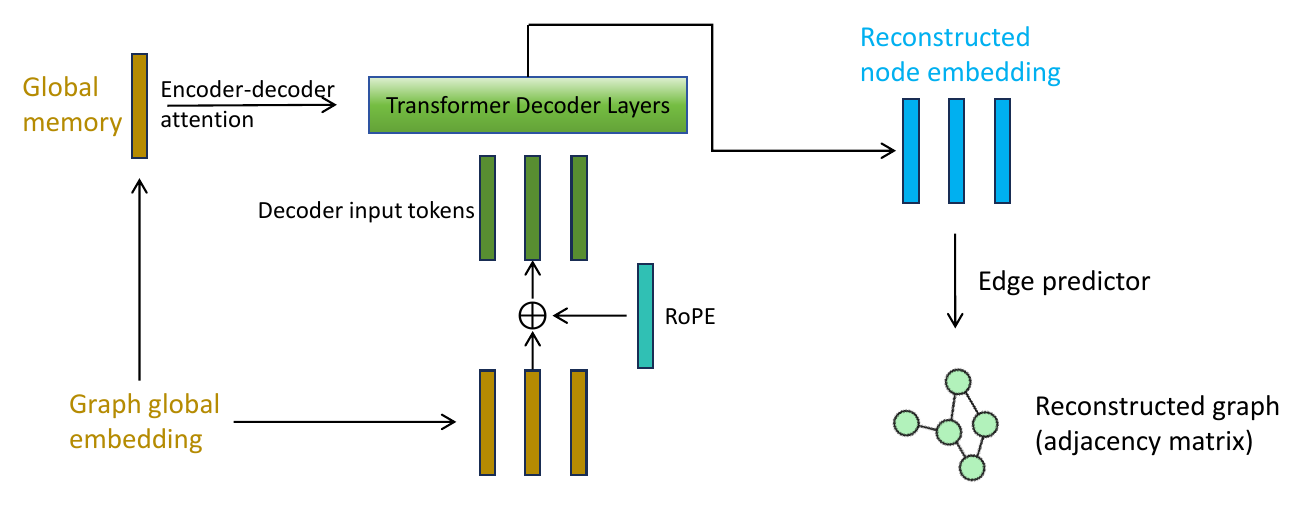}
  \caption{Decoder Structure}
  \label{fig:decoder}
\end{figure*}

\section{Definition of the Graph Metrics}
\label{sec:graph_metrics}
\paragraph{Mean Degree:} 
The mean degree of a graph is the mean total degree of each node $i$:
\begin{equation}
k_i=\sum_{j\in N}a_{ij}.
\end{equation}

\paragraph{Efficiency:}
\begin{equation}E=\frac{1}{n}\sum_{i\in N}E_i=\frac{1}{n}\sum_{i\in N}\frac{\sum_{j\in N,j\neq i}d_{ij}^{-1}}{n-1},\end{equation}
where $E_i$ is the efficiency of node $i$. Directed global efficiency is defined as:
\begin{equation}E^{\rightarrow}=\frac{1}{n}\sum_{i\in N}\frac{\sum_{j\in N,j\neq i}\left(d_{ij}\right)^{-1}}{n-1}.\end{equation}

\paragraph{Clustering coefficient:}
\begin{equation}C=\frac{1}{n}\sum_{i\in N}C_i=\frac{1}{n}\sum_{i\in N}\frac{2t_i}{k_i(k_i-1)},\end{equation}
where $C_i$ is the clustering coefficient of node $i(C_i=0$ \text{for} $k_i<2)$. Directed clustering coefficient is defined as:
\begin{equation}C^\to=\frac{1}{n}\sum_{i\in N}\frac{t_i}{\left(k_i^\mathrm{out}+k_i^\mathrm{in}\right)\left(k_i^\mathrm{out}+k_i^\mathrm{in}-1\right)-2\sum_{j\in N}a_{ij}a_{ji}}.\end{equation}

\paragraph{Transitivity of the network:}
\begin{equation}T=\frac{\sum_{i\in N}2t_i}{\sum_{i\in N}k_i(k_i-1)},\end{equation}
Directed transitivity is defined as:
\begin{equation}T^{\to}=\frac{\sum_{i\in N}t_{i}^{\to}}{\sum_{i\in N}[(k_{i}^{\mathrm{out}}+k_{i}^{\mathrm{in}})(k_{i}^{\mathrm{out}}+k_{i}^{\mathrm{in}}-1)-2\sum_{j\in N}a_{ij}a_{ji}]}.\end{equation}

\paragraph{Modularity:}
\begin{equation}Q=\sum_{u\in M}\left[e_{uu}-\left(\sum_{v\in M}e_{uv}\right)^2\right], \end{equation}
where the network is fully subdivided into a set of non-overlapping modules $M$, and $e_{uv}$ is the proportion of all links that connect nodes in module $u$ with 
nodes in module $v$.

Directed modularity is defined as:
\begin{equation}Q^\to=\frac{1}{l}\sum_{i,j\in N}\left[a_{ij}-\frac{k_i^\mathrm{out}k_i^\mathrm{in}}{l}\right]\delta_{m_i,m_j}.\end{equation}

\section{Reconstruction Results}
\label{sec:recon_results}
\begin{figure*}[htbp]
    \centering

    \begin{subfigure}[b]{0.9\textwidth}
        \centering
        \includegraphics[width=\textwidth]{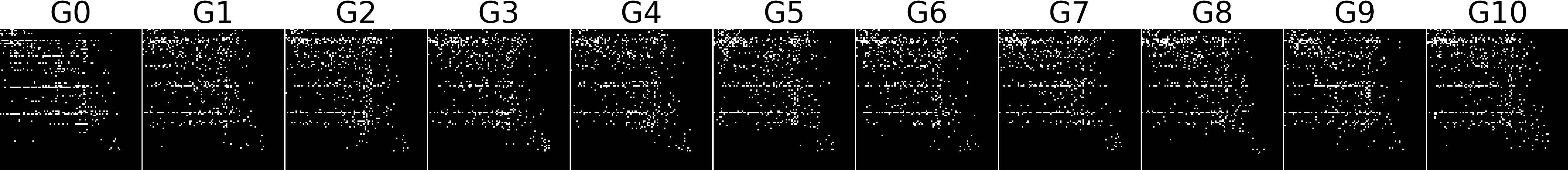}
        \caption{Sample1, Adjacency Matrix}

    \end{subfigure}
    
    \vskip\baselineskip 

    \begin{subfigure}[b]{0.9\textwidth}
        \centering
        \includegraphics[width=\textwidth]{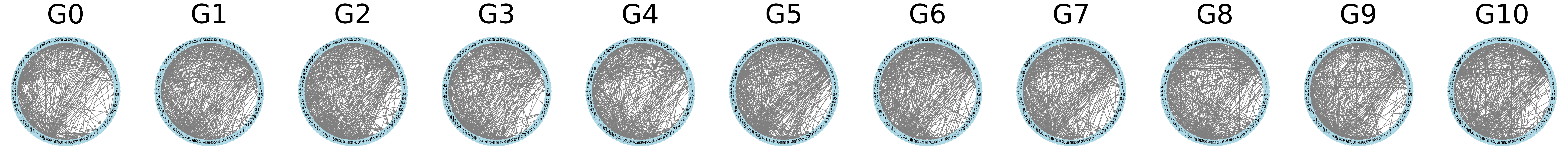}
        \caption{Sample1, Graph Network}

    \end{subfigure}

    \vskip\baselineskip 

    \begin{subfigure}[b]{0.9\textwidth}
        \centering
        \includegraphics[width=\textwidth]{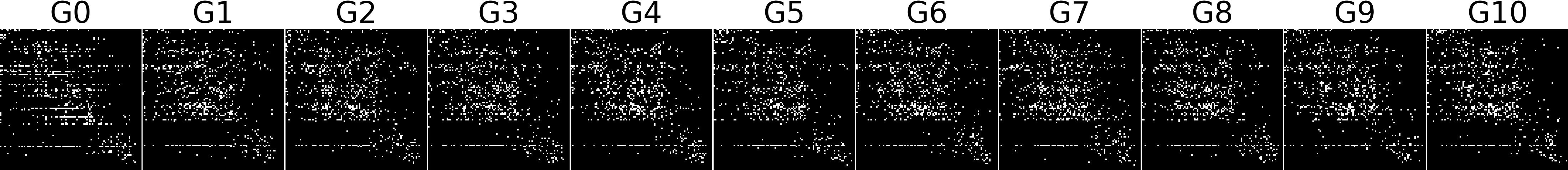}
        \caption{Sample2, Adjacency Matrix}

    \end{subfigure}
    
    \vskip\baselineskip 

    \begin{subfigure}[b]{0.9\textwidth}
        \centering
        \includegraphics[width=\textwidth]{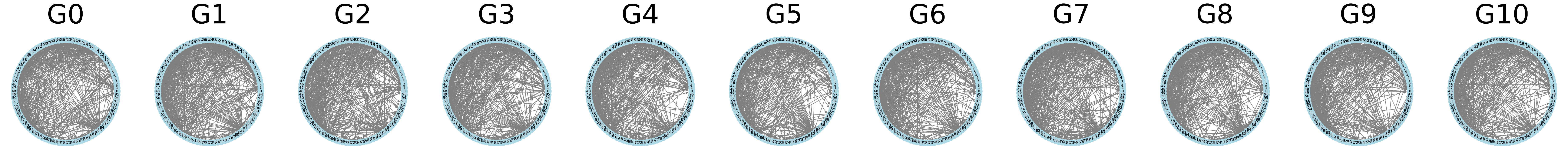}
        \caption{Sample2, Graph Network}

    \end{subfigure}

    \vskip\baselineskip 

    \begin{subfigure}[b]{0.9\textwidth}
        \centering
        \includegraphics[width=\textwidth]{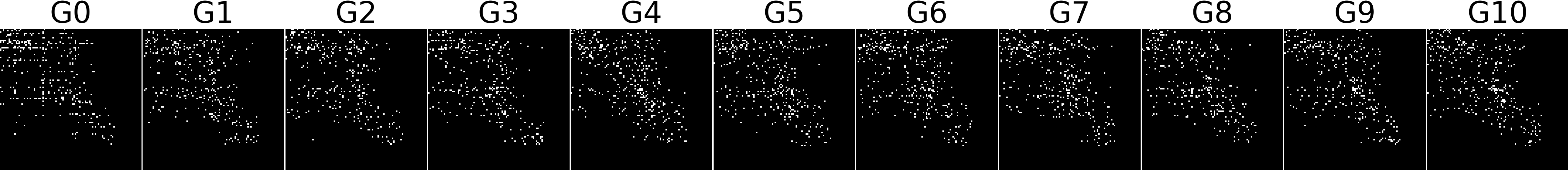}
        \caption{Sample3, Adjacency Matrix}

    \end{subfigure}
    
    \vskip\baselineskip 

    \begin{subfigure}[b]{0.9\textwidth}
        \centering
        \includegraphics[width=\textwidth]{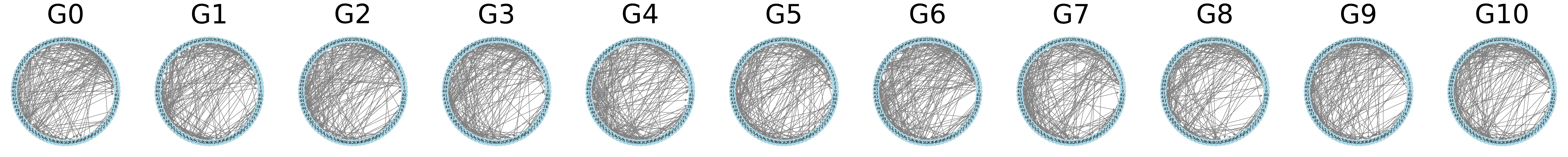}
        \caption{Sample3, Graph Network}

    \end{subfigure}

    \caption{Reconstruction Examples: Original Graphs (G0) and Multiple Decoded Samples (G1-G10). Each row displays one original graph followed by ten distinct reconstructions, resulting from the probabilistic nature of the VAE and the binarization process.}
    \label{fig:reconstruction}
\end{figure*}

The reconstruction results are shown in Figure~\ref{fig:reconstruction}.

\section{Details of SHAP Analysis}
\label{sec:shap_details}

\begin{figure*}[htbp]
    \centering
    \begin{subfigure}[b]{0.45\textwidth}
        \centering
        \includegraphics[width=\textwidth]{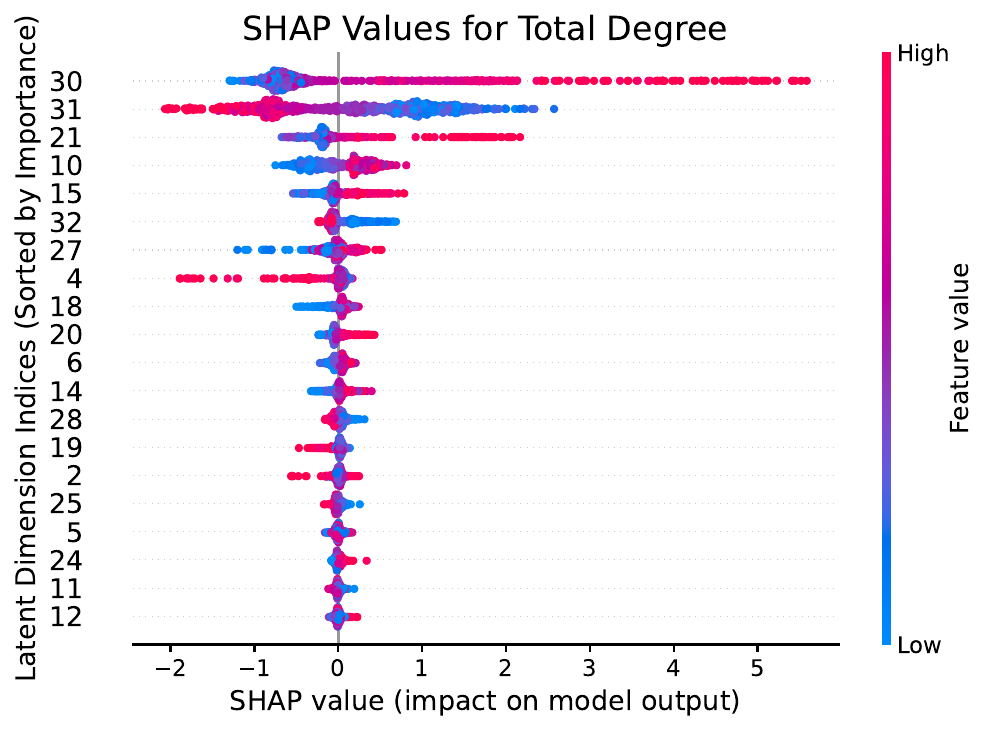}
        \caption{}
        \label{fig:shap_degree}
    \end{subfigure}
    \hfill
    \begin{subfigure}[b]{0.45\textwidth}
        \centering
        \includegraphics[width=\textwidth]{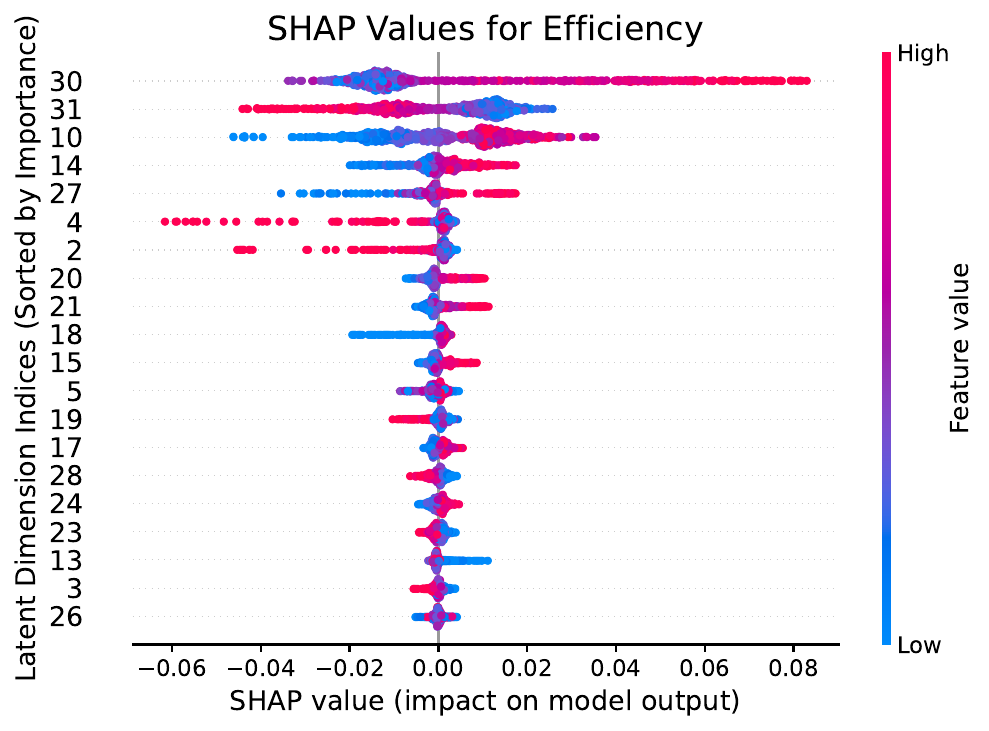}
        \caption{}
        \label{fig:shap_efficiency}
    \end{subfigure}
    
    \vskip\baselineskip 
    
    \begin{subfigure}[b]{0.45\textwidth}
        \centering
        \includegraphics[width=\textwidth]{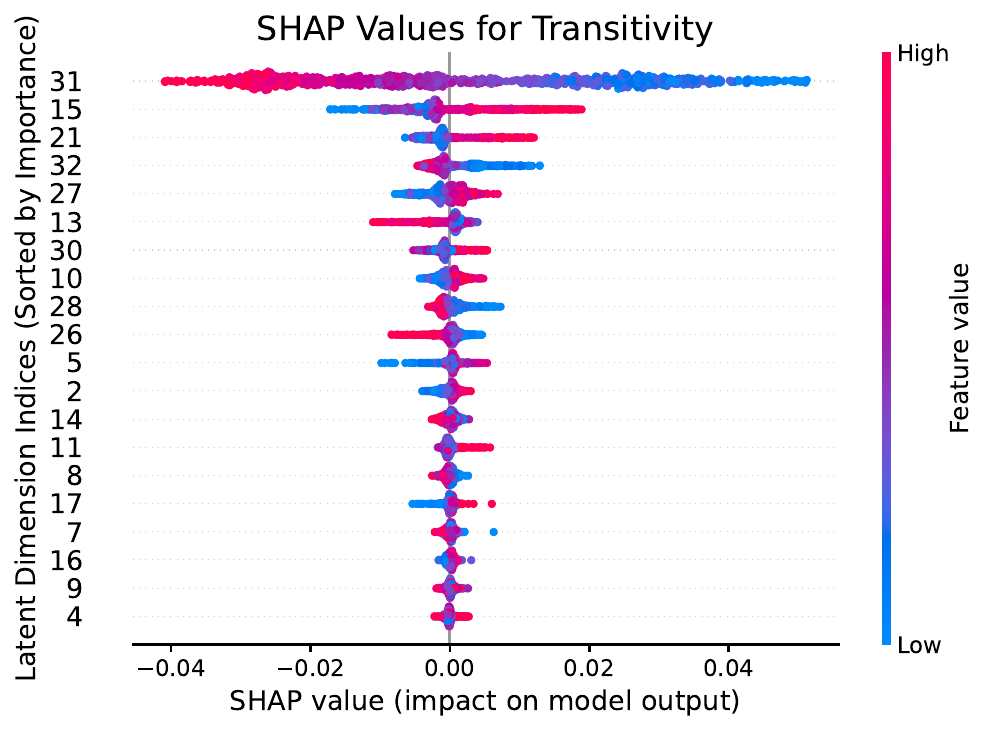}
        \caption{}
        \label{fig:shap_transitivity}
    \end{subfigure}
    \hfill
    \begin{subfigure}[b]{0.45\textwidth}
        \centering
        \includegraphics[width=\textwidth]{reports/nips_lxy/figures/shap/shap_analysis_clustering_coefficient.pdf}
        \caption{}
        \label{fig:shap_clustering}
    \end{subfigure}

    \vskip\baselineskip 
    
    \begin{subfigure}[b]{0.45\textwidth}
        \centering
        \includegraphics[width=\textwidth]{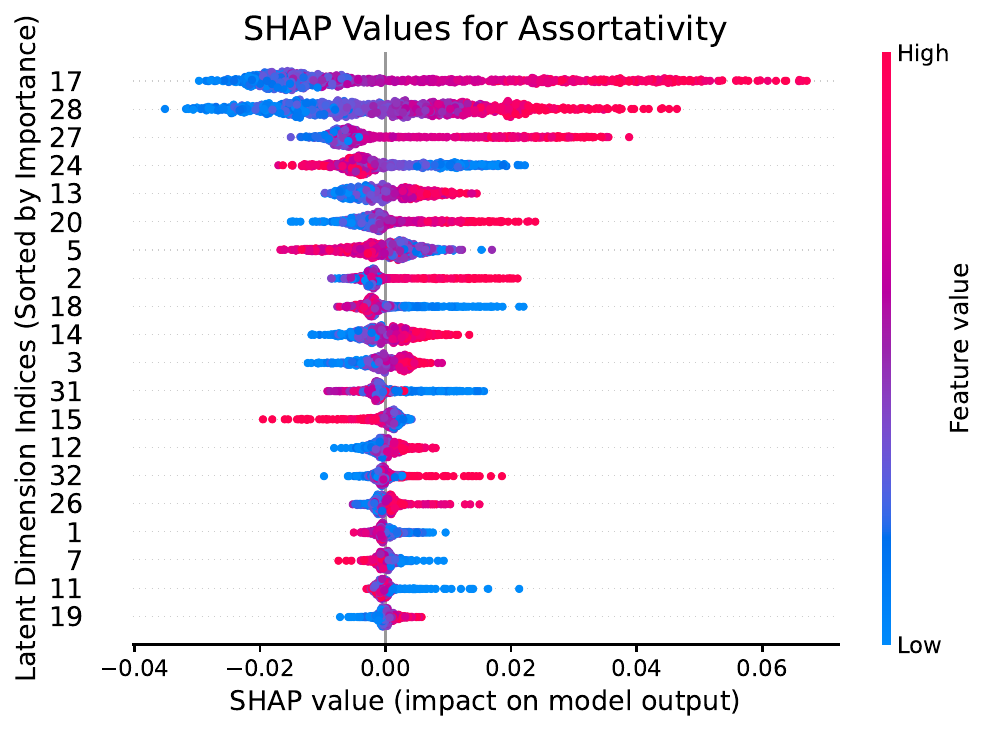}
        \caption{}
        \label{fig:shap_assortivity}
    \end{subfigure}
    \hfill
    \begin{subfigure}[b]{0.45\textwidth}
        \centering
        \includegraphics[width=\textwidth]{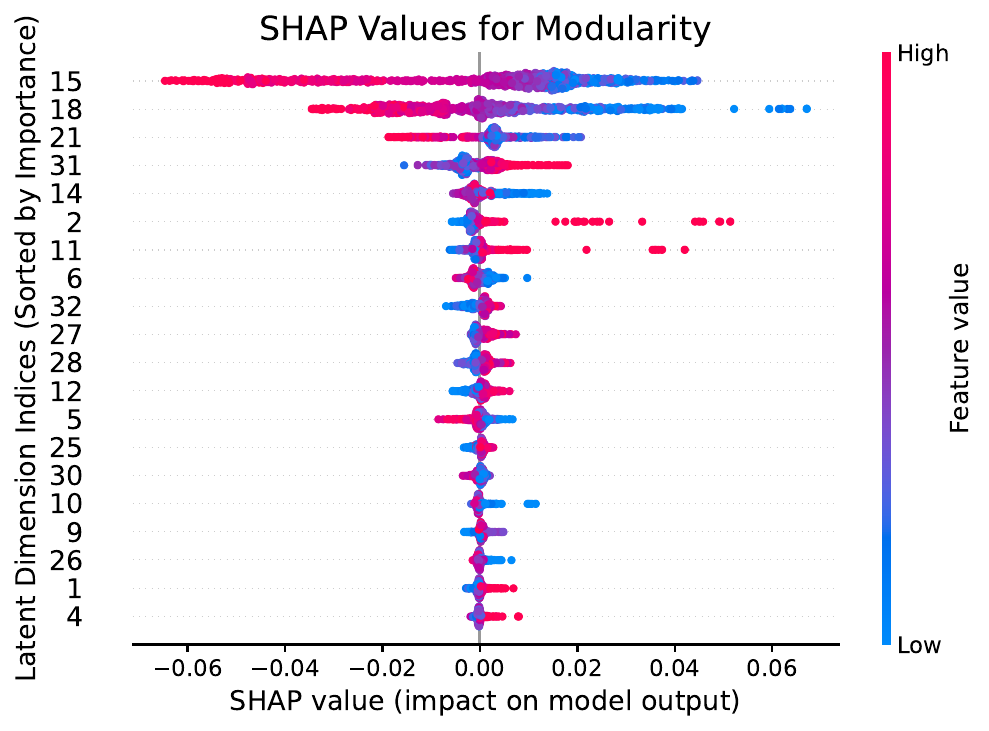}
        \caption{}
        \label{fig:shap_modularity}
    \end{subfigure}

    \caption{SHAP analysis for different metrics. The dimensions are sorted by their importance of contribution and only the top 20 are displayed due to limited space.}
    \label{fig:SHAP_detail}
\end{figure*}

To understand the relationship between the latent space and the generated graph properties, we trained a random forest regression model to predict each graph metric of the generated graphs based on the 32-dimensional latent vectors. Subsequently, we performed standard SHAP analysis and visualized the SHAP values for each latent dimension. The SHAP analysis details for different metrics is shown in Figure~\ref{fig:SHAP_detail}.

For instance, the SHAP analysis for assortativity indicates that a few dimensions significantly influence the assortativity. 

\section{Details of the Linear Regression Model}
\label{sec:linear_regression}
Formally, let $y\in \{0,1,2,...,19\}$ represent the bin index for a given metric, corresponding to the 20 bins of the metric's values. Given $N$ graphs in our test set, we first compute their latent codes: $\mathbf{z}^{(i)*}=E(\mathcal{G}_\pi^{(i)})$, where $\mathbf{z}^{(i)*}$ is a 32-dimensional vector representing the latent code for the $i$-th graph $\mathcal{G}_\pi^{(i)}$ obtained from the encoder $E$. We normalize the latent codes to have zero mean and unit variance dimension-wise by $\mathbf{z}^{(i)}=\displaystyle\frac{\mathbf{z}^{(i)*}-\mathbf{\mu}_{\mathbf{z}}}{\mathbf{\sigma}_{\mathbf{z}}}$. The linear regression model uses these normalized latent codes $\mathbf{z}^{(i)}$ to predict the corresponding bin index $\hat y$:
\begin{equation}
    \hat y=f(\mathbf{z})=\mathbf{w}^T\mathbf{z}+b.
\end{equation}
We fit this linear equation using Ridge regression. The gradient of the predicted bin index 
$\hat y$ with respect to the normalized latent vector $\mathbf{z}$ is then given by: $\nabla f(\mathbf{z})=\mathbf{w}$. Finally, we consider the direction of the gradient by normalization: $\mathbf{w}_0 = \displaystyle \frac{\mathbf{w}}{|\mathbf{w}|}$.

\begin{figure*}[htbp]
\centering
\includegraphics[width=1.0\textwidth]{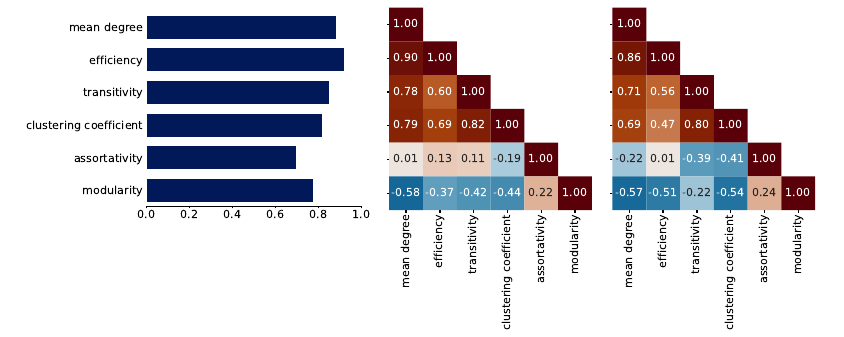} 
\caption{Left: $\text{R}^2$ scores of the linear regression models predicting each of the six graph metrics from the 32-dimensional latent embeddings, indicating the predictive strength of the latent space. Middle: Spearman's rank correlation coefficient matrix between the six graph metric descriptors, calculated directly from the test dataset. Right: Cosine distance matrix between the gradient directions in the 32-dimensional latent space for each of the six graph metrics, illustrating the similarity in how different metrics are encoded in the latent space.}
\label{fig:correlations}
\end{figure*}

The $R^2$ scores of the linear regression, Spearman's rank correlation coefficient matrix between the six graph metric descriptors and cosine distance matrix between the gradient directions in the 32-dimensional latent space are shown in Fig~\ref{fig:correlations}

\section{Details of Gradient Direction Moving}
\label{sec:gradient_direction_moving}

\begin{figure*}[h!]
    \centering
    \begin{subfigure}[b]{0.3\textwidth}
        \centering
        \includegraphics[width=\textwidth]{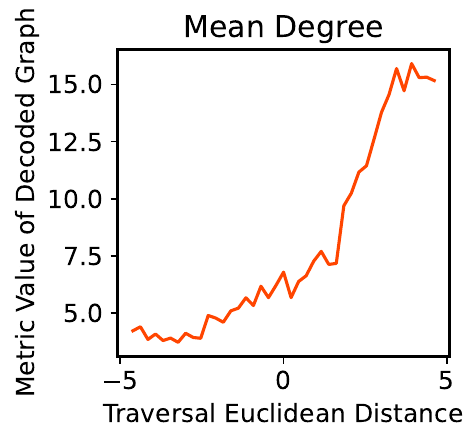}
        \caption{}
        \label{fig:grad_mean_degree}
    \end{subfigure}
    \hfill
    \begin{subfigure}[b]{0.3\textwidth}
        \centering
        \includegraphics[width=\textwidth]{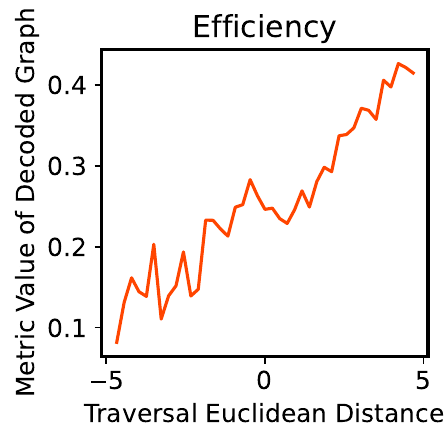}
        \caption{}
        \label{fig:grad_efficiency}
    \end{subfigure}
    \hfill
    \begin{subfigure}[b]{0.3\textwidth}
        \centering
        \includegraphics[width=\textwidth]{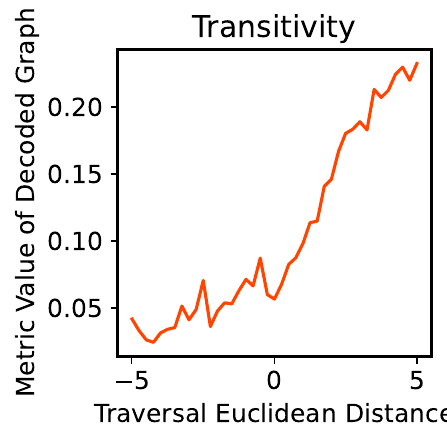}
        \caption{}
        \label{fig:grad_transitivity}
    \end{subfigure}
    
    \vskip\baselineskip 

    \begin{subfigure}[b]{0.3\textwidth}
        \centering
        \includegraphics[width=\textwidth]{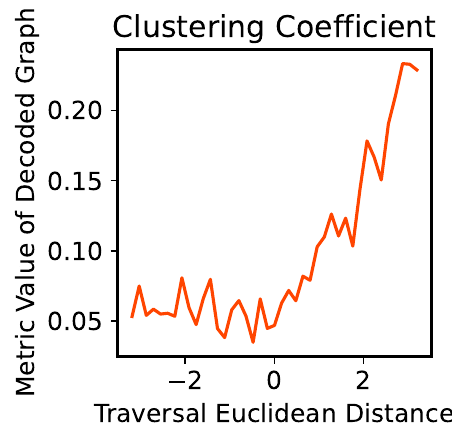}
        \caption{}
        \label{fig:grad_clustering}
    \end{subfigure}
    \hfill
    \begin{subfigure}[b]{0.3\textwidth}
        \centering
        \includegraphics[width=\textwidth]{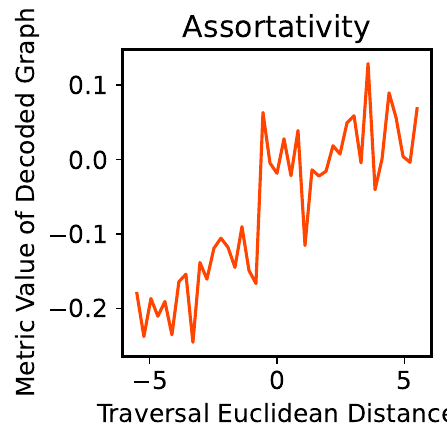}
        \caption{}
        \label{fig:grad_assortivity}
    \end{subfigure}
    \hfill
    \begin{subfigure}[b]{0.3\textwidth}
        \centering
        \includegraphics[width=\textwidth]{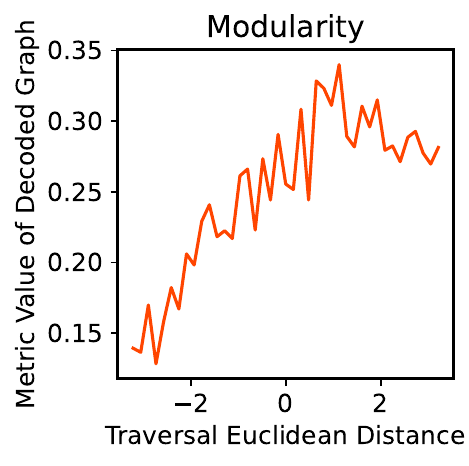}
        \caption{}
        \label{fig:grad_modularity}
    \end{subfigure}

    \caption{Metric variation of decoded graphs along the directions of different gradient vectors in the latent space.}
    \label{fig:move_along_grad}
\end{figure*}

To validate whether these identified directions truly reflect changes in the corresponding metrics, we performed a traversal experiment. Starting from the mean latent vector of the test set, we moved along each metric's identified gradient direction (both positive and negative shifts) and decoded the resulting latent vectors to observe the change in the metric's value (Figure~\ref{fig:move_along_grad}). The moving region was carefully chosen to ensure that the shifted latent vector with the maximum offset remained within the range of $[\mathbf{\mu_z}-2\mathbf{\sigma_z},\mathbf{\mu_z}+2\mathbf{\sigma_z}]$ across all dimensions, where $\mu_z$ and $\sigma_z$ are the mean and standard deviation of the latent vectors in the test set, respectively.

\begin{figure*}[tbp] 
    \centering

    \begin{subfigure}[b]{0.7\textwidth}
        \centering
        \includegraphics[width=\textwidth]{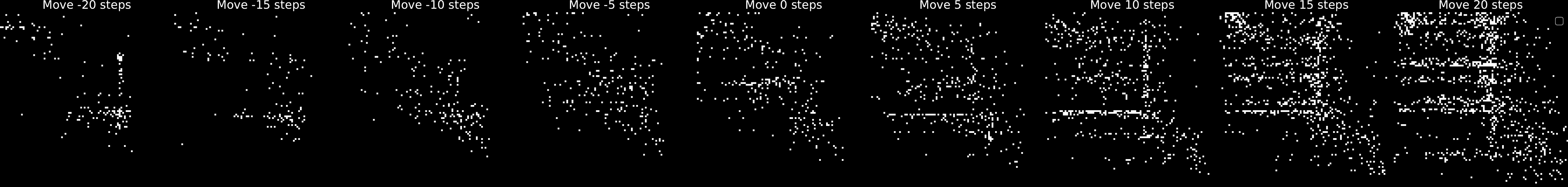}
        \caption{Adjacency matrix along gradient of mean degree}
    \end{subfigure}
    
    \vskip\baselineskip 

    \begin{subfigure}[b]{0.7\textwidth}
        \centering
        \includegraphics[width=\textwidth]{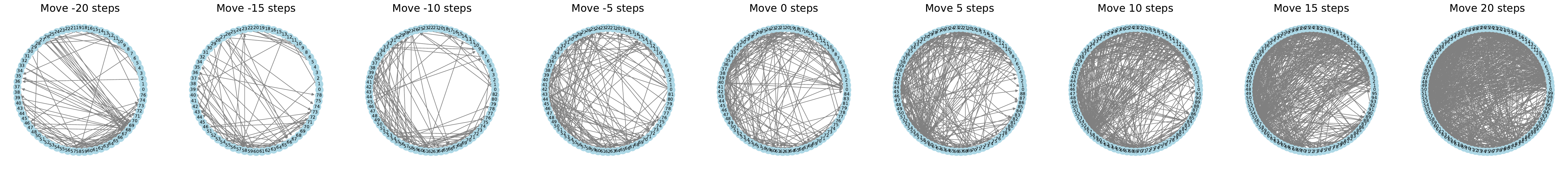}
        \caption{Network structure along gradient of mean degree}
    \end{subfigure}

    \vskip\baselineskip 

    \begin{subfigure}[b]{0.7\textwidth}
        \centering
        \includegraphics[width=\textwidth]{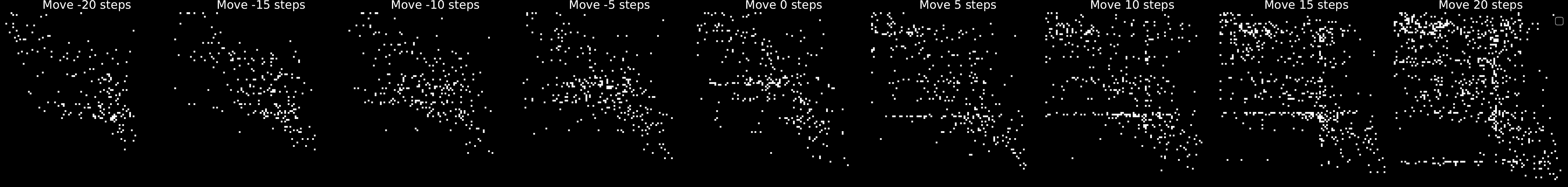}
        \caption{Adjacency matrix along gradient of efficiency}
    \end{subfigure}
    
    \vskip\baselineskip 

    \begin{subfigure}[b]{0.7\textwidth}
        \centering
        \includegraphics[width=\textwidth]{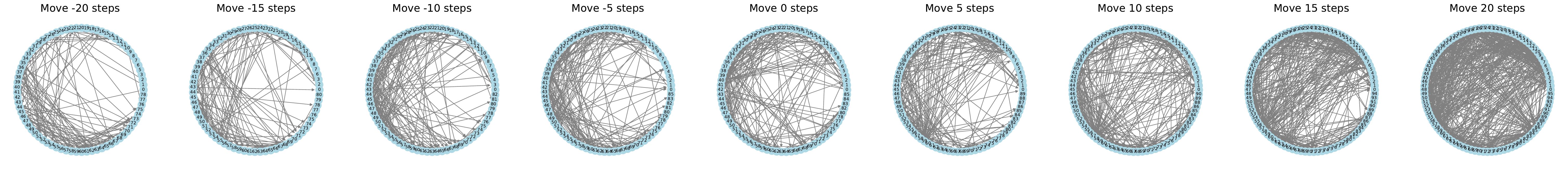}
        \caption{Network structure along gradient of efficiency}
    \end{subfigure}

    \vskip\baselineskip 

    \begin{subfigure}[b]{0.7\textwidth}
        \centering
        \includegraphics[width=\textwidth]{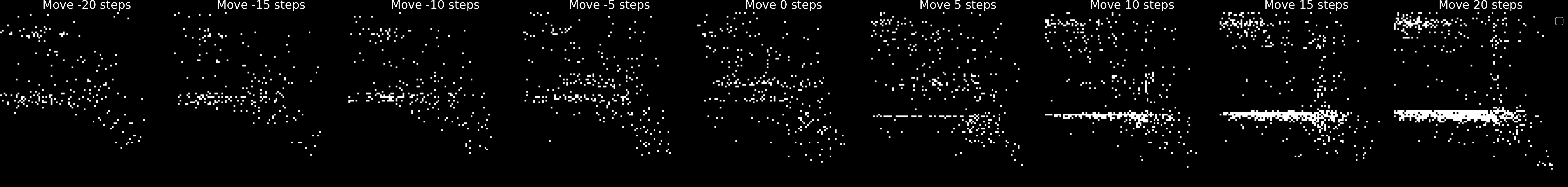}
        \caption{Adjacency matrix along gradient of transitivity}
    \end{subfigure}
    
    \vskip\baselineskip 

    \begin{subfigure}[b]{0.7\textwidth}
        \centering
        \includegraphics[width=\textwidth]{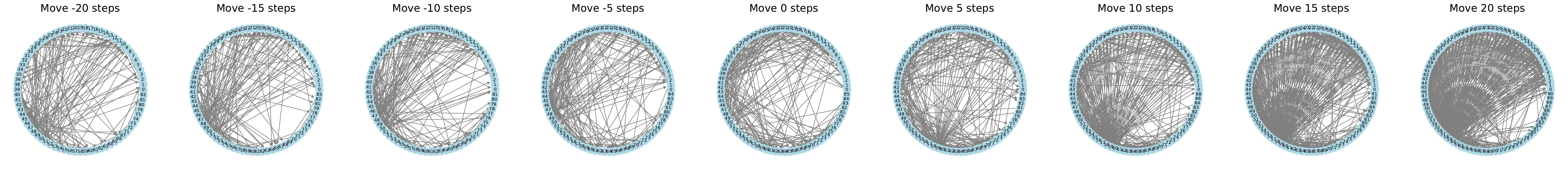}
        \caption{Network structure along gradient of transitivity}
    \end{subfigure}

    \caption{Generated samples by traversing the latent space along the gradient direction of different metrics (Part 1). Images show decoded graphs as the latent vector moves from a center point (middle) towards the positive (right) and negative (left) gradient directions. Each number indicates the Euclidean distance from the center point of the dataset.}
    \label{fig:move_along_grad_samples_part1}
\end{figure*}

\begin{figure*}[tbp] 
    \centering
    \begin{subfigure}[b]{0.7\textwidth}
        \centering
        \includegraphics[width=\textwidth]{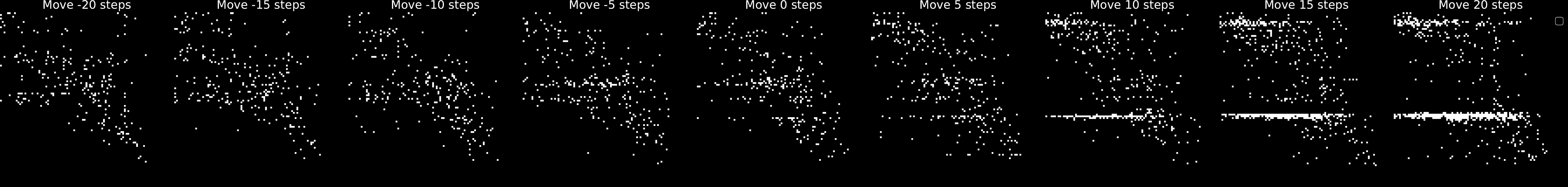}
        \caption{Adjacency matrix along gradient of clustering coefficient}
    \end{subfigure}
    
    \vskip\baselineskip 

    \begin{subfigure}[b]{0.7\textwidth}
        \centering
        \includegraphics[width=\textwidth]{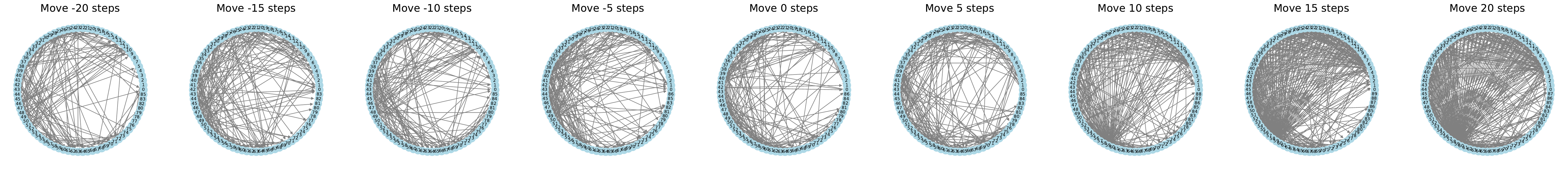}
        \caption{Network structure along gradient of clustering coefficient}
    \end{subfigure}

    \vskip\baselineskip 

    \begin{subfigure}[b]{0.7\textwidth}
        \centering
        \includegraphics[width=\textwidth]{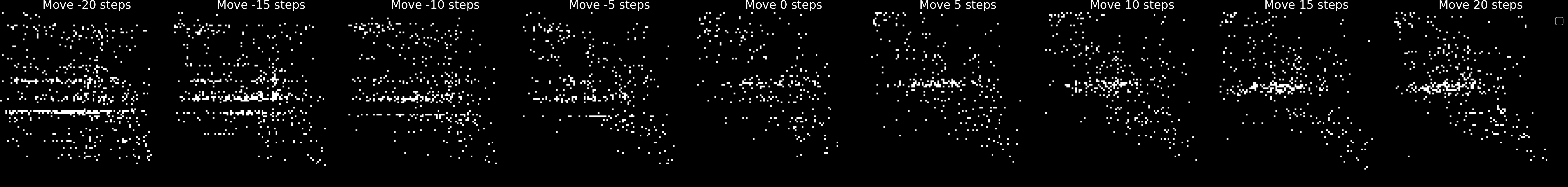}
        \caption{Adjacency matrix along gradient of assortativity}
    \end{subfigure}
    
    \vskip\baselineskip 

    \begin{subfigure}[b]{0.7\textwidth}
        \centering
        \includegraphics[width=\textwidth]{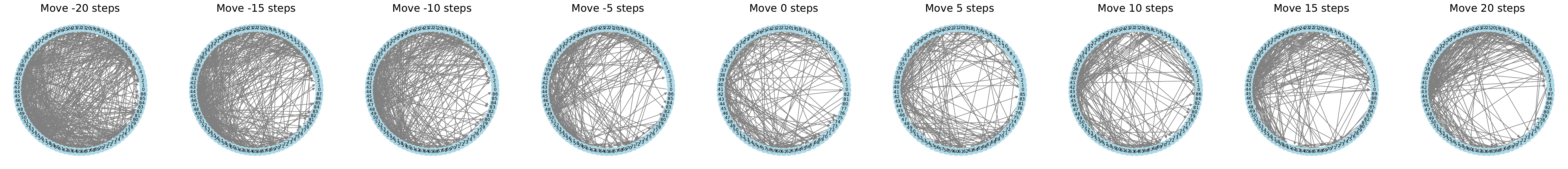}
        \caption{Network structure along gradient of assortativity}
    \end{subfigure}

    \vskip\baselineskip 

    \begin{subfigure}[b]{0.7\textwidth}
        \centering
        \includegraphics[width=\textwidth]{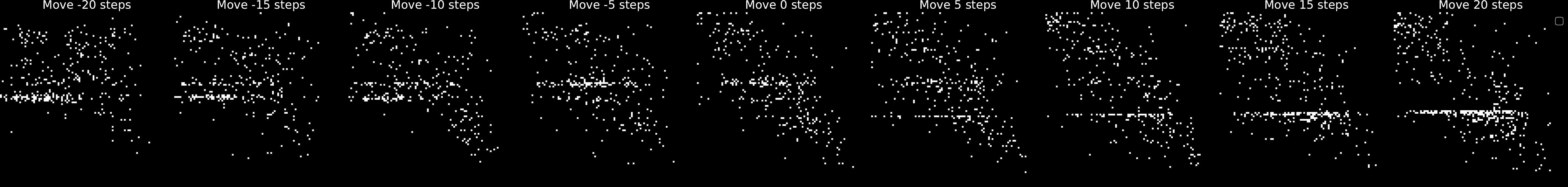}
        \caption{Adjacency matrix along gradient of modularity}
    \end{subfigure}
    
    \vskip\baselineskip 

    \begin{subfigure}[b]{0.7\textwidth}
        \centering
        \includegraphics[width=\textwidth]{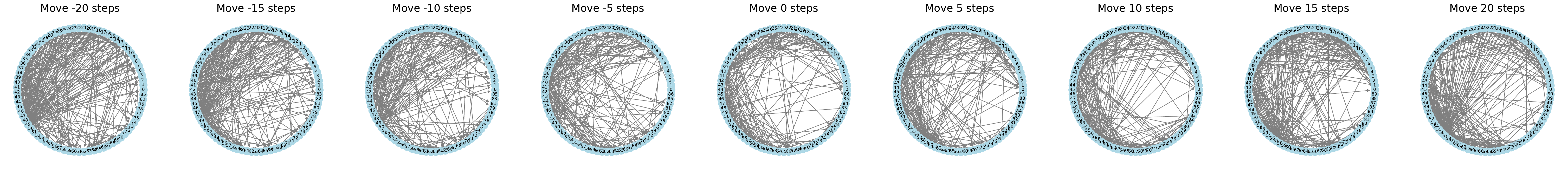}
        \caption{Network structure along gradient of modularity}
    \end{subfigure}

    \caption{Generated samples by traversing the latent space along the gradient direction of different metrics (Part 2). Images show decoded graphs as the latent vector moves from a center point (middle) towards the positive (right) and negative (left) gradient directions. Each number indicates the Euclidean distance from the center point of the dataset.}
    \label{fig:move_along_grad_samples_part2}
\end{figure*}

We show the generated graphs when we move along the direction of gradient of different metrics in Figure~\ref{fig:move_along_grad_samples_part1} and Figure~\ref{fig:move_along_grad_samples_part2}.

\section{Details of Correlations Between Graph Metrics}
\label{sec:correlations_details}
\begin{figure*}[p] 
    \centering
    \subfloat[]{\includegraphics[width=0.3\textwidth]{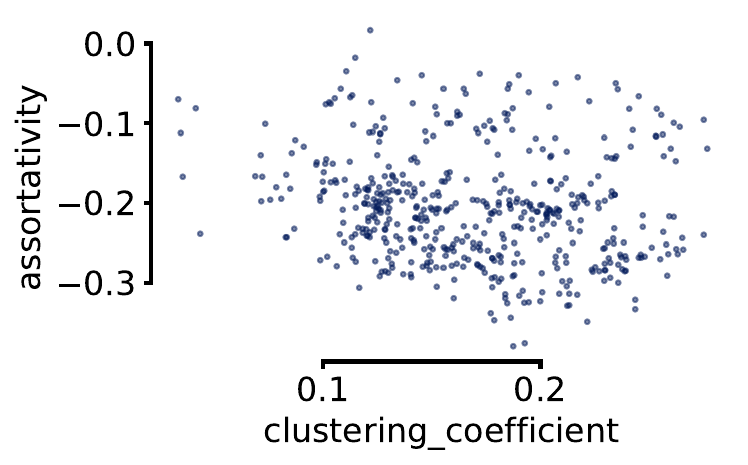}}
    \subfloat[]{\includegraphics[width=0.3\textwidth]{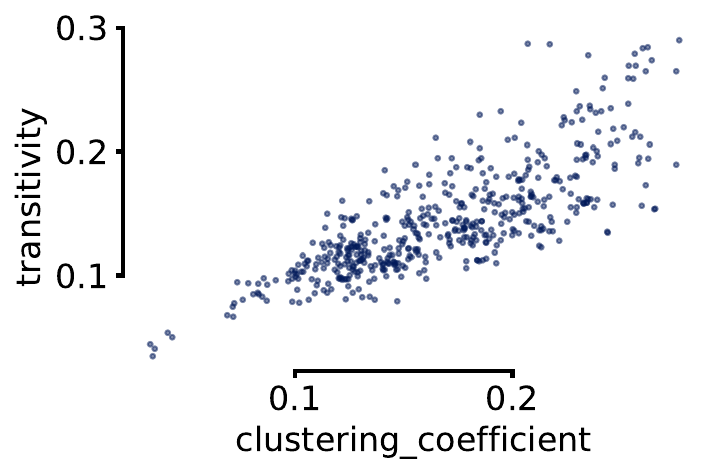}}
    \subfloat[]{\includegraphics[width=0.3\textwidth]{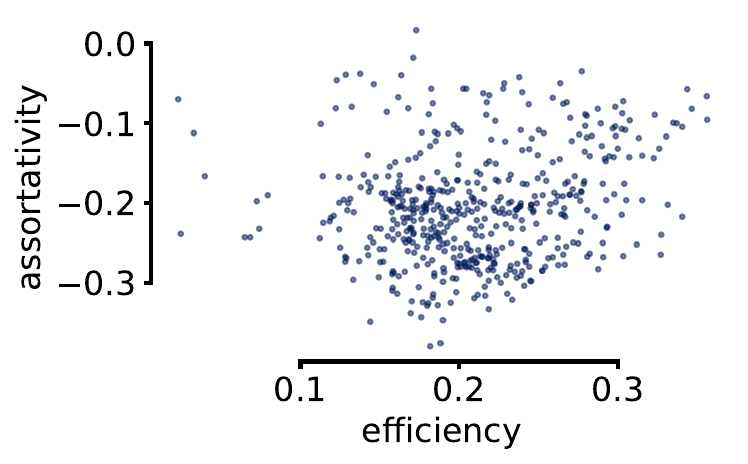}}
    \\[0.5\baselineskip] 

    \subfloat[]{\includegraphics[width=0.3\textwidth]{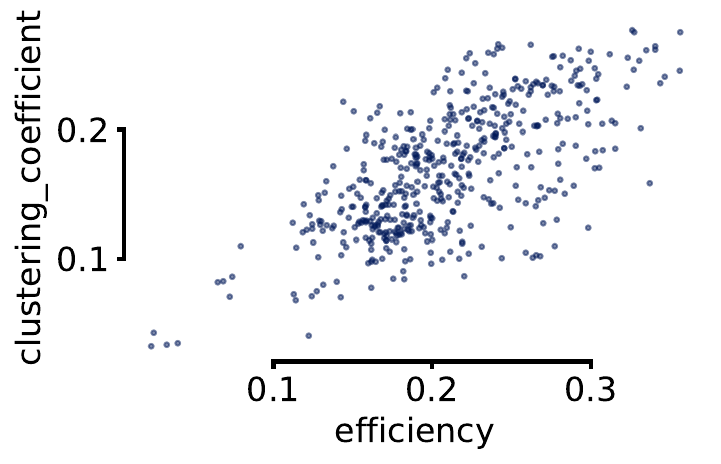}}
    \subfloat[]{\includegraphics[width=0.3\textwidth]{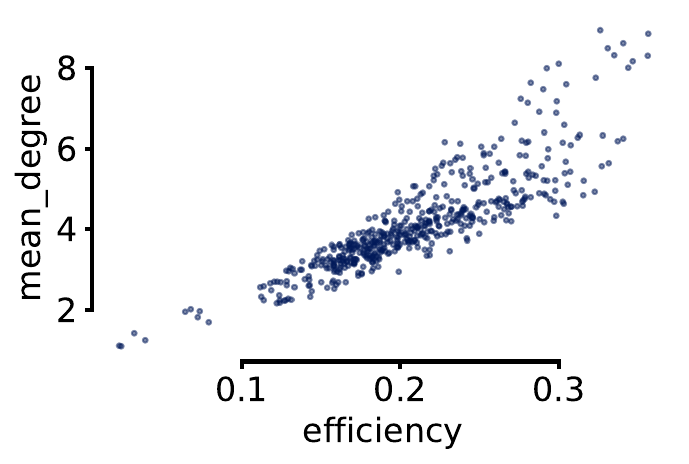}}
    \subfloat[]{\includegraphics[width=0.3\textwidth]{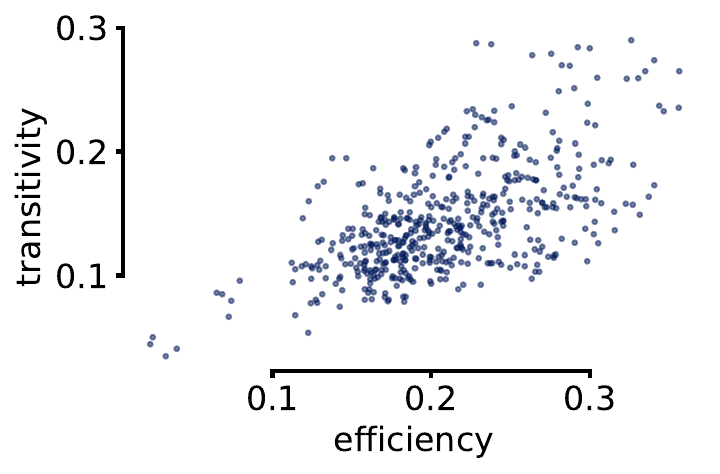}}
    \\[0.5\baselineskip]

    \subfloat[]{\includegraphics[width=0.3\textwidth]{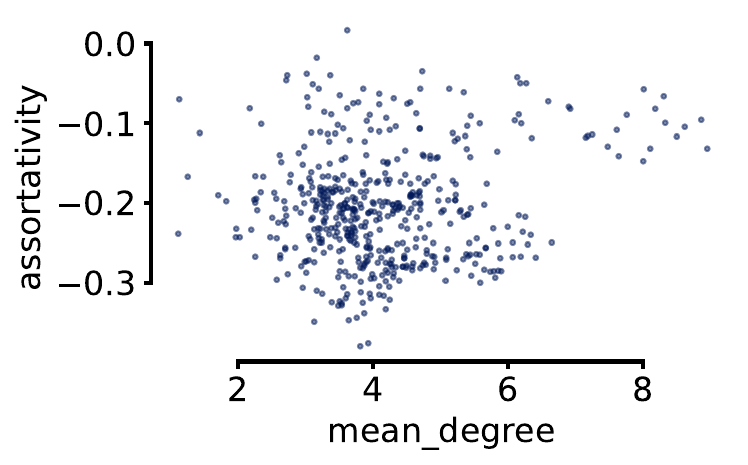}}
    \subfloat[]{\includegraphics[width=0.3\textwidth]{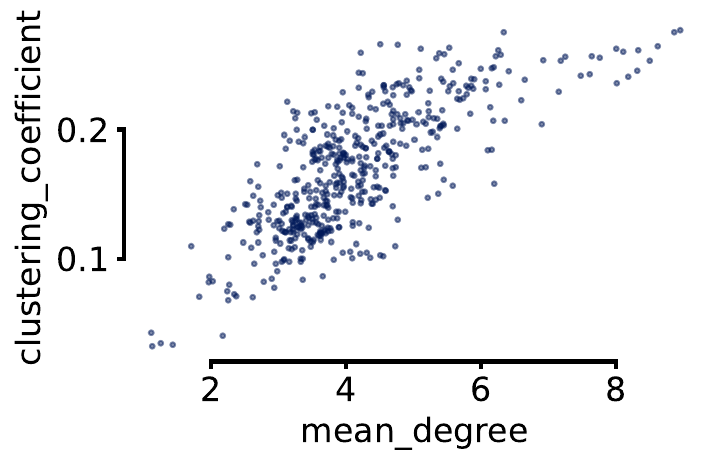}}
    \subfloat[]{\includegraphics[width=0.3\textwidth]{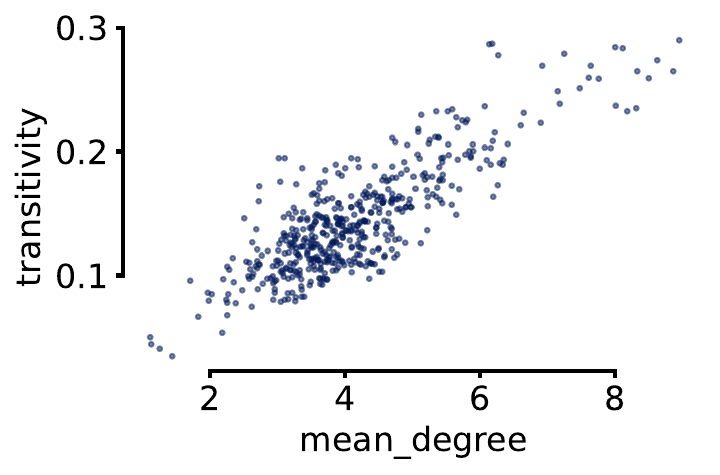}}
    \\[0.5\baselineskip]

    \subfloat[]{\includegraphics[width=0.3\textwidth]{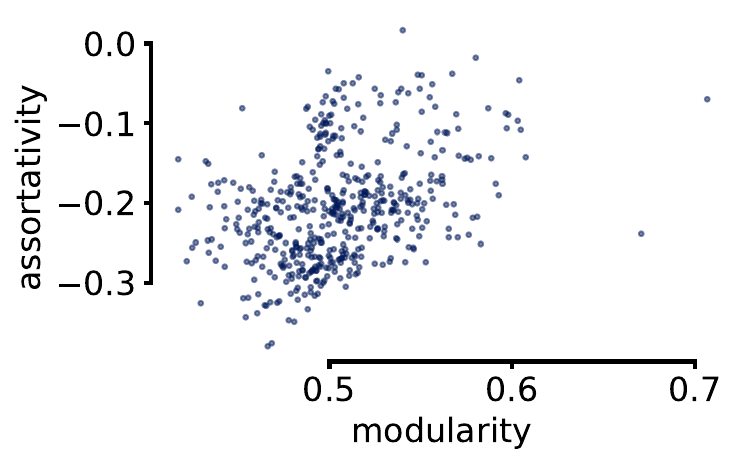}}
    \subfloat[]{\includegraphics[width=0.3\textwidth]{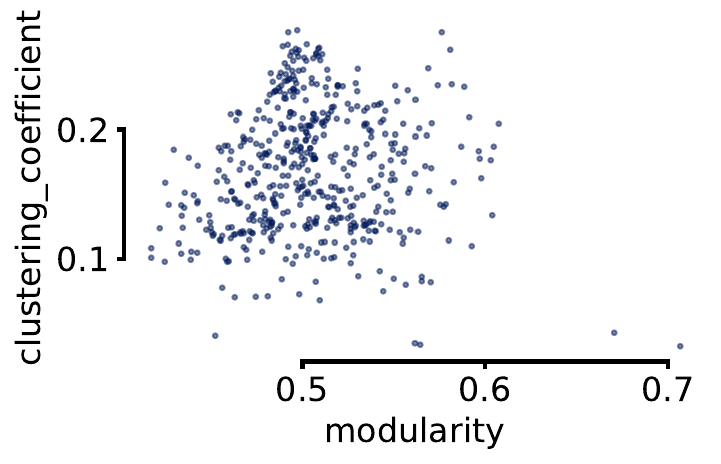}}
    \subfloat[]{\includegraphics[width=0.3\textwidth]{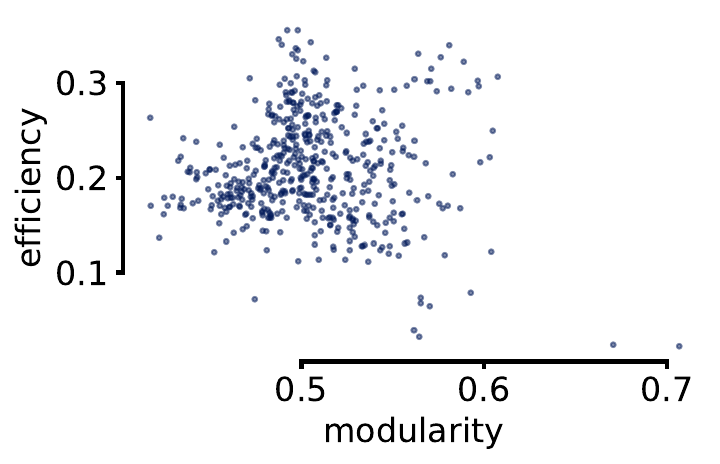}}
    \\[0.5\baselineskip]


    \subfloat[]{\includegraphics[width=0.3\textwidth]{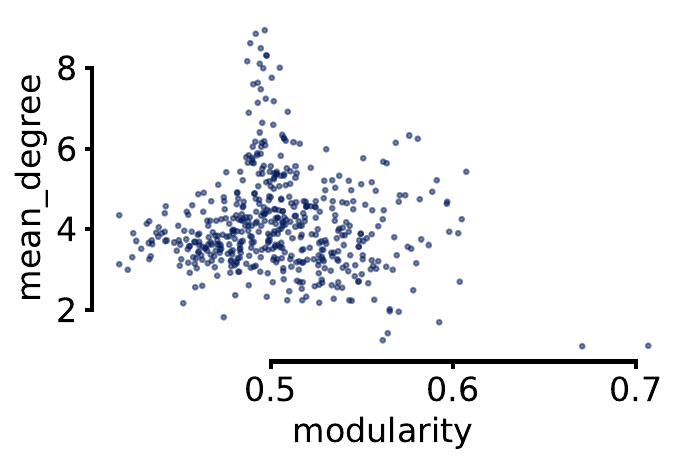}}
    \subfloat[]{\includegraphics[width=0.3\textwidth]{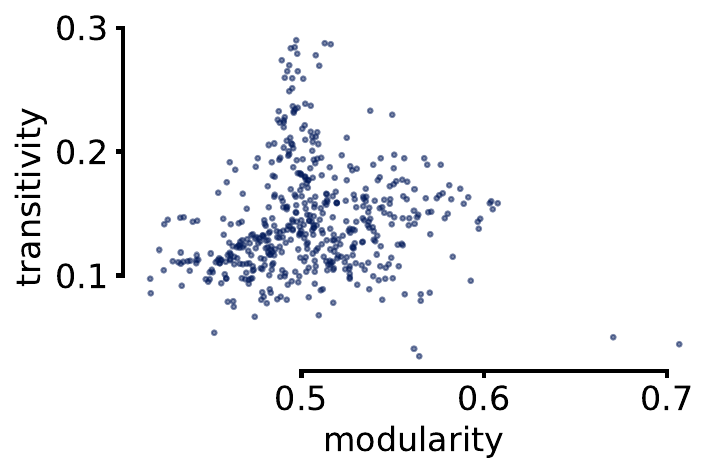}}
    \subfloat[]{\includegraphics[width=0.3\textwidth]{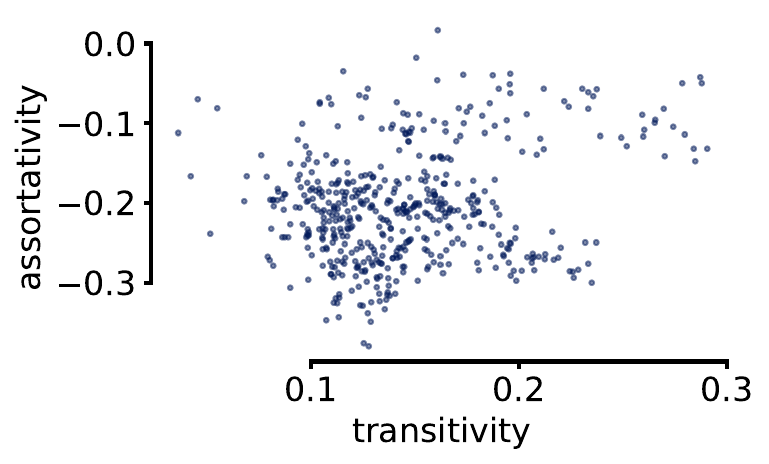}}

    \caption{Correlation between Pairs of Graph Metrics}
    \label{fig:pairs_of_correlations}
\end{figure*}

The correlations between pairs of graph metrics are shown in Figure~\ref{fig:pairs_of_correlations}.

\section{MCMC Sampling Details}
\label{sec:mcmc_details}
For features $\mathbf{z^*}=(z_1, ..., z_{32})$, we first normalize it to zero-mean and unit-variance $\mathbf{z}$. Then we fit a linear regression model from the 32-dimensional feature to the percentile of certain property, for example, mean degree. Denote the linear regression model as:
\begin{equation}
    y=f(\mathbf{z})=\mathbf{w}^T\mathbf{z}+b
\end{equation}
We rewrite the LTD function as:
\begin{equation}
    LTD(\mathbf{z}) = 
\begin{cases}
  w\text{log}p(\textbf{z}) & \text{if } \mathbf{z}\in \Omega_\mathcal{T}\\
  -\infty & \text{else }  \\
\end{cases}
\end{equation}

where $p(\textbf{z})$ is estimated by fitting a multivariate Gaussian distribution according to the correlations between dimensions of the normalized latent vectors:
\begin{equation}f(\mathbf{x};\boldsymbol{\mu},\boldsymbol{\Sigma})=\frac{1}{\sqrt{(2\pi)^{D}|\boldsymbol{\Sigma}|}}\exp\left(-\frac{1}{2}(\mathbf{x}-\boldsymbol{\mu})^{T}\boldsymbol{\Sigma}^{-1}(\mathbf{x}-\boldsymbol{\mu})\right)\end{equation}
and $w=\displaystyle\frac{1}{\tau}$ is the weight to control the degree how the sampled points concentrated around the high density area of the latent space. 
Let the target value to be $t$
We find the initial feasible point $\mathbf{z_0}$ by equation:
\begin{equation}
    k = \displaystyle \frac{t-\bar y}{|\mathbf{w}|^2},
\end{equation}
\begin{equation}
    \mathbf{z_0}=k\mathbf{w}.
\end{equation}
This point must satisfy the condition $|f(\mathbf{z})-t|<\epsilon$ because the linear regression plane pass the center point of the dataset, where $\mathbf{z}=\mathbf{0}$ and $y=\bar y$.

From this initial feasible solution, we run the Metropolis-Hastings Algorithm:

\begin{algorithm*}
\caption{Metropolis-Hastings Algorithm}
\label{alg:1}
\begin{algorithmic}[1]
\State \textbf{Input:} 
\State \quad $t$: Target value;
\State \quad $\mathbf{z_0}$: Initial feasible point;
\State \quad $\text{Samples}$: The list of sampled points, initialized as an empty list;
\State \quad $\mathbf{\Sigma}, \mathbf{\mu}$: The covariance matrix and mean vector of the fitted multivariate Gaussian distribution;
\State \quad $f(\mathbf{z})$: Linear regression function;
\State \quad $w$: Prior probability weight;
\State \quad $\sigma$: Proposal standard deviation;
\State \quad $N$: Total sample number;
\State \quad $burn$: burn in round;
\State \quad $thin$: Sampling interval.

\State \textbf{Output:} A list of sampled latent vectors: $\mathbf{z}_0, ..., \mathbf{z}_N$.
\Statex
\State $c=\mathbf{x}_0$
\State $\text{Samples}.append(c)$
\State $\text{total iteration} = burn + N * thin$
\For{$i = 1$ \textbf{to} $\text{total iteration}$}
    \State Propose a new point $c'$ by disturbing $c$ a small step: $c'\sim \mathcal{N}(c, \sigma^2)$
    \State $\text{log acceptance ratio}=\text{LTD}(c')-\text{LTD}(c)$
    \State $r \sim \mathcal{U}(0, 1)$
    \If {$\text{log}r<\text{log acceptance ratio}$}
        $c=c'$
    \EndIf
    \If {$i \geq burn$ and $(i-burn) \text{ mod } thin==0$}
        \State $\text{Samples}.append(c)$
    \EndIf
\EndFor
\State \Return $\text{Samples}$
\end{algorithmic}
\end{algorithm*}

In the specific setting, we adopt $burn=0$, $w=10$, $\epsilon=0.1$, $\sigma=0.01$ and $thin=1$. 

\section{Generated Graphs for Different Targets}
\label{sec:controlled_generation_results}

\begin{figure*}[htbp]
    \centering
    \begin{subfigure}[b]{0.27\textwidth}
        \centering
        \includegraphics[width=\textwidth]{reports/nips_lxy/figures/fig_controllable_gene/curves/target_name=mean_degree_curves_with_percentile.pdf}
        \caption{}
        \label{fig:mean_degree_apdx}
    \end{subfigure}
    \hfill
    \begin{subfigure}[b]{0.27\textwidth}
        \centering
        \includegraphics[width=\textwidth]{reports/nips_lxy/figures/fig_controllable_gene/curves/target_name=efficiency_curves_with_percentile.pdf}
        \caption{}
        \label{fig:efficiency_apdx}
    \end{subfigure}
    \hfill
    \begin{subfigure}[b]{0.27\textwidth}
        \centering
        \includegraphics[width=\textwidth]{reports/nips_lxy/figures/fig_controllable_gene/curves/target_name=transitivity_curves_with_percentile.pdf}
        \caption{}
        \label{fig:transitivity_apdx}
    \end{subfigure}
    
    \vskip\baselineskip 

    \begin{subfigure}[b]{0.27\textwidth}
        \centering
        \includegraphics[width=\textwidth]{reports/nips_lxy/figures/fig_controllable_gene/curves/target_name=clustering_coefficient_curves_with_percentile.pdf}
        \caption{}
        \label{fig:clustering_apdx}
    \end{subfigure}
    \hfill
    \begin{subfigure}[b]{0.27\textwidth}
        \centering
        \includegraphics[width=\textwidth]{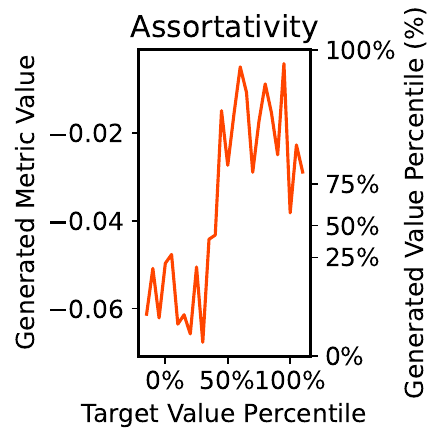}
        \caption{}
        \label{fig:assortivity_apdx}
    \end{subfigure}
    \hfill
    \begin{subfigure}[b]{0.27\textwidth}
        \centering
        \includegraphics[width=\textwidth]{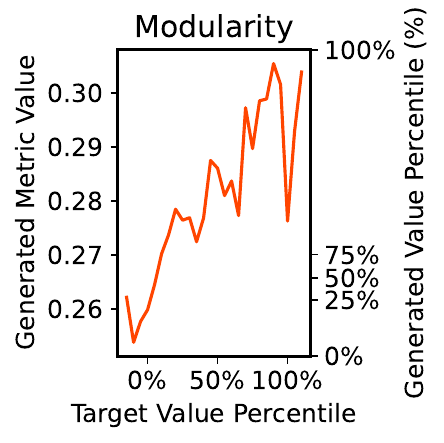}
        \caption{}
        \label{fig:modularity_apdx}
    \end{subfigure}

    \caption{Metrics of generated graphs when setting different targets.}
    \label{fig:control_gene_curves_full}
\end{figure*}

\begin{figure*}[tbp]
    \centering
    \begin{subfigure}[b]{0.9\textwidth}
        \centering
        \includegraphics[width=\textwidth]{reports/nips_lxy/figures/fig_controllable_gene/mean_degree/adj_samples_mean_degree_target=0.png}
        \caption{Target mean degree: 0\%-5\% percentile, Adjacency Matrix} 
    \end{subfigure}

    \vskip\baselineskip 

    \begin{subfigure}[b]{0.9\textwidth}
        \centering
        \includegraphics[width=\textwidth]{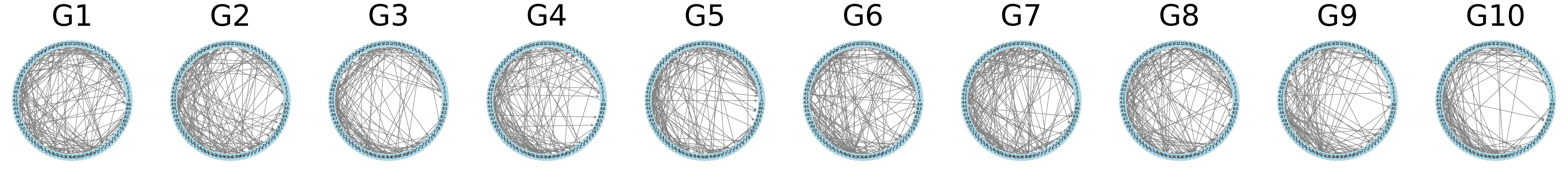}
        \caption{Target mean degree: 0\%-5\% percentile, Network Visualization} 
    \end{subfigure}

    \vskip\baselineskip 

    \begin{subfigure}[b]{0.9\textwidth}
        \centering
        \includegraphics[width=\textwidth]{reports/nips_lxy/figures/fig_controllable_gene/mean_degree/adj_samples_mean_degree_target=10.png}
        \caption{Target mean degree: 50\%-55\% percentile, Adjacency Matrix} 
    \end{subfigure}

    \vskip\baselineskip 

    \begin{subfigure}[b]{0.9\textwidth}
        \centering
        \includegraphics[width=\textwidth]{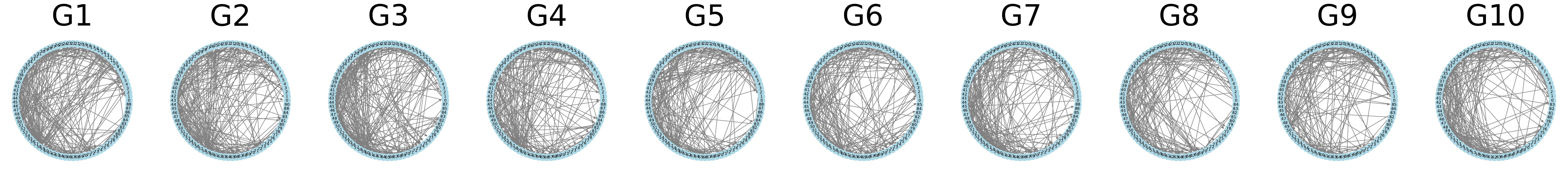}
        \caption{Target mean degree: 50\%-55\% percentile, Network Visualization} 
    \end{subfigure}

    \vskip\baselineskip 

    \begin{subfigure}[b]{0.9\textwidth}
        \centering
        \includegraphics[width=\textwidth]{reports/nips_lxy/figures/fig_controllable_gene/mean_degree/adj_samples_mean_degree_target=20.png}
        \caption{Target mean degree: 95\%-100\% percentile, Adjacency Matrix} 
    \end{subfigure}

    \vskip\baselineskip 

    \begin{subfigure}[b]{0.9\textwidth}
        \centering
        \includegraphics[width=\textwidth]{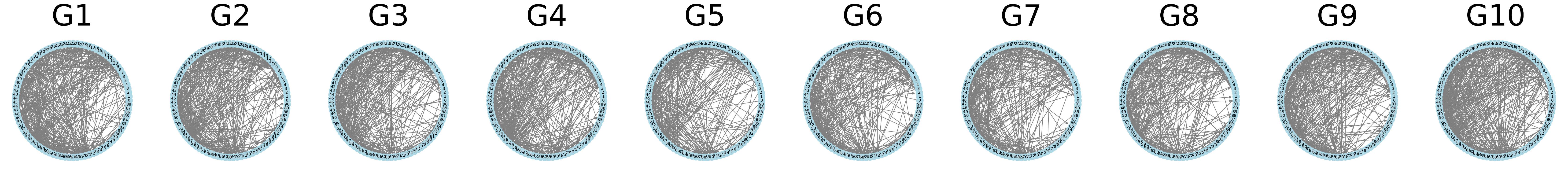}
        \caption{Target mean degree: 95\%-100\% percentile, Network Visualization} 
    \end{subfigure}

    \caption{Generated graph examples targeting different mean degree percentile ranges. For each target range, 10 graphs were randomly selected from 1000 graphs sampled and decoded. Pairs of adjacency matrices and corresponding network visualizations are shown for target mean degrees in the 0\%-5\% (a, b), 50\%-55\% (c, d), and 95\%-100\% (e, f) percentile ranges of the dataset.} 
    \label{fig:control_mean_degree}
\end{figure*}

\begin{figure*}[tbp]
    \centering
    \begin{subfigure}[b]{0.9\textwidth}
        \centering
        \includegraphics[width=\textwidth]{reports/nips_lxy/figures/fig_controllable_gene/efficiency/adj_samples_efficiency_target=0.png}
        \caption{Target efficiency: 0\%-5\% percentile, Adjacency Matrix} 
    \end{subfigure}

    \vskip\baselineskip 

    \begin{subfigure}[b]{0.9\textwidth}
        \centering
        \includegraphics[width=\textwidth]{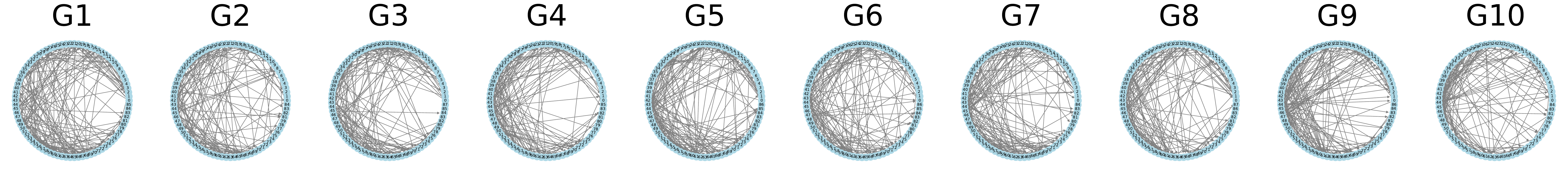}
        \caption{Target efficiency: 0\%-5\% percentile, Network Visualization} 
    \end{subfigure}

    \vskip\baselineskip 

    \begin{subfigure}[b]{0.9\textwidth}
        \centering
        \includegraphics[width=\textwidth]{reports/nips_lxy/figures/fig_controllable_gene/efficiency/adj_samples_efficiency_target=10.png}
        \caption{Target efficiency: 50\%-55\% percentile, Adjacency Matrix} 
    \end{subfigure}

    \vskip\baselineskip 

    \begin{subfigure}[b]{0.9\textwidth}
        \centering
        \includegraphics[width=\textwidth]{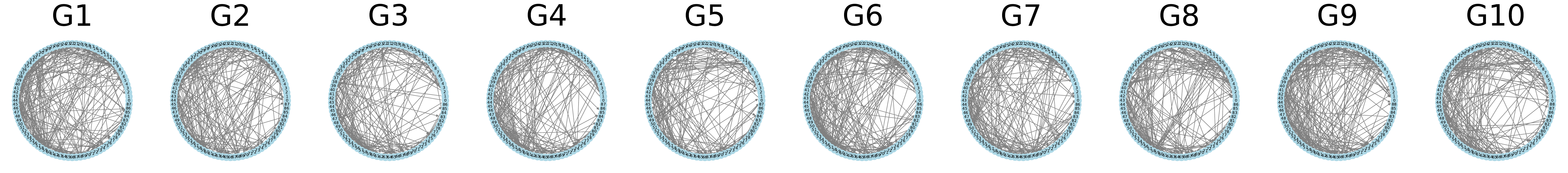}
        \caption{Target efficiency 50\%-55\% percentile, Network Visualization} 
    \end{subfigure}

    \vskip\baselineskip 

    \begin{subfigure}[b]{0.9\textwidth}
        \centering
        \includegraphics[width=\textwidth]{reports/nips_lxy/figures/fig_controllable_gene/efficiency/adj_samples_efficiency_target=20.png}
        \caption{Target efficiency: 95\%-100\% percentile, Adjacency Matrix} 
    \end{subfigure}

    \vskip\baselineskip 

    \begin{subfigure}[b]{0.9\textwidth}
        \centering
        \includegraphics[width=\textwidth]{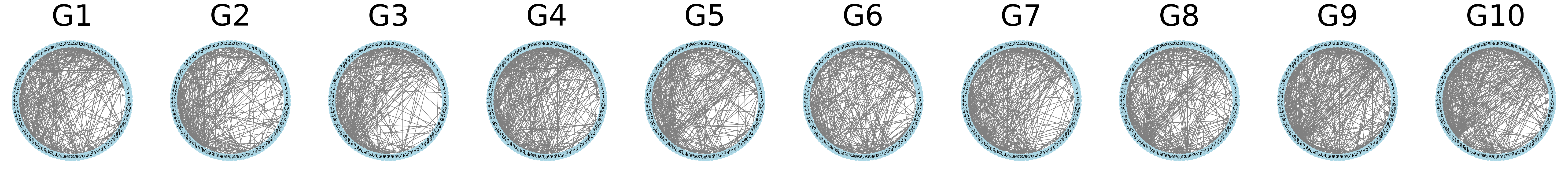}
        \caption{Target efficiency: 95\%-100\% percentile, Network Visualization} 
    \end{subfigure}

    \caption{Generated graph examples targeting different efficiency percentile ranges. For each target range, 10 graphs were randomly selected from 1000 graphs sampled and decoded. Pairs of adjacency matrices and corresponding network visualizations are shown for target efficiency in the 0\%-5\% (a, b), 50\%-55\% (c, d), and 95\%-100\% (e, f) percentile ranges of the dataset.} 
    \label{fig:control_efficiency}
\end{figure*}

\begin{figure*}[tbp]
    \centering
    \begin{subfigure}[b]{0.9\textwidth}
        \centering
        \includegraphics[width=\textwidth]{reports/nips_lxy/figures/fig_controllable_gene/transitivity/adj_samples_transitivity_target=0.png}
        \caption{Target transitivity: 0\%-5\% percentile, Adjacency Matrix} 
    \end{subfigure}

    \vskip\baselineskip 

    \begin{subfigure}[b]{0.9\textwidth}
        \centering
        \includegraphics[width=\textwidth]{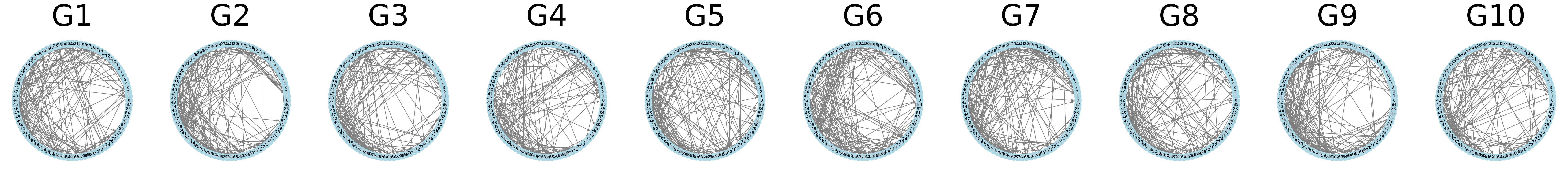}
        \caption{Target transitivity: 0\%-5\% percentile, Network Visualization} 
    \end{subfigure}

    \vskip\baselineskip 

    \begin{subfigure}[b]{0.9\textwidth}
        \centering
        \includegraphics[width=\textwidth]{reports/nips_lxy/figures/fig_controllable_gene/transitivity/adj_samples_transitivity_target=10.png}
        \caption{Target transitivity: 50\%-55\% percentile, Adjacency Matrix} 
    \end{subfigure}

    \vskip\baselineskip 

    \begin{subfigure}[b]{0.9\textwidth}
        \centering
        \includegraphics[width=\textwidth]{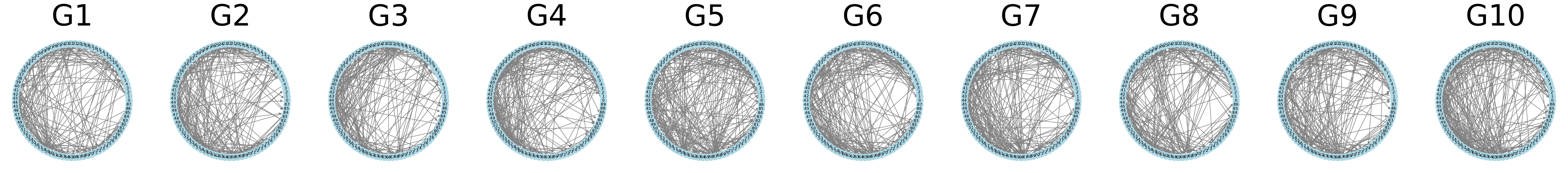}
        \caption{Target transitivity 50\%-55\% percentile, Network Visualization} 
    \end{subfigure}

    \vskip\baselineskip 

    \begin{subfigure}[b]{0.9\textwidth}
        \centering
        \includegraphics[width=\textwidth]{reports/nips_lxy/figures/fig_controllable_gene/transitivity/adj_samples_transitivity_target=20.png}
        \caption{Target transitivity: 95\%-100\% percentile, Adjacency Matrix} 
    \end{subfigure}

    \vskip\baselineskip 

    \begin{subfigure}[b]{0.9\textwidth}
        \centering
        \includegraphics[width=\textwidth]{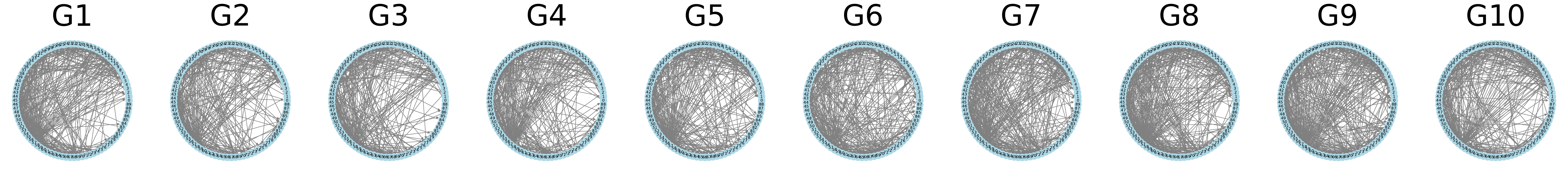}
        \caption{Target transitivity: 95\%-100\% percentile, Network Visualization} 
    \end{subfigure}

    \caption{Generated graph examples targeting different transitivity percentile ranges. For each target range, 10 graphs were randomly selected from 1000 graphs sampled and decoded. Pairs of adjacency matrices and corresponding network visualizations are shown for target transitivity in the 0\%-5\% (a, b), 50\%-55\% (c, d), and 95\%-100\% (e, f) percentile ranges of the dataset.} 
    \label{fig:control_transitivity}
\end{figure*}

\begin{figure*}[tbp]
    \centering
    \begin{subfigure}[b]{0.9\textwidth}
        \centering
        \includegraphics[width=\textwidth]{reports/nips_lxy/figures/fig_controllable_gene/clus/adj_samples_clustering_coefficient_target=0.png}
        \caption{Target clustering coefficient: 0\%-5\% percentile, Adjacency Matrix} 
    \end{subfigure}

    \vskip\baselineskip 

    \begin{subfigure}[b]{0.9\textwidth}
        \centering
        \includegraphics[width=\textwidth]{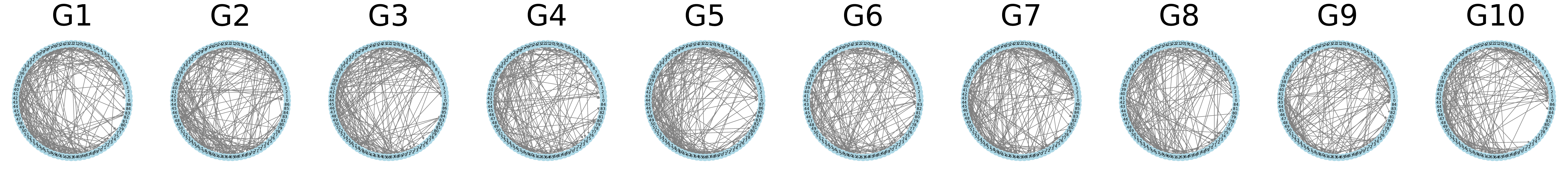}
        \caption{Target clustering coefficient: 0\%-5\% percentile, Network Visualization} 
    \end{subfigure}

    \vskip\baselineskip 

    \begin{subfigure}[b]{0.9\textwidth}
        \centering
        \includegraphics[width=\textwidth]{reports/nips_lxy/figures/fig_controllable_gene/clus/adj_samples_clustering_coefficient_target=10.png}
        \caption{Target clustering coefficient: 50\%-55\% percentile, Adjacency Matrix} 
    \end{subfigure}

    \vskip\baselineskip 

    \begin{subfigure}[b]{0.9\textwidth}
        \centering
        \includegraphics[width=\textwidth]{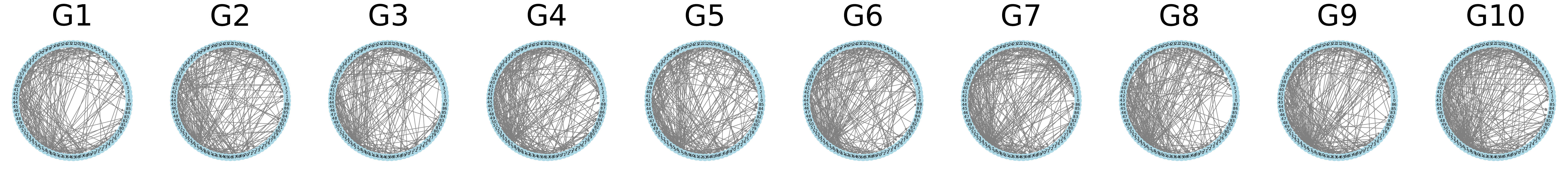}
        \caption{Target clustering coefficient 50\%-55\% percentile, Network Visualization} 
    \end{subfigure}

    \vskip\baselineskip 

    \begin{subfigure}[b]{0.9\textwidth}
        \centering
        \includegraphics[width=\textwidth]{reports/nips_lxy/figures/fig_controllable_gene/clus/adj_samples_clustering_coefficient_target=20.png}
        \caption{Target clustering coefficient: 95\%-100\% percentile, Adjacency Matrix} 
    \end{subfigure}

    \vskip\baselineskip 

    \begin{subfigure}[b]{0.9\textwidth}
        \centering
        \includegraphics[width=\textwidth]{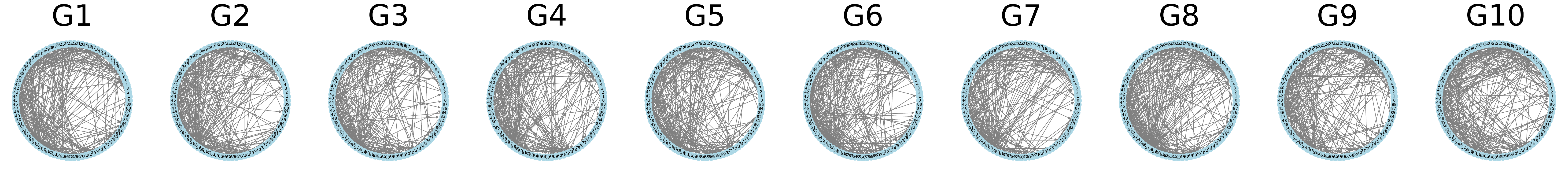}
        \caption{Target clustering coefficient: 95\%-100\% percentile, Network Visualization} 
    \end{subfigure}

    \caption{Generated graph examples targeting different clustering coefficient percentile ranges. For each target range, 10 graphs were randomly selected from 1000 graphs sampled and decoded. Pairs of adjacency matrices and corresponding network visualizations are shown for target clustering coefficient in the 0\%-5\% (a, b), 50\%-55\% (c, d), and 95\%-100\% (e, f) percentile ranges of the dataset.} 
    \label{fig:control_clustering_coefficient}
\end{figure*}

\begin{figure*}[tbp]
\centering
\begin{subfigure}[b]{0.9\textwidth}
\centering
\includegraphics[width=\textwidth]{reports/nips_lxy/figures/fig_controllable_gene/assortativity/adj_samples_assortativity_target=0.png}
\caption{Target assortativity: 0\%-5\% percentile, Adjacency Matrix} 
\end{subfigure}

\vskip\baselineskip 

\begin{subfigure}[b]{0.9\textwidth}
\centering
\includegraphics[width=\textwidth]{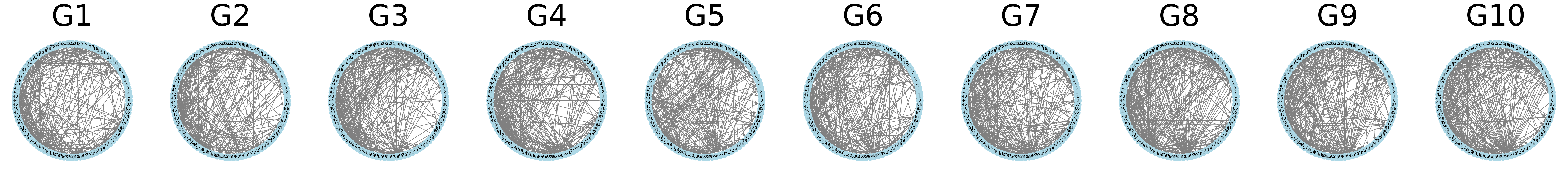} 
\caption{Target assortativity: 0\%-5\% percentile, Network Visualization} 
\end{subfigure}

\vskip\baselineskip 

\begin{subfigure}[b]{0.9\textwidth}
\centering
\includegraphics[width=\textwidth]{reports/nips_lxy/figures/fig_controllable_gene/assortativity/adj_samples_assortativity_target=10.png}
\caption{Target assortativity: 50\%-55\% percentile, Adjacency Matrix} 
\end{subfigure}

\vskip\baselineskip 

\begin{subfigure}[b]{0.9\textwidth}
\centering
\includegraphics[width=\textwidth]{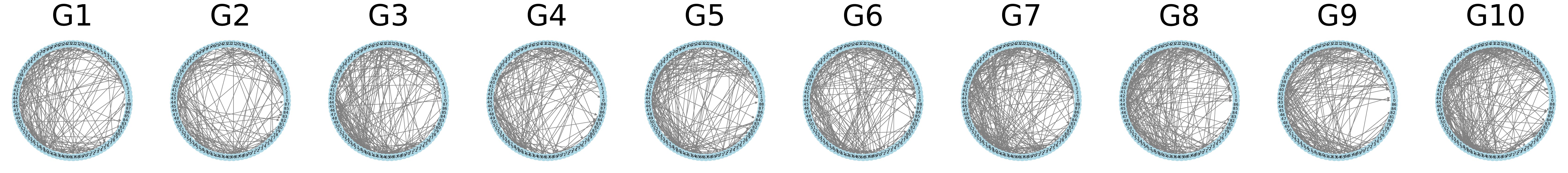}
\caption{Target assortativity 50\%-55\% percentile, Network Visualization} 
\end{subfigure}

\vskip\baselineskip 

\begin{subfigure}[b]{0.9\textwidth}
\centering
\includegraphics[width=\textwidth]{reports/nips_lxy/figures/fig_controllable_gene/assortativity/adj_samples_assortativity_target=20.png}
\caption{Target assortativity: 95\%-100\% percentile, Adjacency Matrix} 
\end{subfigure}

\vskip\baselineskip 

\begin{subfigure}[b]{0.9\textwidth}
\centering
\includegraphics[width=\textwidth]{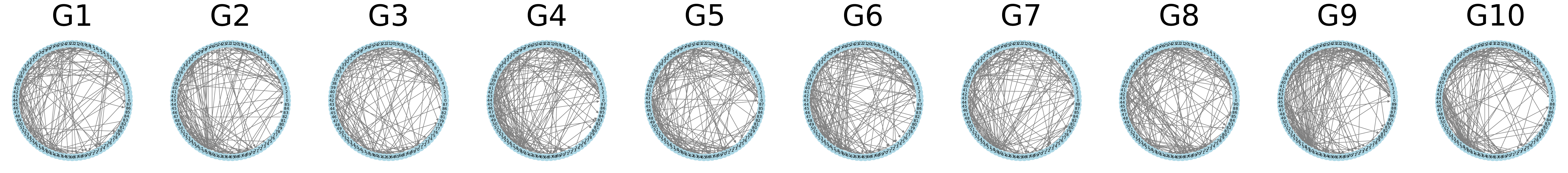}
\caption{Target assortativity: 95\%-100\% percentile, Network Visualization} 
\end{subfigure}

\caption{Generated graph examples targeting different assortativity percentile ranges. For each target range, 10 graphs were randomly selected from 1000 graphs sampled and decoded. Pairs of adjacency matrices and corresponding network visualizations are shown for target assortativity in the 0\%-5\% (a, b), 50\%-55\% (c, d), and 95\%-100\% (e, f) percentile ranges of the dataset.} 
\label{fig:control_assortativity}
\end{figure*}

\begin{figure*}[tbp]
\centering
\begin{subfigure}[b]{0.9\textwidth}
\centering
\includegraphics[width=\textwidth]{reports/nips_lxy/figures/fig_controllable_gene/modularity/adj_samples_modularity_target=0.png}
\caption{Target modularity: 0\%-5\% percentile, Adjacency Matrix} 
\end{subfigure}

\vskip\baselineskip 

\begin{subfigure}[b]{0.9\textwidth}
\centering
\includegraphics[width=\textwidth]{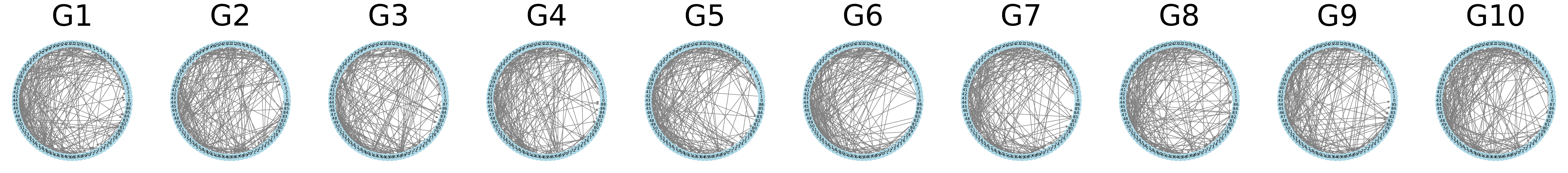} 
\caption{Target modularity: 0\%-5\% percentile, Network Visualization} 
\end{subfigure}

\vskip\baselineskip 

\begin{subfigure}[b]{0.9\textwidth}
\centering
\includegraphics[width=\textwidth]{reports/nips_lxy/figures/fig_controllable_gene/modularity/adj_samples_modularity_target=10.png}
\caption{Target modularity: 50\%-55\% percentile, Adjacency Matrix} 
\end{subfigure}

\vskip\baselineskip 

\begin{subfigure}[b]{0.9\textwidth}
\centering
\includegraphics[width=\textwidth]{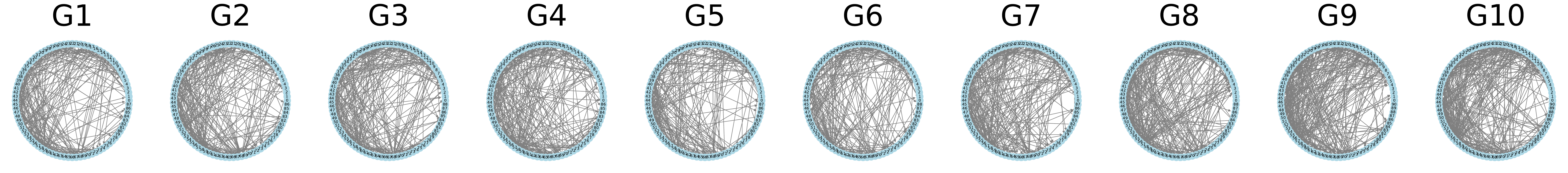}
\caption{Target modularity 50\%-55\% percentile, Network Visualization} 
\end{subfigure}

\vskip\baselineskip 

\begin{subfigure}[b]{0.9\textwidth}
\centering
\includegraphics[width=\textwidth]{reports/nips_lxy/figures/fig_controllable_gene/modularity/adj_samples_modularity_target=20.png}
\caption{Target modularity: 95\%-100\% percentile, Adjacency Matrix} 
\end{subfigure}

\vskip\baselineskip 

\begin{subfigure}[b]{0.9\textwidth}
\centering
\includegraphics[width=\textwidth]{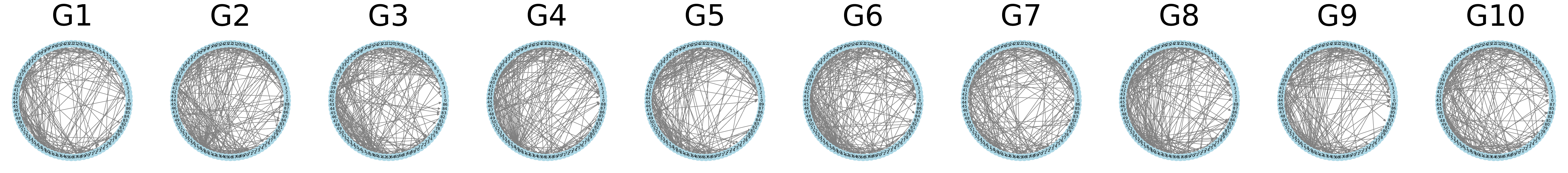}
\caption{Target modularity: 95\%-100\% percentile, Network Visualization} 
\end{subfigure}

\caption{Generated graph examples targeting different modularity percentile ranges. For each target range, 10 graphs were randomly selected from 1000 graphs sampled and decoded. Pairs of adjacency matrices and corresponding network visualizations are shown for target modularity in the 0\%-5\% (a, b), 50\%-55\% (c, d), and 95\%-100\% (e, f) percentile ranges of the dataset.} 
\label{fig:control_modularity}
\end{figure*}

Generated graph examples targeting different metric percentile ranges are shown in Figure~\ref{fig:control_mean_degree} (mean degree), Figure~\ref{fig:control_efficiency} (efficiency), Figure~\ref{fig:control_transitivity} (transitivity), Figure~\ref{fig:control_clustering_coefficient} (clustering coefficient), Figure~\ref{fig:control_assortativity} (assortativity) and Figure~\ref{fig:control_modularity} (modularity).

\section{Details of Reservoir Network Experiments}
\label{sec:detail_esn}

We employed a reservoir network architecture with a reservoir of $N=100$ units. We trained both the input weights ($W_{in}$) and output weights ($W_{out}$), while only the recurrent weights ($W$) within the reservoir were held fixed post-initialization. The neuron types were set to excitatory or inhibitory according to a fixed configuration sampled from the MICRONS dataset distributions, which remained constant across all experiments. The reservoir's connectivity matrix was either generated by our Variational Autoencoder (VAE) to produce structured patterns or formulated as a density-matched random graph for control comparisons. The initial recurrent weights were assigned values of $+1$ (from an excitatory neuron), $-1$ (from an inhibitory neuron), or $0$ (no connection), and the final matrix $W$ was scaled to a specified spectral radius. Both $W_{\text{in}}$ and $W_{\text{out}}$ were initialized with weights from a uniform distribution $U[-1, 1)$ and subsequently trained end-to-end using the Adam optimizer. This design allowed us to investigate the influence of fixed, structured connectivity while permitting the network to adapt its input and output mappings for each task.

\subsection{Copy Task}
\label{sec:detail_copy}
The copy memory task uses sequences of categorical inputs. Each input sequence has a total length of $T + 2t$ and is structured as follows:
\begin{itemize}
    \item The first $t$ tokens are one-hot vectors randomly chosen from a set of $N$ categories, representing the information to be memorized.
    \item These are followed by $T$ "blank" tokens (category 0), which create a delay period where the network must hold the information in memory.
    \item A delimiter token (category $N+1$) is then presented to signal the start of the recall phase.
    \item The sequence concludes with $t-1$ final blank tokens.
\end{itemize}
The corresponding target sequence is of the same length, consisting of blank tokens except for the last $t$ positions, which must reproduce the initial random sequence from the input. For our experiments with a 100-neuron reservoir network, we set the task parameters to $t=5$, $N=3$, and $T=10$.

The networks were trained for 60,000 epochs with a learning rate of 0.001. For the reservoir parameters, we configured the leaking rate to 0.3 and the spectral radius to 0.999.

\begin{figure*}[htbp]
    \centering 

    \begin{subfigure}[b]{0.48\textwidth}
        \centering
        \includegraphics[width=\linewidth]{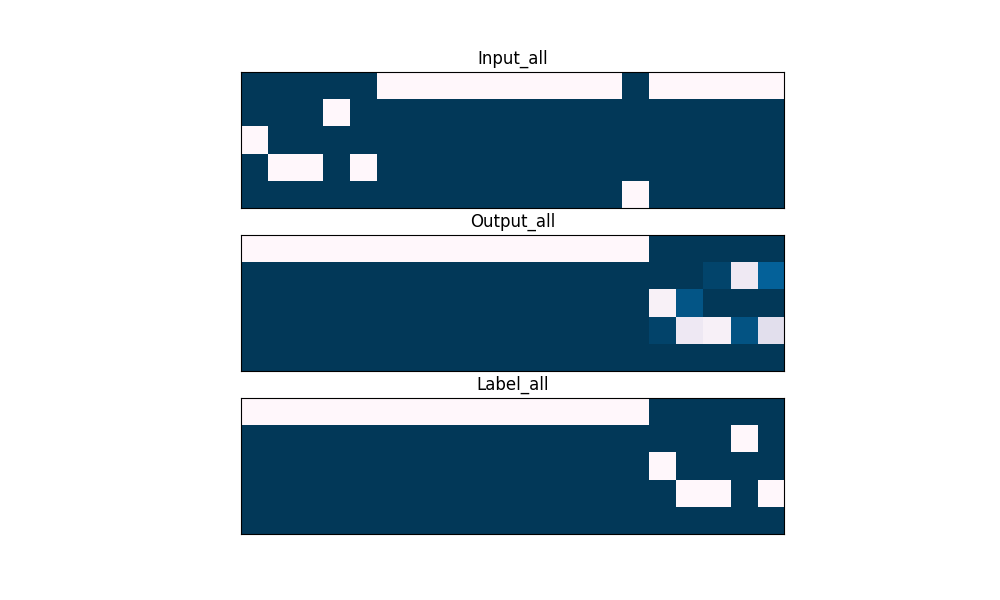}
        \subcaption{VAE-Generated Network (Target: Mean Degree)}
    \end{subfigure}
    \hfill 
    \begin{subfigure}[b]{0.48\textwidth}
        \centering
        \includegraphics[width=\linewidth]{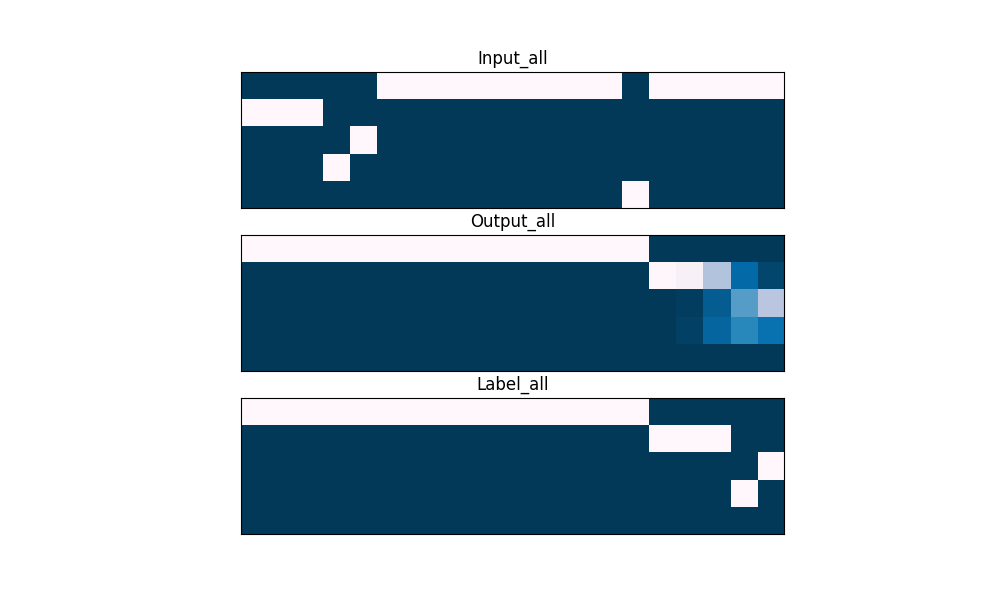}
        \subcaption{Random Network (Target: Mean Degree)}
    \end{subfigure}

    \vspace{1em} 

    \begin{subfigure}[b]{0.48\textwidth}
        \centering
        \includegraphics[width=\linewidth]{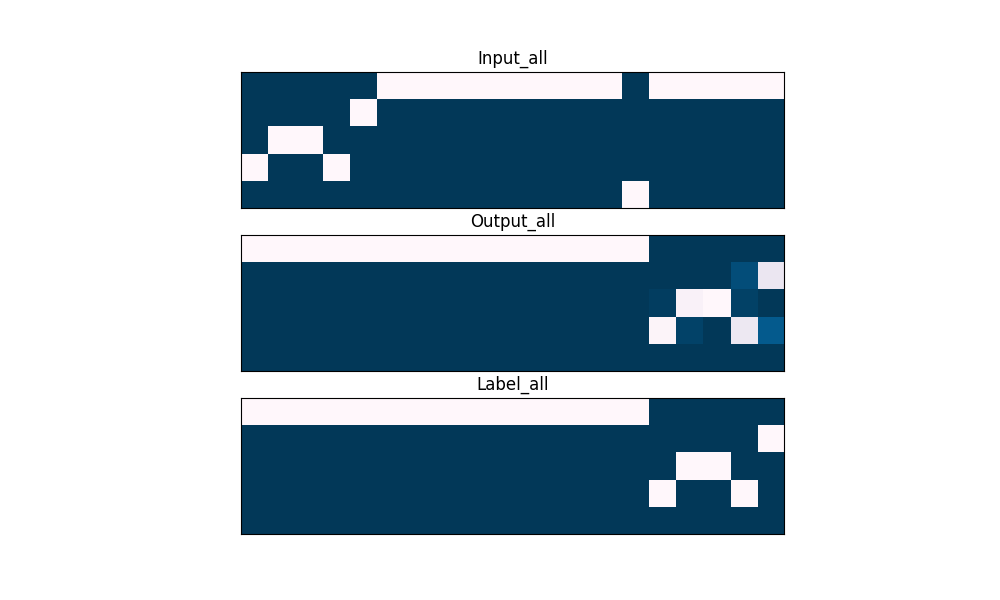}
        \subcaption{VAE-Generated Network (Target: Efficiency)}
    \end{subfigure}
    \hfill
    \begin{subfigure}[b]{0.48\textwidth}
        \centering
        \includegraphics[width=\linewidth]{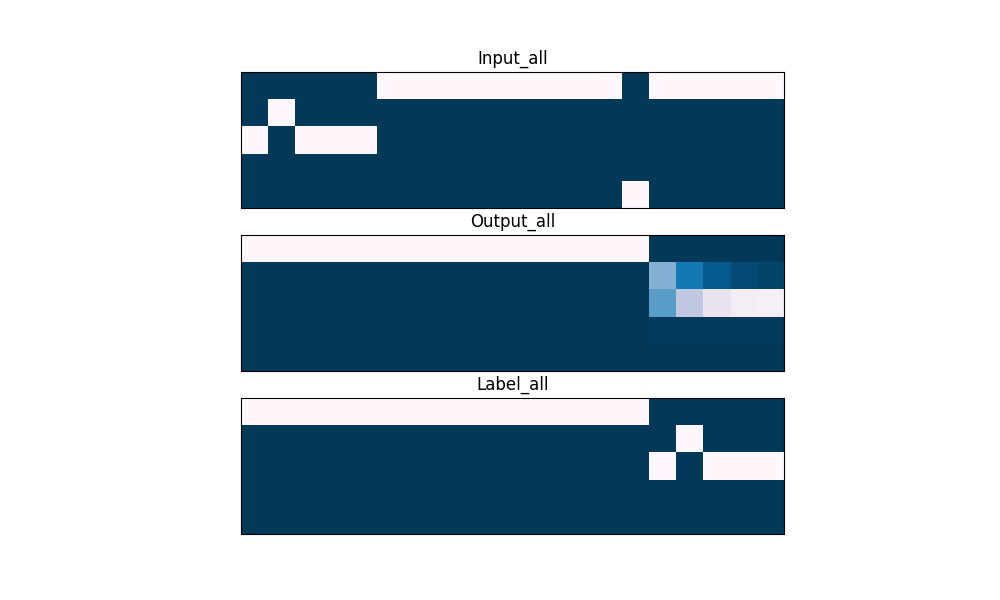} 
        \subcaption{Random Network (Target: Efficiency)}
    \end{subfigure}

    \vspace{1em} 

    \begin{subfigure}[b]{0.48\textwidth}
        \centering
        \includegraphics[width=\linewidth]{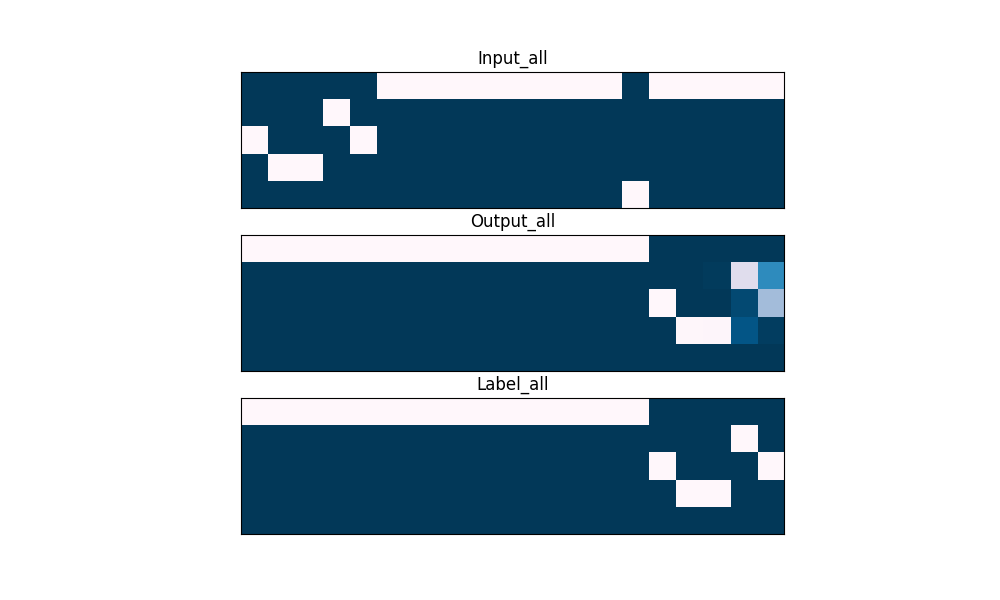} 
        \subcaption{VAE-Generated Network (Target: Transitivity)}
    \end{subfigure}
    \hfill
    \begin{subfigure}[b]{0.48\textwidth}
        \centering
        \includegraphics[width=\linewidth]{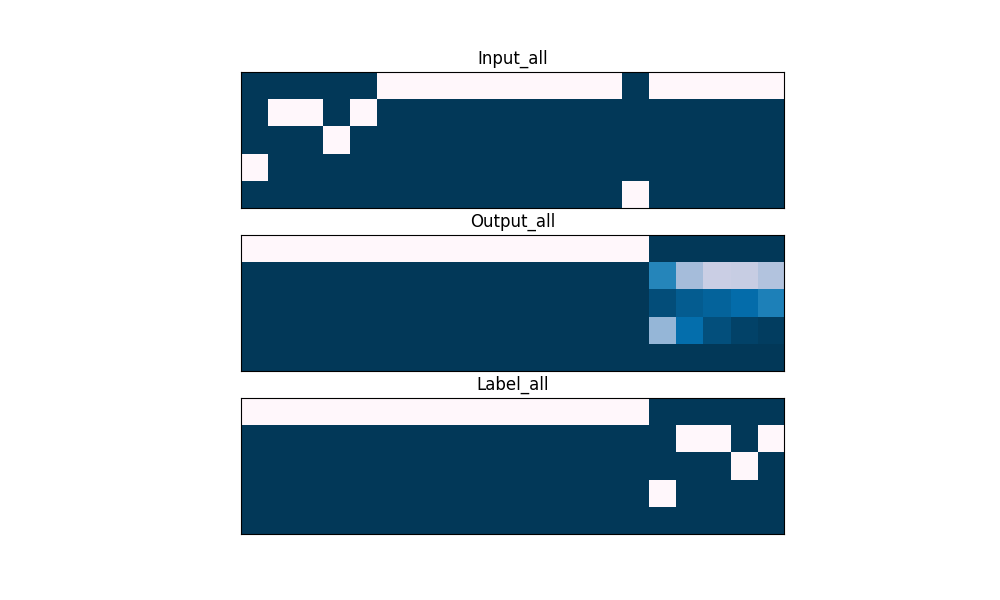}
        \subcaption{Random Network (Target: Transitivity)}
    \end{subfigure}

    \caption{Performance comparison of reservoir networks on the copy task using VAE-generated graphs (left column) versus density-matched random graphs (right column) for three target metrics: Mean Degree, Efficiency, and Transitivity. The comparison is continued in Figure~\ref{fig:copy_compare_part2}. Each subplot visualizes the input pattern to be memorized (top), the model's output (middle), and the ground truth (bottom).}
    \label{fig:copy_compare_part1}
\end{figure*}

\begin{figure*}[htbp]
    \centering 

    \begin{subfigure}[b]{0.48\textwidth}
        \centering
        \includegraphics[width=\linewidth]{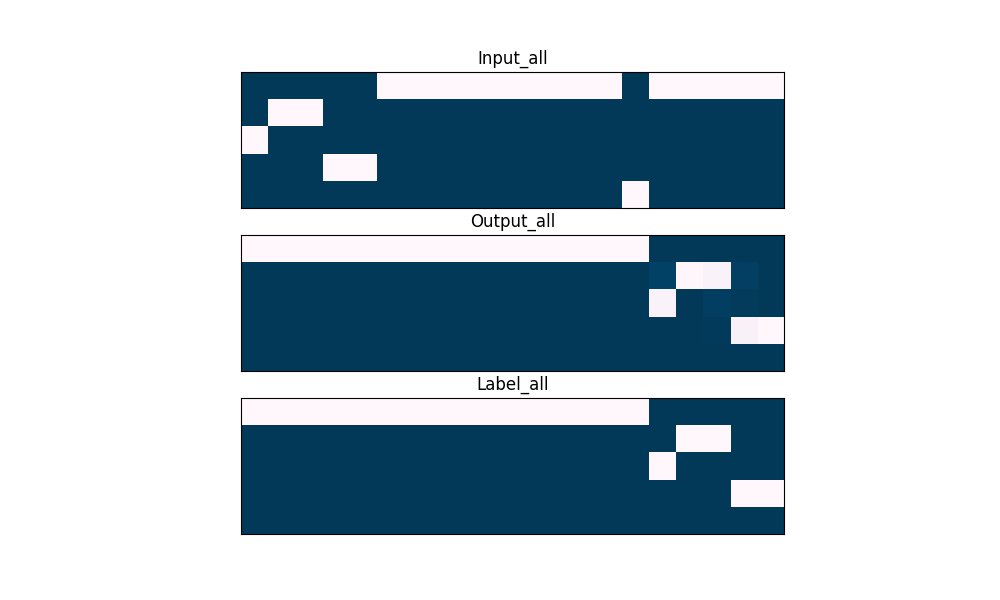}
        \subcaption{VAE-Generated Network (Target: Clustering Coefficient)}
    \end{subfigure}
    \hfill 
    \begin{subfigure}[b]{0.48\textwidth}
        \centering
        \includegraphics[width=\linewidth]{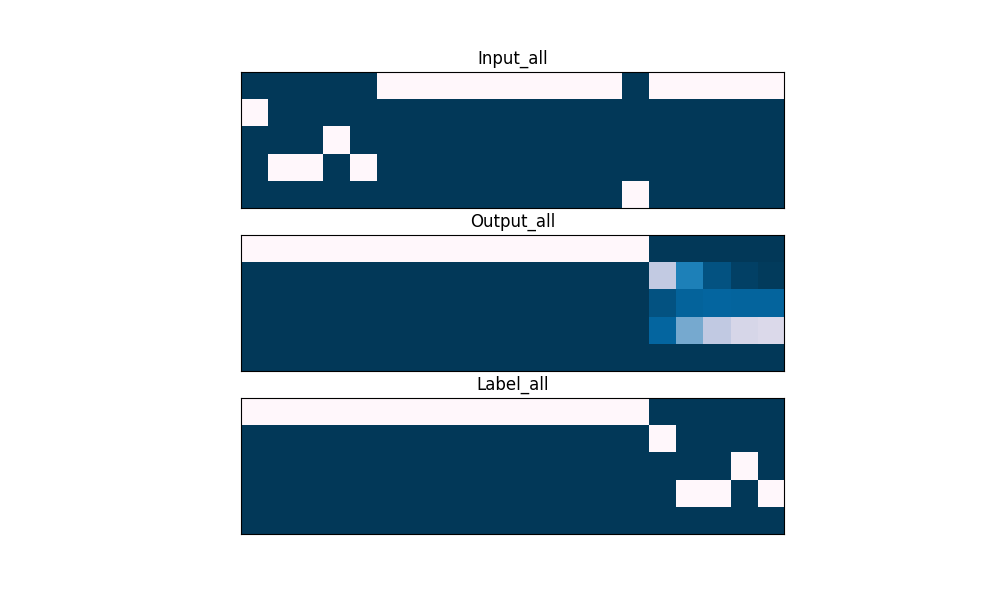}
        \subcaption{Random Network (Target: Clustering Coefficient)}
    \end{subfigure}

    \vspace{1em} 

    \begin{subfigure}[b]{0.48\textwidth}
        \centering
        \includegraphics[width=\linewidth]{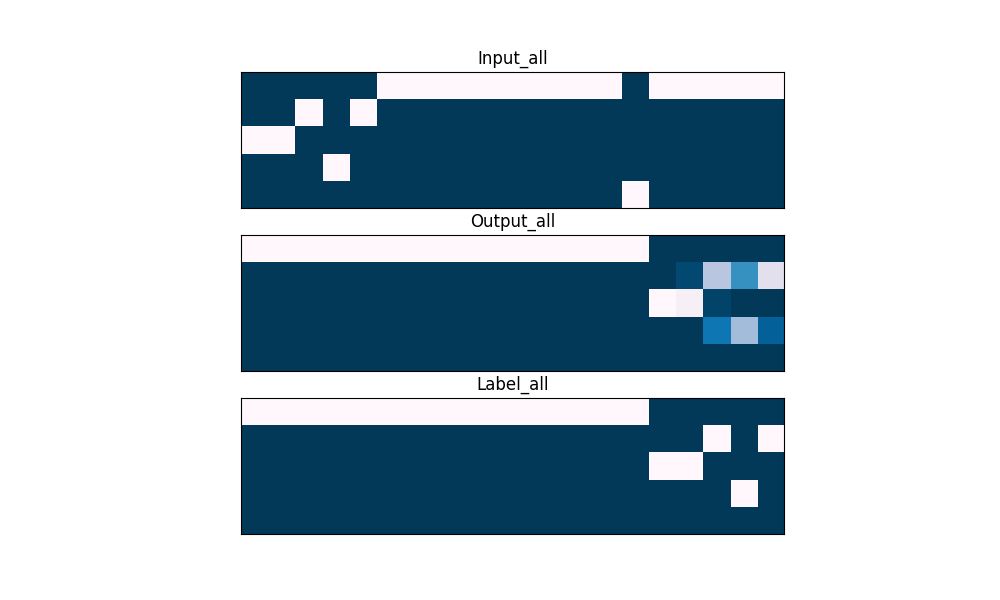}
        \subcaption{VAE-Generated Network (Target: Assortativity)}
    \end{subfigure}
    \hfill
    \begin{subfigure}[b]{0.48\textwidth}
        \centering
        \includegraphics[width=\linewidth]{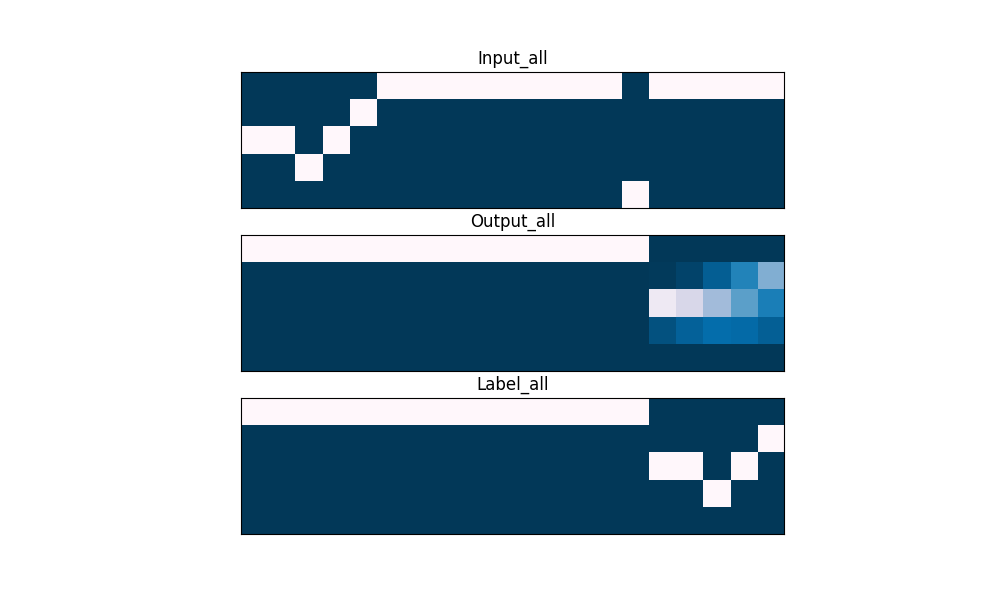} 
        \subcaption{Random Network (Target: Assortativity)}
    \end{subfigure}

    \vspace{1em} 

    \begin{subfigure}[b]{0.48\textwidth}
        \centering
        \includegraphics[width=\linewidth]{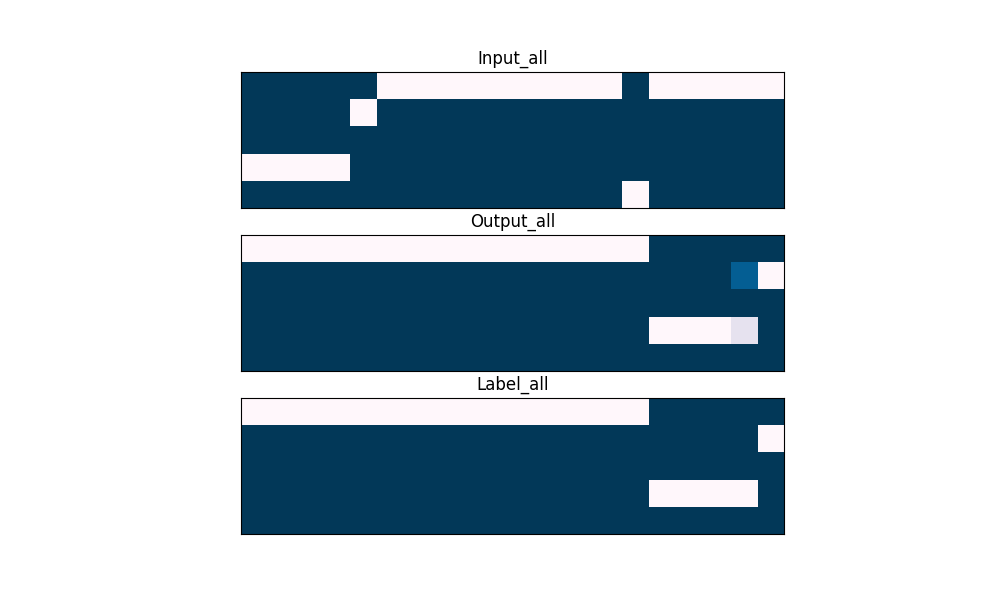} 
        \subcaption{VAE-Generated Network (Target: Modularity)}
    \end{subfigure}
    \hfill
    \begin{subfigure}[b]{0.48\textwidth}
        \centering
        \includegraphics[width=\linewidth]{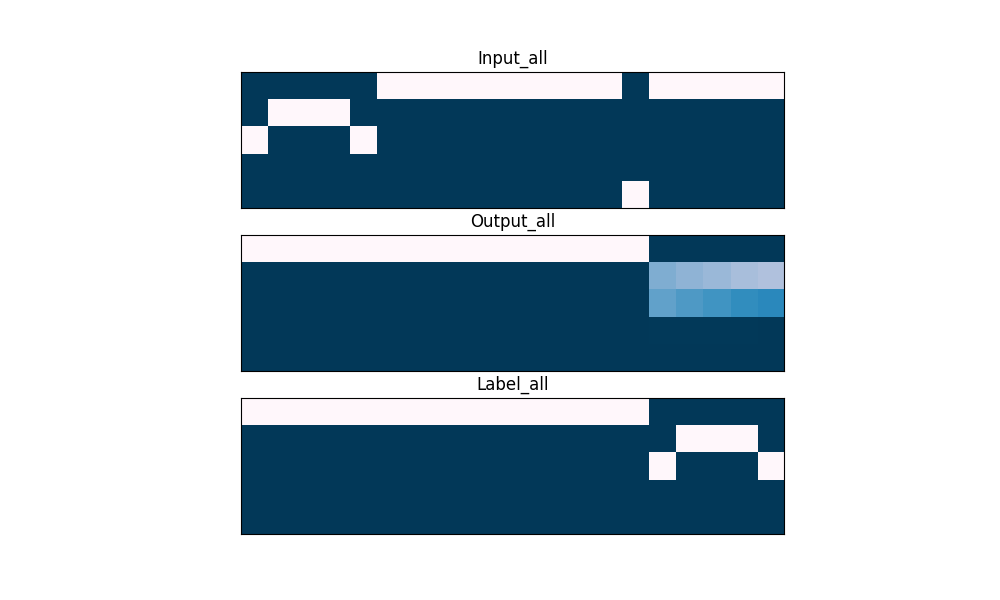}
        \subcaption{Random Network (Target: Modularity)}
    \end{subfigure}

    \caption{Performance comparison of reservoir networks on the copy task using VAE-generated graphs (left column) versus density-matched random graphs (right column). This figure shows the results for the last three target metrics: Clustering Coefficient, Assortativity, and Modularity. Each subplot visualizes the input pattern to be memorized (top), the model's output (middle), and the ground truth (bottom).}
    \label{fig:copy_compare_part2}
\end{figure*}

Figures~\ref{fig:copy_compare_part1} and~\ref{fig:copy_compare_part2} compare the copy task performance of reservoir connectivity built from VAE-generated graphs (left column) versus density-matched random graphs (right column). The VAE graphs were generated by targeting the 50th percentile for various structural metrics.

\subsection{Classification Task}
\label{sec:detail_class}
In the sequential MNIST (sMNIST) task, each image from the MNIST dataset is presented to the reservoir network row-by-row or column-by-column. The objective is to classify the digit based on the resulting hidden states of the reservoir network.

The networks were trained for 300 epochs with an initial learning rate of 0.003, which decayed by a factor of 1/3 every 100 epochs. For the reservoir parameters, we configured the leaking rate to 0.3 and the spectral radius to 0.999.

\begin{figure}[!htbp]
    \centering

    \begin{subfigure}{0.7\columnwidth}
        \centering
        \includegraphics[width=\linewidth]{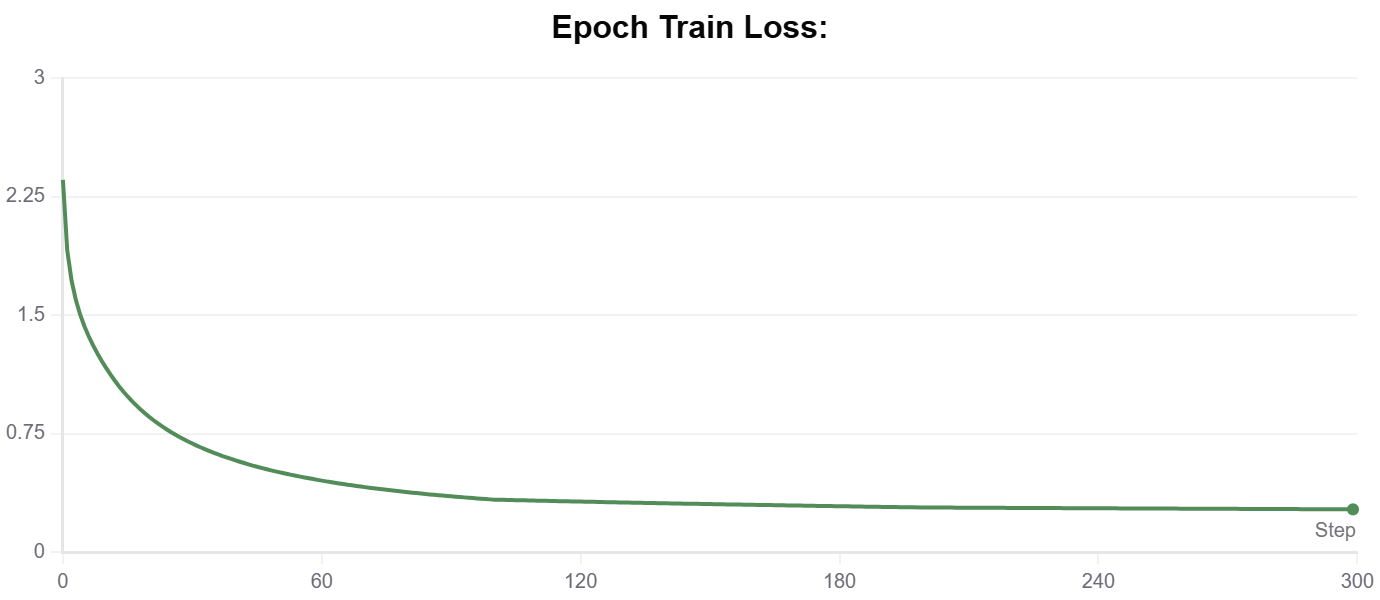}
        \caption{Train Loss}
    \end{subfigure}
    \vspace{1em} 

    \begin{subfigure}{0.7\columnwidth}
        \centering
        \includegraphics[width=\linewidth]{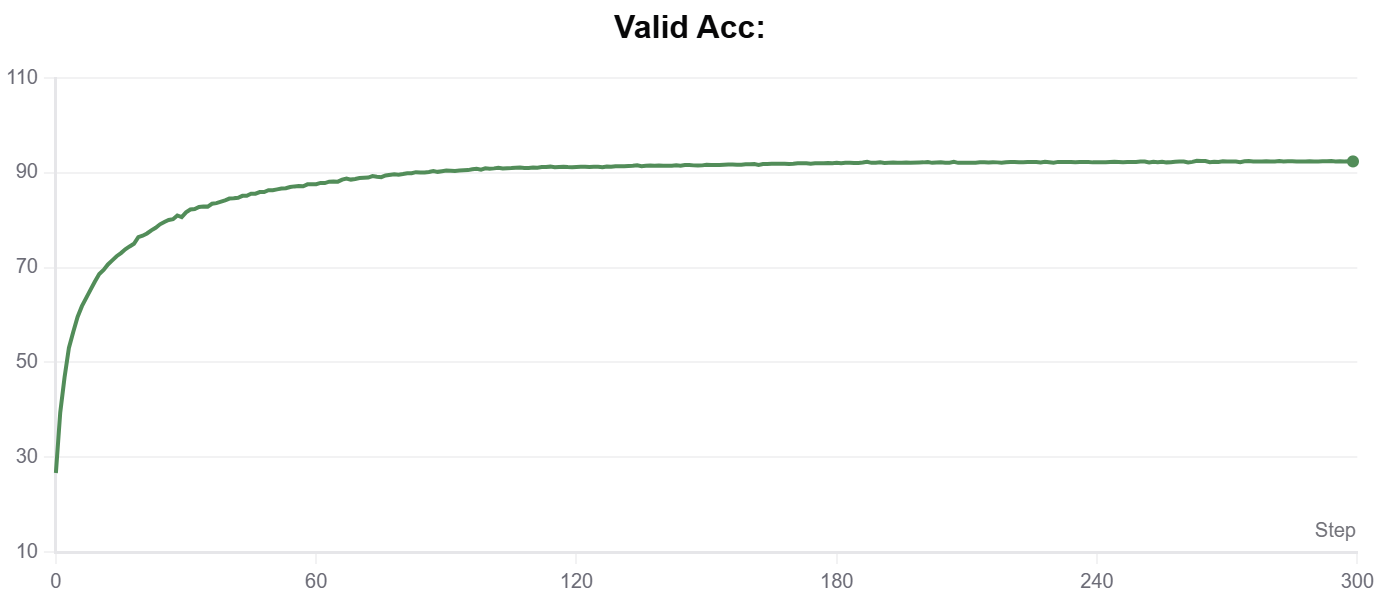}
        \caption{Valid Accuracy}
    \end{subfigure}
    \vspace{1em}

    \begin{subfigure}{0.7\columnwidth}
        \centering
        \includegraphics[width=\linewidth]{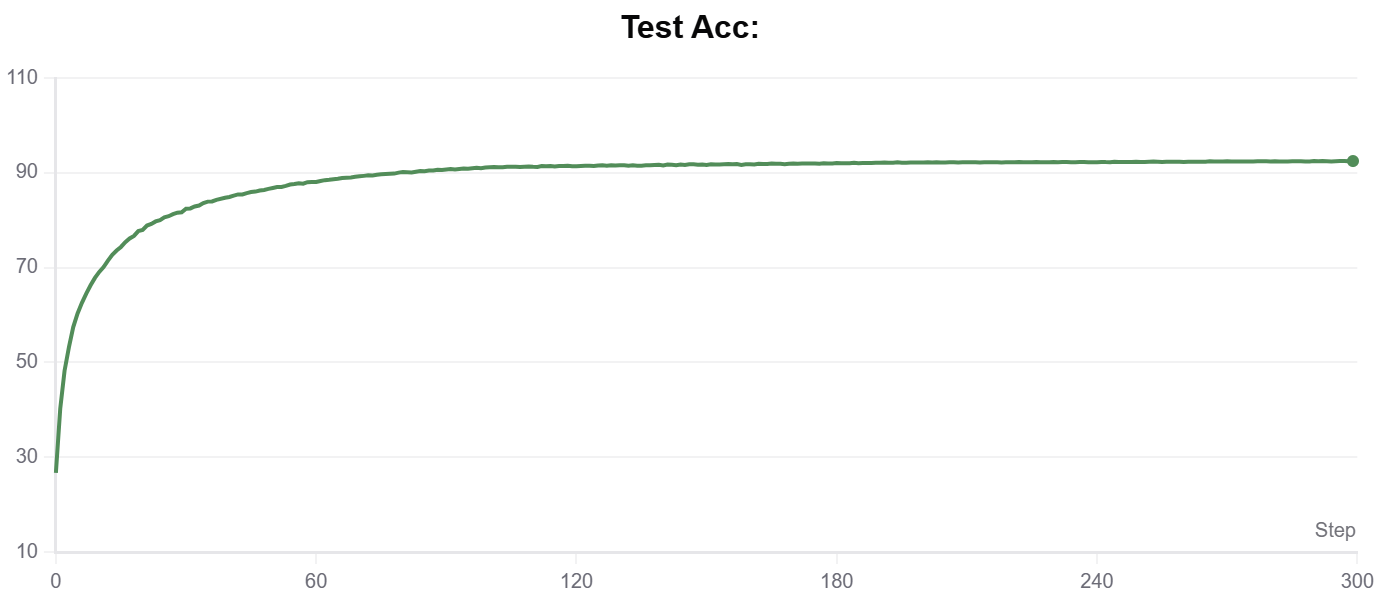}
        \caption{Test Accuracy}
    \end{subfigure}

    \caption{Example training and testing curves for the sMNIST task (Target metric: Transitivity, Value: 25th percentile).}
    \label{fig:sMNIST_curves}
\end{figure}

Figure~\ref{fig:sMNIST_curves} shows example training and testing curves for the sMNIST task, where the objective is the 25th percentile of the transitivity metric.

\section{Another Example of Our Research Paradigm: Studies on FlyWire Dataset}
\label{sec:fly}

To validate our approach on more diverse data, we conducted analogous experiments on the FlyWire~\cite{dorkenwald2022flywire} dataset, a Drosophila (fruit fly) connectome. In this configuration, the graph global encoder and the node feature decoder within the VAE were substituted with Multilayer Perceptrons (MLPs). The node types are considered by grouping neurons by their neurotransmitter types. The evaluation of the generation results is presented in Table~\ref{table:1}.

\begin{table}[!htbp]
\centering
\begin{threeparttable}
\begin{tabular}{lcccc}
\toprule
\multirow{2}{*}{\textbf{Method}} & 
\multicolumn{4}{c}{\textbf{FlyWire}}  \\
\cmidrule(lr){2-5} 
& \textbf{Deg.$\downarrow$} & \textbf{Clus.$\downarrow$} & \textbf{Orbit$\downarrow$}  & \textbf{Avg.$\downarrow$}\\
\midrule
EDGE & 0.009 & 0.099 & 0.038 & \textbf{0.048}\\
DisCo & 0.141 & 0.552 & 0.377 & 0.356\\
GDSS & 0.054 & 0.633 & 0.702 & 0.463 \\
GruM & $\boldsymbol{0.002}$ & 0.165 & $\boldsymbol{0.035}$ & 0.067\\
\midrule
Ours & $\boldsymbol{0.002}$ & $\boldsymbol{0.056}$ & 0.129 & 0.062\\
\bottomrule
\end{tabular}
\caption{Generation results on FlyWire. We compute the Maximum Mean Discrepancy (MMD) between generated and test graphs for three graph features, with the best results shown in bold.}
\label{table:1}
\end{threeparttable}
\end{table}

The model shows particularly strong performance in clustering coefficient, surpassing all comparative methods. This metric reflects the tightness of local circuit connections, and its high fidelity demonstrates our framework's effectiveness in capturing the modular properties of columnar functional units in biological neural networks. In contrast, the reconstruction accuracy for orbit counts is relatively lower, which aligns with expectations: global topological features, being influenced by more stochastic factors, inevitably suffer greater information loss during low-dimensional compression.

\begin{figure*}[!htbp]
    \centering
    \begin{subfigure}[t]{0.155\textwidth}
        \includegraphics[width=\linewidth]{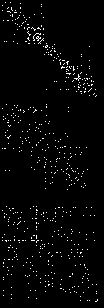}
        \caption{\scriptsize Origin}
    \end{subfigure}
    \hspace{-0.5em}
    \begin{subfigure}[t]{0.155\textwidth}
        \includegraphics[width=\linewidth]{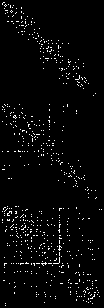}
        \caption{\scriptsize Ours}
    \end{subfigure}
    \hspace{-0.5em}
    \begin{subfigure}[t]{0.155\textwidth}
        \includegraphics[width=\linewidth]{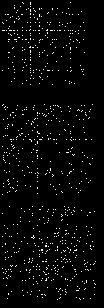}
        \caption{\scriptsize EDGE}
    \end{subfigure}
    \hspace{-0.5em}
    \begin{subfigure}[t]{0.155\textwidth}
        \includegraphics[width=\linewidth]{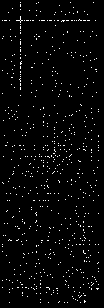}
        \caption{\scriptsize GruM}
    \end{subfigure}
    \hspace{-0.5em}
    \begin{subfigure}[t]{0.155\textwidth}
        \includegraphics[width=\linewidth]{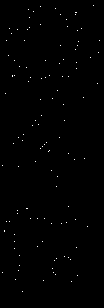}
        \caption{\scriptsize GDSS}
    \end{subfigure}
    \hspace{-0.5em}
    \begin{subfigure}[t]{0.155\textwidth}
        \includegraphics[width=\linewidth]{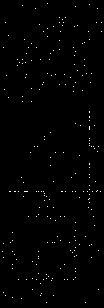}
        \caption{\scriptsize DisCo}
    \end{subfigure}
    
    \caption{Generation results on FlyWire in Adjacency Matrix Format. (a) Original: Shows the sampled graphs from the Drosophila brain connectome with three distinct divergence levels. (b) Ours: Our model's generation results, also exhibiting three divergence levels. (c)-(d) EDGE and GruM results: While performing well on high-divergence graphs, they lack medium and low-divergence samples, indicating limited diversity. (e)-(f) GDSS and DisCo results: Their main limitations are insufficient edge generation and similarly constrained diversity.}
    \label{fig:gen}
\end{figure*}

In Figure~\ref{fig:gen}, we present the generation results of our model alongside comparative models on the FlyWire dataset. The results demonstrate that our model exhibits superior performance in generation diversity compared to baseline methods.

This performance differentiation reveals intrinsic characteristics of our method: while the information bottleneck causes slight distortion in global topology, it effectively filters noise and makes key biological patterns (e.g., local clustering) more salient in the latent space. Notably, although EDGE achieves the lowest average MMD, its high-dimensional representation lacks explicit structural constraints, making subsequent causal analysis substantially more challenging to implement. Our method, through careful balancing of reconstruction accuracy and dimensional compression, establishes interpretable mappings between latent variables and generated graph structural features, providing a solid foundation for analyzing brain connectomes.

In addition to this, we also observe the reconstruction performance of our model on FlyWire.

\begin{table*}[htbp]
\centering

\begin{threeparttable}

\begin{tabular}{lcccc}
\toprule
\multirow{2}{*}{\textbf{Method}} &
\multicolumn{4}{c}{\textbf{FlyWire}} \\  
\cmidrule(lr){2-5}  
& \textbf{Node Acc.$\uparrow$} & \textbf{Node F1$\uparrow$} & \textbf{Edge Acc.$\uparrow$} & \textbf{Edge AUC$\uparrow$}\\ 
\midrule
Ours & 0.982 & 0.979 & 0.978 & 0.959\\ 
\bottomrule
\end{tabular}
\caption{Reconstruction performanc on FlyWire. We evaluate (1) Edge reconstruction performance using AUC and accuracy between original and reconstructed adjacency matrices, and (2) Node category recovery using classification accuracy and F1-score comparing original versus predicted node categories.}
\label{table:2}
\end{threeparttable}
\end{table*}

The reconstruction results in Table~\ref{table:2} demonstrate our model's ability to reasonably recover both node categories and edge connectivity, though with some expected information loss. We achieve 0.982 accuracy in node classification and 0.978 edge reconstruction accuracy, with slightly lower performance in edge AUC (0.959), suggesting the model captures major structural patterns while missing some finer connection details. These metrics reflect the inherent trade-offs of our approach - while the information bottleneck causes some reconstruction imperfection, it enables the low-dimensional analysis that is central to our framework. The performance is encouraging given this constraint, though we recognize these results would benefit from comparison with standard reconstruction baselines in future work. The modest differences between node and edge metrics may represent the necessary compromise between reconstruction fidelity and analytical tractability that defines our method's design philosophy.

\begin{figure}[!htbp]
\centering
\includegraphics[width=\textwidth]{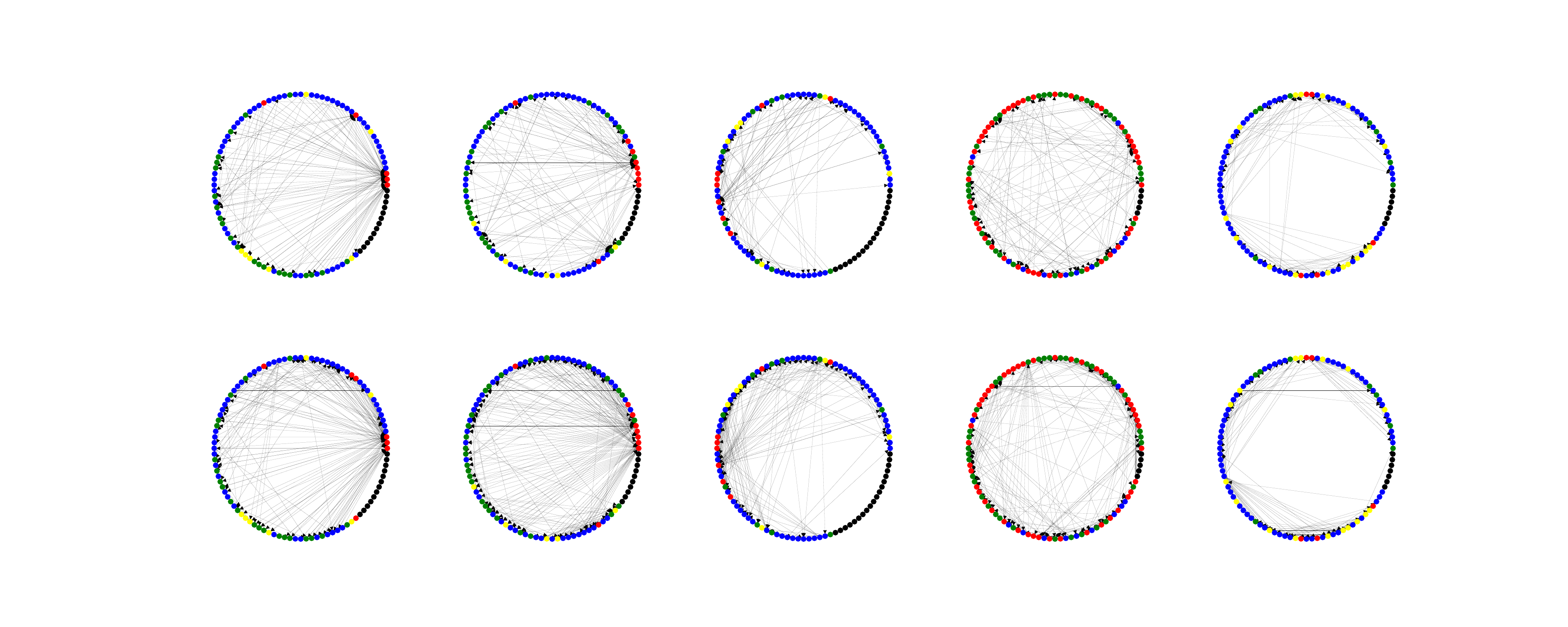}
\caption{Graph reconstruction results: Original samples (top) and their reconstructed counterparts (bottom), with node colors representing type categories.}
\label{fig:recon_fly}
\end{figure}

Figure~\ref{fig:recon_fly} demonstrates the model's graph reconstruction capability by comparing original connectome samples (top row) with their reconstructed counterparts (bottom row). The visual comparison shows that the model successfully preserves the overall topological structure and node type distributions (represented by consistent color patterns), while some subtle differences in local connectivity can be observed upon closer inspection. The reconstructions maintain the characteristic clustering patterns of neural circuits, though with minor variations in edge density and connection specificity that reflect the expected information loss through our model's bottleneck architecture. These results align quantitatively with the reconstruction metrics reported in Table~\ref{table:2}, providing visual confirmation that while the model captures essential connectome features, perfect reconstruction remains challenging due to our framework's emphasis on dimensional reduction and interpretability over absolute fidelity. The preserved node category assignments (colors) are particularly noteworthy, suggesting the model reliably maintains neuron type information during encoding-decoding.

\clearpage

\end{document}